\documentclass[12pt]{article}
\usepackage[utf8]{inputenc}
\usepackage[margin=1in]{geometry}
\usepackage{setspace}
\usepackage{amsmath}
\usepackage{amssymb}
\usepackage{hyperref}
\usepackage{graphicx}
\usepackage{caption}
\usepackage{natbib}
\usepackage{subcaption}
\usepackage{color}
\usepackage{float}
\newcommand{\R}{R}

\renewcommand{\i}{i}

\newcommand{\deleq}{\stackrel{\Delta}{=}}

\newcommand{\loadep}{\delta^1}

\newcommand{\qi}{q_\i}

\newcommand{\rhothree}{\rho^3}
\newcommand{\qit}{\qi^\t}
\newcommand{\alphas}{\alpha^\s}
\newcommand{\betas}{\beta^\s}

\newcommand{\Potentials}{\Psi^s}

\newcommand{\mi}{-\i}

\newcommand{\Rup}{\overline{R}}
\newcommand{\rhoone}{\rho^1}
\newcommand{\rhotwo}{\rho^2}
\newcommand{\Rlow}{\underline{R}}

\newcommand{\Qmi}{Q_{-i}}
\newcommand{\hmax}{h^{max}}
\newcommand{\hmin}{h^{min}}

\newcommand{\chat}{\c}

\newcommand{\Sequiv}{\S^{*}}
\newcommand{\qsran}{q^{*}}
\newcommand{\qdiff}{\tilde{\q}}
\newcommand{\usi}{u^\s_{\i}}
\newcommand{\FP}{\Omega}

\newcommand{\qtk}{\q^{\tk}}
\newcommand{\tk}{k^j}
\renewcommand{\k}{t}
\newcommand{\limitq}{\bar{\Q}}
\newcommand{\neighinitheta}{N_{\thetaep}}
\newcommand{\neighiniload}{N_{\loadep}}
\newcommand{\neighinftheta}{N_{\thetaepfin}}
\newcommand{\neighinfload}{N_{\loadepfin}}

\newcommand{\qj}{q^j}

\newcommand{\Sbar}{\S\setminus [\thetabar]}
\newcommand{\q}{q}

\newcommand{\Q}{Q}
\newcommand{\Qi}{Q_\i}

\newcommand{\thetaep}{\epsilon^1}
\newcommand{\Uhat}{\hat{U}}
\newcommand{\loaddelta}{\delta}

\newcommand{\thetaepfin}{\bar{\epsilon}}
\newcommand{\loadepfin}{\epsilon_{\load}}

\newcommand{\shat}{\hat{\s}}

\newcommand{\atbar}{\bar{a}^\t}

\newcommand{\qmi}{q_{-\i}}
\newcommand{\gbr}{h}

\newcommand{\Qbar}{\bar{\Q}}

\newcommand{\BRi}{\mathrm{BR}_i}
\newcommand{\sbar}{\bar{\s}}
\newcommand{\EQ}{\mathrm{EQ}}

\newcommand{\G}{G}

\newcommand{\gibr}{h_{\i}}

\newcommand{\deltahat}{\hat{\delta}}
\newcommand{\epsilonhat}{\hat{\epsilon}}

\renewcommand{\loadepfin}{\bar{\delta}}
\renewcommand{\loaddelta}{\delta}

\renewcommand{\shat}{\hat{\s}}

\newcommand{\Costhis}{Y^{\t-1}}
\newcommand{\Loadhis}{Q^{\t-1}}

\newcommand{\qbar}{\bar{\q}}

\newcommand{\phibar}{\phi}
\newcommand{\cbar}{\c}
\newcommand{\thetasran}{\theta^{*}}

\newcommand{\gwe}{g}

\newcommand{\load}{x}
\newcommand{\sran}{s^{*}}

\renewcommand{\(}{\left(}
\renewcommand{\)}{\right)}
\renewcommand{\c}{y}

\newcommand{\at}{a^\t}

\newcommand{\phis}{\phi^\s}
\renewcommand{\t}{k}

\renewcommand{\j}{j}

\newcommand{\pro}{\mathrm{Pr}}

\newcommand{\thetat}{\theta^\t}
\newcommand{\thetaj}{\theta^{j+1}}
\newcommand{\cj}{\cbar^j}

\newcommand{\thetatmap}{\theta_{M}^\t}

\newcommand{\Ai}{N_i}
\newcommand{\ai}{n_i}
\newcommand{\actiont}{n^\t}

\newcommand{\ait}{\at_\i}

\newcommand{\thetazero}{\theta^1}
\newcommand{\etaj}{\log\(\frac{\phibar^{\s}(\cj|\q^j)}{\phibar^{\sran}(\cj|\q^j)}\)}

\newcommand{\etabar}{\log\(\frac{\phibar^\s(\cbar|\qbar)}{\phibar^{\sran}(\cbar|\qbar)}\)}

\newcommand{\I}{I}

\newcommand{\iter}{t}

\newcommand{\n}{n}
\newcommand{\N}{N}

\newcommand{\epis}{\epsilon_{\i}^\s}

\newcommand{\abar}{\bar{a}}
\newcommand{\Qhat}{N_{\deltahat}\(\EQ(\thetabar)\)}
\newcommand{\T}{K}
\newcommand{\gwei}{\gwe_\i}
\newcommand{\thetatone}{\theta^{\t+1}}

\newcommand{\Tstar}{\T^{*}}

\newcommand{\BR}{\mathrm{BR}}

\newcommand{\Thetahat}{N_{\epsilonhat}{\(\thetabar\)}}

\newcommand{\Chibarj}{\eta^j}
\newcommand{\s}{s}

\renewcommand{\S}{S}

\newcommand{\thetabar}{\bar{\theta}}
\newcommand{\loadbar}{\bar{x}}
\newcommand{\ct}{\c^\t}

\newcommand{\loadj}{\load^j}
\newcommand{\qt}{\q^\t}

\newcommand{\alphabar}{\bar{\alpha}}

\newcommand{\betabar}{\bar{\beta}}
\usepackage{amsthm}
\usepackage{dsfont}
\usepackage{mathrsfs}
\usepackage{lmodern}

\newtheorem{lemma}{Lemma}
\newtheorem{definition}{Definition}
\newtheorem{theorem}{Theorem}
\newtheorem{proposition}{Proposition}
\newtheorem{corollary}{Corollary}

\newtheorem{example}{Example}
\DeclareMathOperator*{\argmin}{arg\,min}
\providecommand{\keywords}[1]
{
  \small	
  \textbf{\textit{Keywords---}} #1
}
\DeclareMathOperator*{\argmax}{arg\,max}
\newcommand*{\QEDA}{\hfill\ensuremath{\square}}
\usepackage{appendix}
\title{Multi-agent Bayesian Learning with Adaptive Strategies: Convergence and Stability\thanks{This version: October 2020.}}
\author{Manxi Wu, Saurabh Amin, and Asuman Ozdaglar\thanks{M. Wu is with the Institute for Data, Systems, and Society, S. Amin is with Laboratory for Information and Decision Systems, A. Ozdaglar is with Department of Electrical Engineering and Computer Science, Massachusetts Institute of Technology (MIT), Cambridge, MA, USA, \{manxiwu, amins, asuman\}@mit.edu}}
\date{}

\begin{document}

\maketitle
\begin{abstract}
We study learning dynamics induced by strategic agents who repeatedly play a game with an unknown payoff-relevant parameter. In each step, an information system estimates a belief distribution of the parameter based on the players' strategies and realized payoffs using Bayes’s rule. Players adjust their strategies by accounting for an equilibrium strategy or a best response strategy based on the updated belief. We prove that beliefs and strategies converge to a fixed point with probability 1. We also provide conditions that guarantee local and global stability of fixed points. Any fixed point belief consistently estimates the payoff distribution given the fixed point strategy profile. However, convergence to a complete information Nash equilibrium is not always guaranteed. We provide a sufficient and necessary condition under which fixed point belief recovers the unknown parameter. We also provide a sufficient condition for convergence to complete information equilibrium even when parameter learning is incomplete.

\end{abstract}
\keywords{Bayesian learning, Learning in games, Stochastic dynamics, Convergence and Stability Analysis}
%
\section{Problem Setup}\label{sec:model}
We study learning dynamics induced by strategic players belong to a finite set $\I$ who repeatedly play a game $\G$ for an infinite number of steps. Players' payoffs in the game $\G$ depend on an \emph{unknown} (scalar or vector) parameter $\s$ in a finite set $\S$. The true parameter is denoted $\sran \in \S$. Learning is mediated by an information system which repeatedly updates and broadcasts an estimate of the belief $\theta=\(\theta(\s)\)_{\s \in \S} \in \Delta(\S)$ to all players, where $\theta(\s)$ is the probability of parameter $\s$. 

In game $\G$, the strategy of each player $\i \in \I$ is a finite dimensional vector $\qi$ in a convex and continuous strategy set $\Qi$.\footnote{In Appendix \ref{apx:extension}, we extend our learning model to games with discrete strategy sets and players choosing mixed strategies.} The players' strategy profile is denoted $\q=\(\qi\)_{\i \in \I} \in \Q \deleq \prod_{\i \in \I} \Qi$. The payoff of each player is realized randomly from a probability distribution. Specifically, the distribution of players' payoffs $\c=\(\c_i\)_{\i\in \I}$ depends on strategy profile $\q$ and parameter $\s$. Without loss of generality, we write the player $\i$'s payoff $\c_{\i}$ for any $\s \in \S$ as the summation of an average payoff function, denoted  $\usi(\q)$, and a noise term $\epis(\q)$ with zero mean: 
\begin{align}\label{eq:utility}
    \c_i=\usi(\q)+\epis(\q).
\end{align}
The noise terms $\(\epis(\q)\)_{\i \in \I}$ can be correlated across players. Let $\phis(\c|\q)$ denote the probability density function of payoff vector $\c$ for any strategy profile $\q \in \Q$ and any parameter $\s \in \S$. We assume that $\phis(\c|\q)$ is continuous in $\q$ for all $\s \in \S$. 

Our learning model can be described as a discrete-time learning dynamical system of belief estimates of the unknown parameter and players' strategies: In each time step $\t \in \mathbb{N}_{+}$, the information system broadcasts the current belief estimate $\thetat$; the players act according to a strategy profile $\qt=\(\qt_i\)_{\i \in \I}$; and the payoffs $\ct=\(\ct_i\)_{\i \in \I}$ are realized according to $\phi^s(\ct|\qt)$ when the parameter is $\s \in \S$. The state of learning dynamics in step $\t$ is $\(\thetat, \qt\) \in \Delta(\S) \times \Q$. 

We assume that the initial state of the learning dynamics $\(\thetazero, \q^1\)$ satisfies $\thetazero(\s)>0$ for all $\s \in \S$ and $\q^1 \in \Q$; i.e. the initial belief does not exclude any possible parameter and the initial strategy profile is feasible. The evolution of states $\(\thetat, \qt\)_{\t=1}^{\infty}$ is jointly governed by belief and strategy updates, which we introduce next.

\noindent\textbf{Belief update.} In each step~$\t$, the information system observes the players' strategy profile $\qt$ and the realized payoffs $\ct$, and updates the belief $\thetat$ according to the Bayes' rule: 
\begin{align}
    \theta^{\t+1}(\s)&= \frac{\thetat(\s)\phibar^\s(\ct|\qt)}{\sum_{s' \in \S} \thetat(\s') \phibar^{\s'}(\ct|\qt)}, \quad \forall \s \in \S. \tag{$\theta$-update} \label{eq:update_belief}
\end{align}
The information system does not always need to observe the players' strategies and payoffs. In many instances of the problem setup, it is sufficient to update the belief only based on aggregate strategies $\tilde{\q}^\t$ and payoffs $\tilde{\c}^\t$, provided that the tuple $\(\tilde{\q}^\t, \tilde{\c}^\t\)$ is a sufficient statistics of $\(\qt, \ct\)$ in the following sense: one can write $\phibar^\s(\ct|\qt)= \psi\(\qt, \ct|\tilde{\q}^\t, \tilde{\c}^\t\) \tilde{\phi}^\s\(\tilde{\c}^\t|\tilde{\q}^\t\)$, where $\psi\(\qt, \ct|\tilde{\q}^\t, \tilde{\c}^\t\)$ is the $\s$-independent conditional distribution of strategy and payoffs given the aggregate statistics, and $\tilde{\phi}^\s\(\tilde{\c}^\t|\tilde{\q}^\t\)$ is the conditional probability of $\tilde{\c}^\t$ given $\tilde{\q}^\t$ for parameter $\s$. Then, we can re-write \eqref{eq:update_belief} as Bayesian update that only relies on $\(\tilde{\q}^\t, \tilde{\c}^\t\)$: 
\begin{align}\label{eq:equivalent_bayesian}
    \theta^{\t+1}(\s)&= \frac{\thetat(\s)\psi(\qt, \ct|\tilde{\q}^\t, \tilde{\c}^\t) \tilde{\phi}^s(\tilde{\c}^\t|\tilde{\q}^\t)}{\sum_{s' \in \S} \thetat(\s') \psi(\qt, \ct|\tilde{\q}^\t, \tilde{\c}^\t) \tilde{\phi}^{s'}(\tilde{\c}^\t|\tilde{\q}^\t)}= \frac{\thetat(\s) \tilde{\phi}^s(\tilde{\c}^\t|\tilde{\q}^\t)}{\sum_{s' \in \S} \thetat(\s')  \tilde{\phi}^{s'}(\tilde{\c}^\t|\tilde{\q}^\t)}, \quad \forall \s \in \S. 
\end{align}
For simplicity, we assume that the information system observes the players' strategies and payoffs in all steps, and provide examples to illustrate situations in which beliefs are updated based on the aggregate strategies and payoffs.



\vspace{0.2cm}
\noindent \textbf{Strategy update.} Each player updates her strategy $\qi^{\t+1}$ by taking a linear combination of the current strategy $\qit$ and a {\em preferred} strategy in game $\G$ under the updated belief $\thetatone$. The relative weight in the linear combination adopted by each player is determined by the player's stepsize $\ait \in [0, 1]$. We consider the following two types of strategy updates that differ in terms of how the players' preferences are taken into account in each step $\t$:
\begin{enumerate}
\item \emph{Update with equilibrium strategies:} The players' preferred strategy in game $\G$ with belief $\thetatone$ is given by a strategy profile $\gwe(\thetatone)=\(\gwe_i(\thetatone)\)_{\i \in \I}$ in the equilibrium set $\EQ(\thetatone): \Delta(\S) \rightrightarrows \Q$. That is, each player maximizes their expected utility under the updated belief $\thetatone$ assuming that the opponents are also playing an equilibrium strategy. The resulting strategy update can be written as: 
\begin{align}
\qi^{\t+1}&=(1-\ait)\qit+ \ait \gwei(\theta^{\t+1}), \quad \forall \i \in \I, \quad \forall \t. \tag{$\q$-update-EQ} \label{eq:eq_update}
\end{align}
\item \emph{Update with best-response strategies:} In contrast to \eqref{eq:eq_update}, for the updates with best-response strategies, the preferred strategy profile is given by $\gbr(\thetatone, \qt) =$ $ \(\gibr(\thetatone, \qt_{\mi})\)_{\i \in \I}$, which belongs to the best-response correspondence $\BR(\thetatone, \qt): \Delta(\S) \times \Q \rightrightarrows \Q$. That is, each player maximizes their expected utility based on $\thetatone$ while assuming that the opponents' strategies are fixed as $\qmi^{\t}$. In this case, the strategy update is given by: 
\begin{align}
    \qi^{\t+1}&=(1-\ait)\qit+ \ait \gibr(\thetatone, \qt_{\mi}), \quad \forall \i \in \I, \quad \forall \t. \tag{$q$-update-BR} \label{eq:br_update}
\end{align}
\end{enumerate}

In both cases (\eqref{eq:eq_update} and \eqref{eq:br_update}), players update their strategies to move closer to their respective preferences based on the updated belief. The stepsize $\ait$ governs the player $\i$'s ``speed" of strategy update (relative to the belief update) in step $\t$. For example, if $\ait=1$, then player $\i$ entirely adopts an equilibrium or best-response strategy based on the updated belief; thus, the speed of strategy update is same as the belief update. On the other hand, if $\ait \in (0, 1)$, then player $\i$ partially incorporates the updated belief into her strategy, and strategy update is slower. If $\ait=0$, then player $\i$ ignores the updated belief and does not change her strategy in step $\t$. 

In our learning model, the speed of strategy update is \emph{asynchronous}, i.e. the stepsizes $\(\ait\)_{\t=1}^{\infty}$ can be heterogeneous across players. We allow for $\ait$ to be asynchronous and less than or equal 1 to incorporate the external constraints that players may face in updating their strategies. For example, when players are not able to frequently update their strategies, $\ait$ take non-zero values only intermittently. Additionally, the maximum change between $\qi^{\t+1}$ and $\qit$ may be constrained by a certain threshold; then $\ait<1$ for steps in which the difference between the preferred strategy and $\qit$ exceeds this threshold.

We refer the learning dynamics governed by \eqref{eq:update_belief} -- \eqref{eq:eq_update} as \emph{learning with equilibrium strategies}, and \eqref{eq:update_belief} -- \eqref{eq:br_update} as \emph{learning with best-response strategies}. 


\section{Our Contributions and Related Literature}
The above-mentioned problem setup captures many situations in which strategic decision makers (players) repeatedly adjust their strategies in a game, while learning an unknown payoff-relevant parameter. Players have access to a common information system that repeatedly updates and broadcasts a belief estimate of the parameter based on the realized outcomes in each step. A key feature of our learning dynamics is that, in each step, players rely on imperfect information of the unknown parameter to choose their strategies, which impacts the outcomes and the belief estimate in the subsequent steps. 

This model of learning is relevant to a variety of applications. For example, buyers or sellers using an online market platform (e.g., Amazon, eBay, Airbnb) make their transaction decisions based on the information displayed by the platform which aggregates the users' data on previous prices, sales and buyer reviews of products. The users' decisions and realized prices drive the learning of the overall latent market condition, which further impacts the subsequent transactions (\cite{acemoglu2017fast}). Another situation of interest is day-to-day routing in transportation systems, where travelers make individual route choices based on the information provided by a navigation app (Google Maps or Apple Maps). The outcomes resulting from these choices (travel costs and link loads) are then used by the app to update and disseminate traffic congestion information (\cite{wu2020value, wu2019learning, meigs2017learning}). In these examples, the outcomes of players' strategic decisions (prices and sales in online platforms, and congestion costs in transportation networks) are governed by the joint evolution of belief estimates and players' strategies.

In our model, the Bayesian belief update \eqref{eq:update_belief} is similar to the well-known social learning models in \cite{banerjee1992simple, bikhchandani1992theory, smith2000pathological}, although we consider that the belief is updated centrally. In the strategy update \eqref{eq:eq_update} and \eqref{eq:br_update}, players' strategic choices are adjusted in an asynchronous manner based on the beliefs, which are in turn updated using the noisy game outcomes generated by the strategies. The two types of learning models that we study capture the dynamic interplay between Bayesian learning of the payoff parameter, and players' adaptive strategy updates. 


Our main contribution is the analysis of the long-time properties -- convergence and stability -- of beliefs and strategies for both types of learning dynamics. In addition, we identify conditions under which learning leads to complete information Nash equilibrium, and provide extensions to learning with other parameter estimates including Maximum a posteriori estimate (MAP) and Ordinary least squares (OLS).

\textbf{[Convergence]} We prove that the states $\(\thetat, \qt\)_{\t=1}^{\infty}$ in learning with equilibrium strategies converge to a fixed point with probability 1  (Theorems \ref{theorem:convergence_eq}). At a fixed point $\(\thetabar, \qbar\)$, the belief $\thetabar$ consistently estimates the probability distribution of players' payoffs generated by their fixed point strategy profile $\qbar$, and $\qbar$ is an equilibrium of the game $\G$ corresponding to $\thetabar$. That is, Bayesian update based on the realized payoffs no longer changes the belief about the parameter, and no player has an incentive to deviate from her strategy.

Our proof of belief convergence uses the classical martingale property of the Bayesian belief updates, and the proof of strategy convergence relies on a continuity assumption of the chosen equilibrium in \eqref{eq:eq_update} and a mild assumption on the stepsizes. Then, we show that the fixed point belief forms a consistent estimation of payoff distribution by proving that the belief of any parameter that results in a different payoff distribution compared with the true parameter converges to zero exponentially fast.

We obtain an analogous result on the state convergence in learning with best response strategies (Theorem \ref{prop:asynchronous}). In particular, the convergence of Bayesian belief update follows directly from that in learning with equilibrium strategies. Our approach to show convergence of best-response strategy update draws from the rich literature of learning in games. This includes discrete and continuous time best response dynamics (\cite{milgrom1990rationalizability, monderer1996potential, hofbauer2006best}), fictitious play (\cite{fudenberg1993learning, monderer1996fictitious}) and stochastic fictitious play (\cite{benaim1999mixed, hofbauer2002global}). The distinction between our strategy update and the classical best-response dynamics is that players do not know the payoff-relevant parameter in our model, and their strategy updates rely on the Bayesian belief updates. Moreover, the stepsizes used in strategy updates can be heterogeneous across players. 

To analyze how belief updates impact the strategy updates, we express \eqref{eq:br_update} as a sum of discrete-time asynchronous best response dynamics that only depends on the fixed point belief $\thetabar$ and random residuals that depend on the beliefs of each step. We show that these residuals converge to zero as the belief converges to $\thetabar$. The long-term properties of strategies in this discrete-time model can be conveniently evaluated by applying the well-known theory of stochastic approximation (\cite{tsitsiklis1994asynchronous, borkar1998asynchronous, benaim2005stochastic, benaim2006stochastic, perkins2013asynchronous}), leading to a continuous-time differential inclusion involving the fixed point belief. We show that, if the stepsizes satisfy the standard assumptions in stochastic approximation and the strategies in the continuous time differential inclusion converge with the fixed point belief, then the strategies in \eqref{eq:br_update} also converge to the equilibrium set corresponding to $\thetabar$. Consequently, the states $\(\thetat, \qt\)_{\t=1}^{\infty}$ in learning with best response strategies converge to the fixed point set with probability 1.

The conditions for the convergence to fixed point set in learning with best response strategies are satisfied in two classes of games -- potential games (Proposition \ref{prop:potential}) and dominance solvable games (Proposition \ref{prop:supermodular}). Our general convergence results applied to these games contribute to the extensive literature on other types of learning dynamics: log-linear learning (\cite{blume1993statistical}, \cite{marden2012revisiting}, \cite{alos2010logit}), regret-based learning (\cite{hart2003regret}, \cite{foster2006regret}, \cite{marden2007regret}, \cite{daskalakis2011near}), payoff-based learning (\cite{cominetti2010payoff}, \cite{marden2009payoff}), replicator dynamics (\cite{beggs2005convergence}, \cite{hopkins2002two}), and learning in large anonymous games (\cite{kash2011multiagent, adlakha2013mean}). These models can be broadly viewed as prescriptive dynamics that consider how players adjust their strategies based on the randomly realized payoffs in each step. On the other hand, our strategy update reflects a behavioral adjustment of players based on updated belief, which is consistent with the players' rational decision making process.

\textbf{[Stability]} 
We define a fixed point to be locally stable if the states remain close to the fixed point with high probability when the learning starts with an initial state close to that fixed point. A fixed point is globally stable if the state converges to that fixed point with probability 1 given any initial state. These stability notions are defined for the coupled belief-strategy dynamics in a game theoretic setting.\footnote{Our stability criteria are related to the notion of evolutionarily stable state that has been studied extensively in the context of population games and evolutionary dynamics \cite{smith1973logic, taylor1978evolutionary, samuelson1992evolutionary, matsui1992best, hofbauer2009stable, sandholm2010local}. In that literature, a state is defined to be stable if it is robust to local perturbation under the evolutionary dynamics. The evolutionary stability in population games typically studied using (local) Lyapunov functions. We do not take a Lyapunov approach for stability analysis due to the coupled nature of Bayesian belief updates and asynchronous strategy updates. Instead, we develop a first principles approach to analyze the stability of the beliefs and the strategies jointly in an atomic player game.}\textsuperscript{,}\footnote{\cite{frick2020stability} defined a similar stability notion for Bayesian beliefs in a single agent problem under a misspecified learning model. In their problem, the unknown parameters can be ordered and information is endogenously acquired by a single decision maker.} 


We present sufficient conditions that guarantee the local stability of fixed points in learning with equilibrium strategies (Theorem \ref{theorem:stability_eq}) and learning with best response strategies (Theorem \ref{theorem:stability_br}). In particular, we show that the following condition forms part of the set of sufficient conditions for local stability of both types of dynamics: All parameters in the support set of the fixed point belief $\thetabar$ have payoff distributions that are identical to that of the true parameter, given any strategy in a local neighborhood of the equilibrium set $\EQ(\thetabar)$. This condition ensures that Bayesian update eventually excludes all parameters that are not in the support of $\thetabar$, and with high probability the beliefs of the remaining parameters remain close to their corresponding values in $\thetabar$. Consequently, beliefs of all steps remain in a small neighborhood of $\thetabar$ with high probability. By assuming the continuity properties of the equilibrium set and the best response correspondence, we show that the strategies also remain close to the fixed point strategies in each type of learning dynamics -- this leads to local stability of fixed point. Additionally, we find that in both learning dynamics, there exists a set of globally stable fixed points if and only if all fixed points have complete information of the unknown parameter (Proposition \ref{prop:global}). %

\textbf{[Comparison with complete information Nash equilibrium and self-confirming equilibrium]} Clearly, any Nash equilibrium of the game with complete information is a fixed point strategy profile that corresponds to the complete information belief. However, there may exist other fixed points $\(\thetabar, \qbar\)$, where the belief $\thetabar$ forms an incorrect estimate of the payoff distribution for strategies that differ from $\qbar$. Consequently, the fixed point strategy $\qbar$ attained by learning dynamics may not be the same as a complete information equilibrium. 

Thus, the fixed points in our model share common features with the notion of self-confirming equilibrium\footnote{Similar concepts include conjectural equilibrium in \cite{hahn1978exercises} and subjective equilibrium in \cite{kalai1993subjective} and \cite{kalai1995subjective}} introduced in \cite{fudenberg1993self} for extensive games.\footnote{A variety of learning models have been proposed for achieving the self-confirming equilibrium (\cite{fudenberg1993learning} and \cite{fudenberg1993steady}) or the subjective equilibrium (\cite{kalai1993rational} and \cite{kalai1995subjective}). These learning dynamics assume that players maximizes the present value of future payoffs in each step of a repeated game with a fixed discount factor while updating the subjective beliefs of the nature or the opponents' strategies. On the other hand, our learning dynamics is more suited to study the convergence to fixed point states when (i) players' strategies are updated asynchronously and involve either Nash equilibrium or best response strategies corresponding to the most current belief update; and (ii) Bayesian estimation using noisy payoff information affects whether or not the fixed points correspond to complete information equilibrium.} At a self-confirming equilibrium, players maintain consistent beliefs of their opponents' strategies at information sets that are reached, but the beliefs of strategies can be incorrect at unreached information sets. Therefore, each player's self-confirming equilibrium strategy, which maximizes their own payoff based on individual belief of the opponents' strategies, may not be the same as a subgame perfect equilibrium. Both the self-confirming equilibrium and the fixed points in our model can be distinct from a complete information equilibrium due to the incorrect estimates on the unobserved game outcomes formed by the beliefs (i.e. the opponents' strategies on unreached information sets in the case of self-confirming equilibrium, and the payoff distributions of strategies that are different from $\qbar$ in our model). These incorrect estimates are not corrected by the learning dynamics because information of game outcomes is endogenously acquired based on the chosen strategies in each step.\footnote{The phenomenon that endogenous information acquisition leads to incomplete learning is also central to multi-arm bandit problems \cite{rothschild1974two, easley1988controlling} and endogenous social learning \cite{duffie2009information, acemoglu2014dynamics, ali2018herding}.}

We say that a fixed point is a complete information fixed point if the belief assigns probability 1 to the true parameter, and the strategy is a complete information Nash equilibrium. We discover that all fixed points are complete information fixed points if and only if, for any belief with less than perfect information of the unknown parameter, one can distinguish at least one parameter in the support set of that belief given the payoffs of a corresponding equilibrium strategy profile (Corollary \ref{cor:complete_belief}). In this case, all players eventually learn the true parameter and choose the complete information equilibrium with probability 1. 

Moreover, we find that if the payoff equivalent parameter set does not change in a local neighborhood of a fixed point strategy $\qbar$ and each player's payoff function is concave in their own strategy, then the fixed point strategy $\qbar$ must be a complete information Nash equilibrium, even if the belief $\thetabar$ may not provide complete information of the parameter (Proposition \ref{prop:complete}). Essentially, the condition that payoff equivalent parameter set remains the same in local neighborhood of $\qbar$ ensures that the belief $\thetabar$ consistently estimates the payoff distributions for all strategies in a local neighborhood of the fixed point strategy $\qbar$ (instead of just at $\qbar$), and hence each player's fixed point strategy must be a local maximizer of their payoff functions with the true parameter. Moreover, since payoff functions are concave in players' own strategies, the local maximizer of the true payoff function is a global maximizer for the entire strategy set. That is, each player's strategy is a best response to their opponents' strategies with complete information of the parameter, and thus $\qbar$ must be a complete information Nash equilibrium.

\textbf{[Extensions]} We extend our model to situations when the unknown parameter lies in a continuous set. In this extended model, we consider an alternative formulation of belief estimate, in which the information system updates and broadcasts the Maximum a posteriori estimate (MAP) of the unknown parameter instead of full Bayesian belief estimate. Similar to before, players' strategy updates incorporates either an equilibrium strategy or a best response strategy based on the updated MAP estimator. We provide analogous convergence results for both learning with equilibrium strategies and learning with best response strategies (Proposition \ref{prop:map_convergence}). In the special case where the average payoff functions are affine in the strategy profile, we obtain similar convergence result when ordinary least squares (OLS) is used to estimate the unknown parameter (Corollary \ref{cor:OLS}).

Rest of the article is organized as follows: Sec. \ref{sec:eq} and Sec. \ref{sec:br} present the convergence and stability results of learning with equilibrium strategies and learning with best response strategies, respectively. Sec. \ref{sec:discussion} discusses the conditions under which players learn the complete information equilibrium. We extend our results to continuous parameter set and learning with non-Bayesian estimates (MAP and OLS) in Sec. \ref{sec:other_estimate}.

\section{Learning with Equilibrium Strategies}\label{sec:eq}
In this section, we prove that the states $(\thetat, \qt)_{\t=1}^{\infty}$ in learning dynamics~\eqref{eq:update_belief} - \eqref{eq:eq_update} converge to a fixed point (Sec. \ref{subsec:convergence_eq}), and analyze local and global stability (Sec. \ref{subsec:stability_eq}).

\subsection{Convergence}\label{subsec:convergence_eq}

We first introduce two definitions.

\begin{definition}[Kullback–Leibler (KL)-divergence]\label{def:KL}
For a strategy profile $\q\in\Q$, the KL divergence between the distributions of observed payoffs $\chat$ with parameters $\s$  and $\sran\in\S$ is given by:
\begin{align*}
     D_{KL} \left(\phibar^{\sran}(\cbar|\q)||\phibar^{\s}(\cbar|\q)\right) =\left\{
     \begin{array}{ll}
     \int_{\chat} \phibar^{\sran}(\chat|\q) \log\left(\frac{\phibar^{\sran}(\chat|\q)}{\phibar^{\s}(\chat|\q)}\right) d\chat, & \quad \text{if $\phibar^{\sran}(\chat|\q) \ll \phibar^{\s}(\chat|\q)$}, \\
     \infty & \quad \text{otherwise.}
     \end{array}
     \right.
\end{align*}
\end{definition}
Here $\phibar^{\sran}(\chat|\q) \ll \phibar^{\s}(\chat|\q)$ means that the distribution $\phibar^{\sran}(\chat|\q)$ is absolutely continuous with respect to $\phibar^{\s}(\chat|\q)$, i.e. $\phibar^{\s}(\chat|\q)=0$ implies $\phibar^{\sran}(\chat|\q)=0$ with probability 1.

\begin{definition}[Payoff-equivalent parameters]\label{def:payoff_equivalence}
A parameter $\s\in\S$ is payoff-equivalent to the true parameter $\sran$ given the strategy profile $\q\in\Q$ if $D_{KL} \left(\phibar^{\sran}(\cbar|\q)||\phibar^{\s}(\cbar|\q)\right)=0$. Then, for a given strategy profile $\q\in Q$, the set of parameters that are payoff-equivalent to $\sran$ is:  
\begin{align*}
    \Sequiv(\q)\deleq \{\S|D_{KL} \left(\phibar^{\sran}(\cbar|\q)||\phibar^{\s}(\cbar|\q)\right)=0\}.
\end{align*}
\end{definition}

The KL-divergence between any two distributions is non-negative, and is equal to zero if and only if the two distributions are identical. Thus, for a given strategy profile $\q$, if a parameter $\s$ is in the payoff-equivalent parameter set $\Sequiv(\q)$, then the distributions of the observed payoffs are identical for parameters $\s$ and $\sran$, i.e. $\phibar^{\sran}(\chat|\q)=\phibar^{\s}(\chat|\q)$ with probability 1. Therefore, the observed payoffs cannot be used by the information system to distinguish $\s$ and $\sran$ in the belief update \eqref{eq:update_belief} (because the belief ratio $\frac{\thetat(\s)}{\thetat(\sran)}$ remains unchanged with probability 1). Also note that $\Sequiv(\q)$ can vary with the strategy profile $\q$, and hence a payoff-equivalent parameter for a given strategy profile may not be payoff-equivalent for another strategy profile.

In proving our convergence theorem, we assume that the following conditions hold:

\emph{\textbf{(A1)} 
Equilibrium strategy profile $\gwe(\theta) \in \EQ(\theta)$: For any $\theta \in \Delta(\S)$, the function $g(\theta)$ is continuous in $\theta$.}

\emph{\textbf{(A2)} 
Stepsizes $\(\ait\)_{\i \in \I}$: For any $\i \in \I$, $ \prod_{\t=m}^{\infty} \(1-\ait\)=0, ~ \forall ~ m \geq 1.$}

For any $\theta \in \Delta(\S)$, if the game $\G$ has a unique equilibrium (i.e. $\EQ(\theta)$ is a singleton set for all $\theta$), then the assumption \textbf{(A1)} requires that the unique equilibrium is a continuous function of $\theta$. On the other hand, if $\G$ has multiple equilibria, then \textbf{(A1)} requires that there exists at least one equilibrium strategy profile $g(\theta)$ for each $\theta \in \Delta(\S)$ such that $\gwe(\theta)$ is a continuous function of $\theta$. Moreover, in each step $\t$, the players use the updated belief $\thetatone \in \Delta(\S)$, and perform strategy update \eqref{eq:eq_update} to account for $\gwe(\thetatone)$.

On the other hand, Assumption \textbf{(A2)} ensures that the strategy updates continue to incorporate the players' preference of equilibrium behavior based on the {\em updated beliefs} as opposed to solely relying on the initial strategy $q^1$. This assumption trivially holds when $\ait$ is lower-bounded by a small positive number infinitely often (i.o.). However, when $\ait$ indeed converges to zero (i.e. no player updates her strategy eventually), the assumption imposes a mild restriction that $\ait$ does not converge to zero exponentially fast. In particular, $\ait \geq 1-e^{-\frac{1}{\t}}$ is sufficient to ensure that the stepsizes satisfy \textbf{(A2)} since $ \prod_{\t=m}^{\infty} \(1-\ait\) \leq \prod_{\t=m}^{\infty} e^{-\frac{1}{\t}}=0$.

We now present the convergence theorem for learning with equilibrium strategies.
\begin{theorem}\label{theorem:convergence_eq}
For any initial state~$(\theta^1, \q^1)\in\Delta(S)\times \Q$, under Assumptions \textbf{(A1)} -- \textbf{(A2)}, the sequence of states $(\thetat, \qt)_{\t=1}^{\infty}$ converges to a fixed point $(\thetabar, \qbar)\in\Delta(S)\times \Q$ with probability 1. The fixed point $\(\thetabar, \qbar\)$ satisfies: 
\begin{subequations}\label{eq:fixed_point_def}
\begin{align}
    [\thetabar] &\subseteq \Sequiv(\qbar), \label{eq:exclude_distinguished}\\
    \qbar&=\gwe(\thetabar) \in \EQ(\theta), \label{subeq:eq_fixed}
\end{align}
\end{subequations}
where $[\thetabar] \deleq \{\S|\thetabar(\s)>0\}$ is the support set of the fixed point belief $\thetabar$. 

\end{theorem}

We prove Theorem \ref{theorem:convergence_eq} in three lemmas. Firstly, Lemma \ref{lemma:theta_eq} establishes the convergence of beliefs $\(\thetat\)_{\t=1}^{\infty}$ by showing that both sequences $\left(\frac{\thetat(\s)}{\thetat(\sran)}\right)_{\t=1}^\infty$ and $\left(\thetat(\sran)\right)_{\t=1}^{\infty}$ are non-negative martingales, and hence converge.  

\begin{lemma}\label{lemma:theta_eq}
$\lim_{\t \to \infty} \thetat=\thetabar$, where $\thetabar = \gwe(\thetabar) \in\Delta(S)$.
\end{lemma}
\noindent\textbf{\emph{Proof of Lemma \ref{lemma:theta_eq}.}} \\
First, we show that for any parameter $\s \in \S$, the sequence $\(\frac{\theta^\t(\s)}{\theta^\t(\sran)}\)_{\t=1}^{\infty}$ is a non-negative martingale, and hence converges with probability 1. Note that for any $\t=1, 2, \dots $, and any parameter $\s \in \S \setminus \{\sran\}$, we have the following from \eqref{eq:update_belief}: 
\begin{align*}
\frac{\theta^{\t+1}(\s)}{\theta^{\t+1}(\sran)}= \frac{\theta^{\t}(\s) \cdot \phibar^{\s}(\ct|\qt)}{\theta^{\t}(\sran) \cdot \phibar^{\sran}(\ct|\qt)}.
\end{align*}
Now starting from any initial belief $\thetazero$, consider a sequence of strategies $\Loadhis\deleq \(\qj\)_{j=1}^{\t-1}$ and a sequence of realized outcomes $\Costhis\deleq \(\cj\)_{\j =1}^{\t-1}$ before step $\t$. Then, the expected value of $\frac{\theta^{\t+1}(\s)}{\theta^{\t+1}(\sran)}$ conditioned on $\thetazero$, $\Loadhis$ and $\Costhis$ is as follows: 
\begin{align}\label{conditional_iterate}
\mathbb{E}\left[\left.\frac{\theta^{\t+1}(\s)}{\theta^{\t+1}(\sran)}\right\vert \thetazero, \Loadhis, \Costhis\right]&= \frac{\thetat(\s)}{\thetat(\sran)} \cdot \mathbb{E}\left[\left. \frac{\phibar^{\s}(\ct|\qt)}{\phibar^{\sran}(\ct|\qt)}\right\vert \thetazero, \Loadhis, \Costhis \right], 
\end{align}
where $\thetat$ is the repeatedly updated belief from $\thetazero$ based on $\Loadhis$ and $\Costhis$ using \eqref{eq:update_belief}. Note that
\begin{align*}
\mathbb{E}\left[\left. \frac{\phibar^{\s}(\ct|\qt)}{\phibar^{\sran}(\ct|\qt)}\right\vert \thetazero, \Loadhis, \Costhis \right]
=&\int_{\ct} \(\frac{\phibar^{\s}(\ct|\qt)}{\phibar^{\sran}(\ct|\qt)} \) \cdot  \phibar^{\sran}(\ct|\qt) d \ct=1.
\end{align*}
Hence, for any $\t=1, 2, \dots$,  
\begin{align*}
\mathbb{E}\left[\left.\frac{\theta^{\t+1}(\s)}{\theta^{\t+1}(\sran)}\right\vert \thetazero, \Loadhis, \Costhis\right]= \frac{\thetat(\s)}{\thetat(\sran)}, \quad \forall \s \in \S.
\end{align*}
Again, from \eqref{eq:update_belief} we know that $\frac{\theta^\t(\s)}{\theta^\t(\sran)} \geq 0$. Hence, the sequence $\(\frac{\theta^\t(\s)}{\theta^\t(\sran)}\)_{\t=0}^{\infty}$ is a non-negative martingale for any $\s \in \S$. From the martingale convergence theorem, we conclude that $\frac{\theta^\t(\s)}{\theta^\t(\sran)}$ converges with probability 1. 

Next we show that the sequence $\left(\log\thetat(\sran)\right)_{\t=1}^{\infty}$ is a submartingale, and hence converges with probability 1. We define $\mu (\ct|\theta^{\t}, \qt)\deleq\sum_{\s \in \S} \theta^{\t}(\s)\phibar^{\s}(\ct|\qt)$. From \eqref{eq:update_belief}, we have:
\begin{align*}
    &\mathbb{E}\left[\left.\log\theta^{\t+1}(\sran)\right\vert \thetazero, \Loadhis, \Costhis\right]= \mathbb{E}\left[\left.\log\(\frac{\theta^{\t}(\sran)\phibar^{\sran}(\ct|\qt)}{\mu(\ct|\theta^{\t}, \qt)}\)\right\vert \thetazero, \Loadhis, \Costhis\right]\\
  &= \log\theta^{\t}(\sran) +\mathbb{E}\left[\left.\log\(\frac{\phibar^{\sran}(\ct|\qt)}{\mu(\ct|\theta^{\t}, \qt)}\)\right\vert \thetazero, \Loadhis, \Costhis\right]\\
   & =\log\theta^{\t}(\sran) +\int_{\ct} \phibar^{\sran}(\ct|\qt) \log\(\frac{\phibar^{\sran}(\ct|\qt)}{\mu(\ct|\theta^{\t}, \qt)}\) d \ct \\
    &= \log\theta^{\t}(\sran)+  D_{KL}\(\phibar^{\sran}(\ct|\qt)||\mu(\ct|\theta^{\t}, \qt)\) \geq \log\theta^{\t}(\sran),
\end{align*}
where the last inequality is due to the non-negativity of KL divergence between $\phibar^{\sran}(\ct|\qt)$ and $\mu(\ct|\theta^{\t}, \qt)$. Therefore, the sequence $\left(\log\thetat(\sran)\right)_{\t=1}^{\infty}$ is a submartingale. Additionally, since $\log\thetat(\sran)$ is bounded above by zero, by the martingale convergence theorem $\log\thetat(\sran)$ converges with probability 1. Hence, $\thetat(\sran)$ must also converge with probability 1.

From the convergence of $\frac{\theta^\t(\s)}{\theta^\t(\sran)}$ and $\thetat(\sran)$, we conclude that $\thetat(\s)$ converges with probability 1 for any $\s \in \S$. Let the convergent vector be denoted as $\thetabar= \(\thetabar(\s)\)_{\s \in \S}$. We can check that for any $\t$, $\thetat(\s) \geq 0$ for all $\s \in \S$ and $\sum_{\s \in \S}\thetat(\s)=1$. Hence, $\thetabar$ must satisfy $\thetabar(\s) \geq0$ for all $\s \in \S$ and $\sum_{\s \in \S}\thetabar(\s)=1$, i.e. $\thetabar$ is a feasible belief vector. \QEDA

Secondly, Lemma \ref{lemma:q_eq} shows that the sequence of strategies $\(\qit\)_{\t=1}^{\infty}$ converges to an equilibrium strategy of game $\G$ for belief $\thetabar$ as in \eqref{subeq:eq_fixed}. 
\begin{lemma}\label{lemma:q_eq}
$\lim_{\t \to \infty} \qt = \qbar$, where $\qbar \in \EQ(\thetabar)$.
\end{lemma}
By using iterative updates \eqref{eq:eq_update}, we can write $\qt$ as a linear combination of the initial strategy profile $\q^1$ and equilibrium sequence $\(\gwe(\theta^j)\)_{j=2}^{\t}$. The convergence of strategies follow from the convergence of $\(\thetat\)_{\t=1}^{\infty}$ in Lemma \ref{lemma:theta_eq} and Assumptions \textbf{(A1)}--\textbf{(A2)}.  

\vspace{0.2cm}
\noindent\textbf{\emph{Proof of Lemma \ref{lemma:q_eq}.}} \\
By iteratively applying \eqref{eq:eq_update}, we obtain that for any player $\i \in \I$: 
\begin{align}\label{eq:input_calculation}
\qit= \prod_{\j=1}^{\t-1} (1-a_i^{\j}) \q^1_i+ \sum_{\j=1}^{\t-1} \(\prod_{\iter=\j+1}^{\t-1} (1-a_i^{\iter}) a_i^{\j} \gwei(\thetaj)\), \quad \forall \t =2,3, \dots
\end{align}
and, 
\begin{align*}
\|\qit-\gwei(\thetabar)\|&
\leq   \prod_{\j=1}^{\t-1} (1-a_i^{\j}) \|\qi^1-\gwei(\thetabar)\| + \sum_{\j=1}^{\t-1} \(\prod_{\iter=\j+1}^{\t-1} (1-a_i^{\iter}) a_i^{\j}  \|\gwei(\thetaj)-\gwei(\thetabar)\|\)\\
&=\prod_{\j=1}^{\t-1} (1-a_i^{\j}) \|\qi^1-\gwei(\thetabar)\| + \sum_{\j=1}^{m-1} \(\prod_{\iter=\j+1}^{\t-1} (1-a_i^{\iter}) a_i^{\j}  \|\gwei(\thetaj)-\gwei(\thetabar)\|\) \\
&\quad + \sum_{\j=m}^{\t-1} \(\prod_{\iter=\j+1}^{\t-1} (1-a_i^{\iter}) a_i^{\j}  \|\gwei(\thetaj)-\gwei(\thetabar)\|\),
\end{align*}
where $m$ is any integer between 1 and $\t-1$, and $\|\cdot\|$ is the Euclidean norm. 

Since $\lim_{\t \to \infty} \prod_{\j=1}^{\t-1}  \(1-a_i^{\j}\)=0$ \textbf{(A2)} and $\|\qi^1-\gwei(\thetabar)\|$ is finite, for any $\epsilon>0$, we can find an integer $K_1>0$ such that any $\t>K_1$ satisfies: 
\begin{align}\label{eq:one_epsilon}
 \prod_{\j=1}^{\t-1} (1-a_i^{\j}) \|\qi^1-\gwei(\thetabar)\|<\frac{\epsilon}{3}.
 \end{align}
Additionally, since $\lim_{\t \to \infty} \thetat=\thetabar$ with probability 1 and $\gwei(\theta)$ is continuous in $\theta$ \textbf{(A1)}, we have $\lim_{\t \to \infty}\gwei(\thetat)=\gwei(\thetabar)$ with probability 1 for any $\i \in \I$. Therefore, we can find a second integer $K_2>0$ such that for any $\t>K_2$, $\|\gwei(\thetat)-\gwei(\thetabar)\|<\frac{\epsilon}{3}$. Then, for any $m>K_2$ and any $\t>m$:
\begin{align}\label{eq:two_epsilon}
\sum_{\j=m}^{\t-1} \(\prod_{\iter=\j+1}^{\t-1} (1-a_i^{\iter}) a_i^{\j}  \|\gwei(\thetaj)-\gwei(\thetabar)\|\) < \frac{\epsilon}{3}\(\sum_{\j=m}^{\t-1} \prod_{\iter=\j+1}^{\t-1} (1-a_i^{\iter}) a_i^{\j} \) \leq \frac{\epsilon}{3},
\end{align} where we use the fact that $\sum_{\j=m}^{\t-1} \prod_{\iter=\j+1}^{\t-1} (1-a_i^{\iter}) a_i^{\j} =1 - \prod_{i=m}^{\t-1}(1-a_i^{\iter}) \leq 1$. 

Finally, for any fixed $m$, since $\ait \in [0, 1]$ for any $\t$ and $ \|\gwei(\thetaj)-\gwei(\thetabar)\|$ is finite for any step $j$, we have:
\begin{align*}
\sum_{\j=1}^{m-1} \(\prod_{\iter=\j+1}^{\t-1} (1-a_i^{\iter}) a_i^{\j}  \|\gwei(\thetaj)-\gwei(\thetabar)\|\) \leq  \max_{j=1, \dots m-1} \|\gwei(\thetaj)-\gwei(\thetabar)\| (m-1) \prod_{\iter=m}^{\t-1}  (1-a_i^{\iter})  . 
\end{align*}
Again from \textbf{(A2)}, we have $\lim_{\t\to \infty} \prod_{\iter=m}^{\t-1}  (1-a_i^{\iter}) =0$. Then, we can find the third integer $K_3>m$ such that for any $\t>K_3$, \begin{align}\label{eq:three}
\sum_{\j=1}^{m-1} \(\prod_{\iter=\j+1}^{\t-1} (1-a_i^{\iter}) a_i^{\j}  \|\gwei(\thetaj)-\gwei(\thetabar)\|\)  < \frac{\epsilon}{3}. 
\end{align}
From \eqref{eq:one_epsilon} -- \eqref{eq:three}, for any $\epsilon>0$, we can find an integer $K= \max \{K_1, K_2, K_3\}$ such that for any $\t>K$,
\begin{align*}
\|\qit-\gwei(\thetabar)\|< \frac{\epsilon}{3}+\frac{\epsilon}{3}+\frac{\epsilon}{3}=\epsilon, \quad w.p.~1.
\end{align*}
We can thus conclude that $\lim_{\t \to \infty} \|\qit-\gwei(\thetabar)\|=0$, i.e. $\lim_{\t \to \infty} \qit=\gwei(\thetabar)$ w.p.~1. \QEDA

Thirdly, using the convergence of both $\(\thetat\)_{\t=1}^{\infty}$ and $\(\qit\)_{\t=1}^{\infty}$, Lemma \ref{lemma:consistency} shows that the belief of any $\s$ that is not payoff-equivalent to $\sran$ given $\qbar$ must converge to $0$, i.e. $\thetabar$ satisfies \eqref{eq:exclude_distinguished}. This concludes proof of Theorem \ref{theorem:convergence_eq}. Besides, Lemma \ref{lemma:consistency} also provides a convergence rate of beliefs. 
\begin{lemma}\label{lemma:consistency}
Any fix point $\(\thetabar, \qbar\)$ of learning dynamics satisfies \eqref{eq:exclude_distinguished}. Furthermore, for any $\s \in \S\setminus \Sequiv(\qbar)$, if $\phibar^{\sran}(\chat|\qbar) \ll \phibar^{\s}(\chat|\qbar)$, then $\thetat(\s)$ converges to 0 exponentially fast: 
\begin{align}
    \lim_{\t \to \infty} \frac{1}{\t} \log(\thetat(\s))=-D_{KL}(\phibar^{\sran}(\cbar|\qbar)||\phibar^{\s}(\cbar|\qbar)). \quad  w.p.~1\label{eq:rate}
\end{align}
Otherwise, there exists a finite positive integer $\Tstar$ such that $\thetat(\s)=0$ for all $\t>\Tstar$ w.p. 1. 
\end{lemma}

\vspace{0.2cm}

\noindent\emph{\textbf{Proof of Lemma \ref{lemma:consistency}.}} By iteratively applying the belief update in \eqref{eq:update_belief}, we can write:
\begin{align}\label{eq:sequence_update}
    \thetat(\s)= \frac{\theta^1(\s) \prod_{\j=1}^{\t-1}\phibar^{s}(\cj|\qj)}{\sum_{\s' \in \S} \theta^1(\s') \prod_{\j=1}^{\t-1}\phibar^{s'}(\cj|\qj)}, \quad \forall \s \in \S.
\end{align}
We define $\Phi^\s(\Costhis|\Loadhis)$ as the probability density function of the history of the realized outcomes $\Costhis=\(\c^j\)_{j=1}^{\t-1}$ conditioned on the history of strategies $\Loadhis= \(\q^j\)_{j=1}^{\t-1}$ prior to step $\t$, i.e. $\Phi^\s(\Costhis|\Loadhis) \deleq \prod_{\j=1}^{\t-1}\phibar^{s}(\cj|\q^j)$. We rewrite \eqref{eq:sequence_update} as follows: 
\begin{align}
    \thetat(\s)&= \frac{\theta^1(\s) \Phi^\s(\Costhis|\Loadhis)}{\sum_{\s' \in \S} \theta^1(\s') \Phi^{\s'}(\Costhis|\Loadhis)} \leq \frac{\theta^1(\s) \Phi^\s(\Costhis|\Loadhis)}{ \theta^1(\s) \Phi^{\s}(\Costhis|\Loadhis)+ \theta^1(\sran) \Phi^{\sran}(\Costhis|\Loadhis)}\notag\\
    &=\frac{\theta^1(\s) \frac{\Phi^\s(\Costhis|\Loadhis)}{\Phi^{\sran}(\Costhis|\Loadhis)}}{ \theta^1(\s) \frac{\Phi^{\s}(\Costhis|\Loadhis)}{\Phi^{\sran}(\Costhis|\Loadhis)}+ \theta^1(\sran)}.\label{eq:bound_theta}
\end{align}
For any $\s \in \S \setminus \Sequiv(\qbar)$, if we can show that the ratio $ \frac{\Phi^\s(\Costhis|\Loadhis)}{\Phi^{\sran}(\Costhis|\Loadhis)}$ converges to 0, then $\theta(\s)$ must also converge to 0. Now, we need to consider two cases:\\
\noindent\emph{Case 1:} $\phi^{\sran}(\cbar|\qbar) \ll \phi^{\s}(\cbar|\qbar)$: In this case, the log-likelihood ratio can be written as: 
\begin{align}\label{eq:log-likelihood}
    \log \(\frac{\Phi^\s(\Costhis|\Loadhis)}{\Phi^{\sran}(\Costhis|\Loadhis)}\)=   \sum_{\j=1}^{\t-1} \log\(\frac{\phibar^{\s}(\cj|\qj)}{\phibar^{\sran}(\cj|\qj)}\).
\end{align}
For any $\s \in \S$, since $\phibar^{s}(\cj|\qj)$ is continuous in $\q^{j}$, the probability density function of $\etaj$ is also continuous in $\q^{j}$. In Lemma \ref{lemma:q_eq}, we proved that $\(\qt\)_{\t=1}^{\infty}$ converges to $\qbar$. Then, the distribution of $\etaj$ must converge to the distribution of $\log\(\frac{\phibar^\s(\cbar|\qbar)}{\phibar^{\sran}(\cbar|\qbar)}\)$. Note that for any $ \s \in \S \setminus \Sequiv(\qbar)$, the expectation of $\etabar$ can be written as: 
\begin{align*}
\mathbb{E}\left[\etabar\right]= \int_{\cbar}\phibar^{\sran}(\cbar|\qbar) \cdot  \log\(\frac{\phibar^\s(\cbar|\qbar)}{\phibar^{\sran}(\cbar|\qbar)}\)  d \cbar=-D_{KL}\(\phibar^{\sran}(\cbar|\qbar)||\phibar^\s(\cbar|\qbar)\)<0.
\end{align*}  
If we can show that the equation \eqref{eq:hold} below holds, then we can conclude that the log-likelihood sequence defined by~\eqref{eq:log-likelihood} converges to $-\infty$; this would in turn imply that the sequence of likelihood ratios $\frac{\Phi^\s(\Costhis|\Loadhis)}{\Phi^{\sran}(\Costhis|\Loadhis)}$ for all $k=2,3,\dots$ must converge to 0. But first we need to show: 
\begin{align}\label{eq:hold}
\lim_{\t \to \infty}\frac{1}{\t-1} \log \(\frac{\Phi^\s(\Costhis|\Loadhis)}{\Phi^{\sran}(\Costhis|\Loadhis)}\)&=\lim_{\t \to \infty}\frac{1}{\t-1} \sum_{j=1}^{\t-1} \etaj \notag\\
&= \mathbb{E}\left[\etabar\right], \quad w.p.~1.
\end{align}

We denote the cumulative distribution function of $\etaj$ as $F^{j}(z): \mathbb{R} \to [0,1]$, i.e. $F^{j}(z)=\pro\(\etaj \leq z\)$. The cumulative distribution function of $\etabar$ is denoted $\bar{F}(z): \mathbb{R} \to [0,1]$, i.e. $\bar{F}(z)=\pro\(\etabar \leq z\)$. Then, 
\begin{align}\label{ind}
    \lim_{j \to \infty}F^j(z)=\bar{F}(z), \quad \forall z \in \mathbb{R}.
\end{align}
For any sequence of realized outcomes $(\cj)_{j=1}^\infty$, we define a sequence of random variables $\Delta =\(\Delta^{j}\)_{j=1}^{\infty}$, where $\Delta^{j}= F^j\(\etaj\)$. Then, we must have $\Delta^{j} \in [0,1]$, and for any $\delta \in [0,1]$, $\pro(\Delta^{j} \leq \delta)=\pro\(F^j\(\etaj\)\leq\delta\)=\delta$. That is, $\Delta^j$ is independently and uniformly distributed on $[0,1]$. Consider another sequence of random variables $\(\Chibarj\)_{j=1}^{\infty}$, where $\Chibarj\deleq \(\bar{F}\)^{-1}(\Delta^j)$. Since $\Delta^j$ is independently and identically distributed (i.i.d.) with uniform distribution, $\Chibarj$ is also i.i.d. distributed with the same distribution as $\etabar$. Additionally, since each $\Delta^j$ is generated from the realized outcome $\cj$, $\(\Chibarj\)_{j=1}^{\infty}$ is in the same probability space as $\etaj$. From \eqref{ind}, we know that as $j \to \infty$, $F^j$ converges to $\bar{F}$. Therefore, with probability 1,
\begin{align*}
    \lim_{j\to \infty} \left\vert\etaj-\Chibarj\right\vert=\lim_{j \to \infty}\left\vert\etaj-(\bar{F})^{-1}F^{j}\(\etaj\)\right\vert=0.
\end{align*}
Consequently, with probability 1, 
\begin{align}\label{eq:close}
\lim_{\t \to \infty}\left \vert\frac{1}{\t-1} \sum_{j=1}^{\t-1}\( \etaj-  \Chibarj\)\right\vert \leq \lim_{\t \to \infty}\frac{1}{\t-1}  \sum_{j=1}^{\t-1}\left\vert\etaj-\Chibarj\right\vert=0. 
\end{align}
Since $\(\Chibarj\)_{j=1}^{\infty}$ is independently and identically distributed according to the distribution of $\etabar$, from strong law of large numbers, we have:
\begin{align*}
 \lim_{\t \to \infty} \frac{1}{\t-1}\sum_{j=1}^{\t-1}\Chibarj = \mathbb{E}\left[\etabar\right]=-D_{KL}\(\phibar^{\sran}(\cbar|\qbar)||\phibar^\s(\cbar|\qbar)\), \quad w.p.~1.
\end{align*}

From \eqref{eq:close}, we obtain the following: 
\begin{align}\label{eq:sum_chi}
\lim_{\t \to \infty} \frac{1}{\t-1} \sum_{j=1}^{\t-1} \etaj= \lim_{\t \to \infty} \frac{1}{\t-1}\sum_{j=1}^{\t-1}\Chibarj=-D_{KL}\(\phibar^{\sran}(\cbar|\qbar)||\phibar^\s(\cbar|\qbar)\), \quad w.p.~1
\end{align}
Hence, \eqref{eq:hold} holds. Then, for any $\s \in \S \setminus \Sequiv(\qbar)$, $\lim_{\t \to \infty} \frac{\Phi^\s(\Costhis|\Loadhis)}{\Phi^{\sran}(\Costhis|\Loadhis)}=0$. Thus, from \eqref{eq:bound_theta}, we know that $\lim_{\t \to \infty} \thetat(\s)=0$ for all $\s \in \S \setminus \Sequiv(\qbar)$. 

Finally, since $\thetazero(\s)>0$ for all $\s \in \S$, the true parameter $\sran$ is never excluded from the belief. Therefore, $\lim_{\t \to \infty} \frac{1}{\t} \log\(\thetat(\sran)\)=0$. For any $\s \in \S \setminus \Sequiv(\qbar)$, we have the following:
\begin{align*}
    &\lim_{\t \to \infty} \frac{1}{\t}\log\(\thetat(\s)\)=   \lim_{\t \to \infty} \frac{1}{\t} \log\(\thetat(\sran)\) +\lim_{\t \to \infty} \frac{1}{\t}\log\(\frac{\thetat(\s)}{\thetat(\sran)}\)\\
    &=\lim_{\t \to \infty} \frac{1}{\t}\log\(\frac{\thetat(\s)}{\thetat(\sran)}\) = \lim_{\t \to \infty} \frac{1}{\t}\log \(\frac{\theta^1(\s)}{\theta^1(\sran)}\)+\lim_{\t \to \infty} \frac{1}{\t} \log\(\frac{\Phi^\s(\Costhis|\Loadhis)}{\Phi^{\sran}(\Costhis|\Loadhis)}\)\\
    &= \lim_{\t \to \infty} \frac{1}{\t} \sum_{\j=1}^{\t-1} \log\(\frac{\phi^\s(\cj|\qj)}{\phi^{\sran}(\cj|\qj)}\)=\lim_{\t \to \infty} \frac{1}{\t-1} \sum_{\j=1}^{\t-1} \log\(\frac{\phibar^\s(\cj|\qj)}{\phibar^{\sran}(\cj|\qj)}\)\\
    &= \mathbb{E} \left[\log\(\frac{\phibar^{\s}(\cbar|\qbar)}{\phibar^{\sran}(\cbar|\qbar)}\)\right]
    =-D_{KL}\(\phibar^{\sran}(\cbar|\qbar)||\phibar^\s(\cbar|\qbar)\), \quad w.p.~1. 
\end{align*}

\noindent\emph{Case 2:} $\phi^{\sran}(\cbar|\qbar)$ is not absolutely continuous in $\phi^{\s}(\cbar|\qbar)$. \\
In this case, $\phi^\s(\cbar|\qbar)=0$ does not imply $\phi^{\sran}(\cbar|\qbar)=0$ with probability 1, i.e. $\pro\(\phi^\s(\cbar|\qbar)=0\)>0$, where $\pro\(\cdot\)$ is the probability of $\cbar$ with respect to the true distribution $\phi^{\sran}(\cbar|\qbar)$. Since the distributions $\phi^\s(\cbar|\q)$ and $\phi^{\sran}(\cbar|\q)$ are continuous in $\q$, the probability $\pro\(\phi^\s(\cbar|\q)=0\)$ must also be continuous in $\q$. Therefore, for any $\epsilon \in \(0, \pro\(\phi^\s(\cbar|\qbar)=0\)\)$, there exists $\delta>0$ such that $\pro\(\phi^\s(\c|\q)=0\) > \epsilon$ for all $\q \in \{\q|\|\q-\qbar\|<\delta\}$. 

From Lemma \ref{lemma:q_eq}, we know that $\lim_{\t \to \infty} \qt = \qbar$. Hence, we can find a positive number $K_1>0$ such that for any $\t > K_1$, $\|\qt - \qbar\|<\delta$, and hence $\pro\(\phi^\s(\ct|\qt)=0\) > \epsilon$. We then have $\sum_{\t=1}^{\infty}\pro\(\phi^\s(\ct|\qt)=0\) = \infty$. Moreover, since the event $\phi^\s(\ct|\qt)=0$ is independent from the event $\phi^\s(\c^{\t'}|\q^{\t'})=0$ for any $\t, \t'$, we can conclude that $\pro\(\phi^\s(\ct|\qt)=0, \text{infinitely often}\)=1$ based on the second Borel-Cantelli lemma. Hence, $\pro\(\phi^\s(\ct|\qt)>0, ~\forall \t\)=0$. From the Bayesian update \eqref{eq:update_belief}, we know that if $\phi^\s(\ct|\qt)=0$ for some step $\t$, then the belief $\theta^{\t+1}(\s)=0$. Therefore, we can conclude that $\pro\(\thetat(\s)>0, ~\forall \t\)=0$ with probability 1, i.e. there exists a positive number $\Tstar > K_1$ with probability 1 such that $\thetat(\s)=0$ for any $\t > \Tstar$. 
    \QEDA
    
    \vspace{0.2cm}
From theorem \ref{theorem:convergence_eq}, we know that the states of the learning dynamics $\(\thetat, \qt\)_{\t=1}^{\infty}$ converges to a fixed point $\(\thetabar, \qbar\)$ with probability 1, and the fixed point must satisfy two properties:\footnote{In the proof of the theorem, Assumption \textbf{(A1)} -- which requires that the players choose a continuous equilibrium function in all steps -- ensures that the strategy $\qt$ converges as the belief $\thetat$ converges (Lemma \ref{lemma:q_eq}). Theorem \ref{theorem:convergence_eq} holds in the alternative setting of learning with equilibrium strategies when the set $\EQ(\theta)$ is convex and upper-hemicontinuous in $\theta$, and the players may choose {\em any} equilibrium strategy profile $\gwe(\thetat) \in \EQ(\thetat)$ in each step $\t$.}

\begin{itemize}
    \item[(1)] The belief identifies the true parameter $\sran$ in the payoff-equivalent set $\Sequiv(\qbar)$ given the fixed point strategy $\qbar$. As a result, the belief forms a consistent estimate of the payoff distribution at the fixed point. To see this, let us denote the estimated distribution of the observed payoff $\chat$ as $\mu(\chat|\thetabar, \qbar)$. Then, 
\begin{align}\label{eq:marginal}
    \mu(\chat|\thetabar, \qbar)\deleq\sum_{\s \in \S} \thetabar(\s) \phibar^\s(\chat|\qbar)\stackrel{\eqref{eq:exclude_distinguished}}{=}\sum_{\s \in \Sequiv(\qbar)}\thetabar(\s) \phibar^{\s}(\cbar|\qbar)=\sum_{\s \in \Sequiv(\qbar)} \thetabar(\s) \phibar^{\sran}(\cbar|\qbar)= \phibar^{\sran}(\cbar|\qbar).
 \end{align}
 \item[(2)] Players have no incentive to deviate from fixed point strategy profile $\qbar$ because it is an equilibrium of the game $\G$ with the fixed point belief $\thetabar$.
\end{itemize}

Following Theorem \ref{theorem:convergence_eq}, the set of all fixed points, denoted as $\FP$, can be written as follows: 
\begin{align}\label{eq:FP}
    \FP = \left\{\(\thetabar, \qbar\)\left\vert [\theta] \subseteq \Sequiv\(\qbar\), ~ \qbar \in \EQ(\thetabar)\right.\right\}.
\end{align}

We denote $\thetasran$ with $\thetasran(\sran)=1$ as the \emph{complete information belief}, and any strategy $\qsran\in \EQ(\thetasran)$ as a \emph{complete information equilibrium}. Since $[\thetasran]=\{\sran\} \subseteq \Sequiv(\qsran)$, the state $\(\thetasran, \qsran\)$ is a fixed point in the set $\FP$, and has the property that all players have complete information of the true parameter $\sran$ and choose a complete information equilibrium. Therefore, we say that $\(\thetasran, \qsran\)$ is a \emph{complete information fixed point}.


Indeed, the set $\FP$ may contain other fixed points $\(\thetabar, \qbar\)$ that are not equivalent to the complete information environment, i.e. $\thetabar \neq \thetasran$. Such belief $\thetabar$ must assign positive probability to at least one parameter $\s \neq \sran$. The equation \eqref{eq:exclude_distinguished} ensures that $\s$ is payoff-equivalent to $\sran$ given the fixed point strategy profile $\qbar$, and hence the average payoff function in \eqref{eq:utility} satisfies $\usi(\qbar)=u^{\sran}_i(\qbar)$ for all $\i \in \I$. However, for $\q \neq \qbar$, the value of $\usi(\q)$ may be different from $u^{\sran}_i(\q)$ for some players $\i \in \I$. That is, belief $\thetabar$ consistently estimates the payoff at a fixed point strategy $\qbar$ but not necessarily at all $\q \in \Q$. If one or more players had access to complete information of the true parameter $\sran$, they may have an incentive to deviate from the fixed point strategy; thus the fixed point strategy profile $\qbar$ does not correspond to a complete information equilibrium.



We now present three examples to illustrate our convergence result: Example \ref{ex:potential} is a quasi-linear Cournot game, in which there exists a fixed point with less than complete information; Example \ref{ex:coordinate} is a coordination game with non-linear payoff functions, in which the fixed point set is a continuous set; Example \ref{ex:supermodular} is a public good investment game with a unique complete information fixed point.  

\begin{example}[Cournot competition]\label{ex:potential}
 {\normalfont A set of 2 firms $\I=\{1, 2\}$ produce an identical product in a market. In each step $\t$, firm $\i$'s strategy is the production level $\qit \geq 0$. The price of the product is $p^{\t}=\alpha^\s - \beta^\s \(\q_1+\q_2\)+ \epsilon^s$, where $\s=\(\alpha^s, \beta^s\)$ is the parameter vector that represents the unknown market condition, and $\epsilon^s$ is the noise term with Normal distribution $N(0. 0.5)$. The parameter set is $\S =\{\s_1, \s_2\}$, where $\s_1 = \(2, 1\)$ and $\s_2= \(4, 3\)$. The true parameter is $\sran=\s_1$.  The marginal cost of each firm is 0. Therefore, the payoff of firm $\i$ in step $\t$ is $\ct_i=\qit \(\alpha^\s - \beta^\s \(\qt_1+\qt_2\)+ \epsilon^s\)$ for each $\s \in \S$.
 
 The information system updates belief $\thetat$ based on the total production $\tilde{q}^\t=\qt_1+\qt_2$ and the realized price $\tilde{\c}^\t= p^\t =\ct_i/\qt_i$ rather than the production $\qit$ and payoff $\ct_i$ of each firm $\i \in \I$. We can check that $\phi^s(\ct|\qt) = \tilde{\phi}^s(\tilde{\c}^\t|\tilde{\q}^\t)$ for all $\s \in \S$, where $\tilde{\phi}^s(\tilde{\c}^\t|\tilde{\q}^\t)$ is the probability density function of the price given the total production. Thus, from \eqref{eq:equivalent_bayesian}, we know that $\(\tilde{\q}^\t, \tilde{\c}^\t\)$ is a sufficient statistic of $\(\qt, \ct\)$, and the belief update based on the total production and price is equivalent to that based on the strategy profile and payoffs.

For any $\theta \in \Delta(\S)$, the game has a unique equilibrium strategy profile $\gwe(\theta)= \(\frac{\alphabar(\theta)}{3\betabar(\theta)}, \frac{\alphabar(\theta)}{3\betabar(\theta)}\)$, where $ \alphabar(\theta) = \sum_{\s \in \S} \theta(\s)\alpha^\s$ and $\betabar(\theta) = \sum_{\s \in \S} \theta(\s)\beta^\s$. The complete information fixed point is $\thetasran=\(1,0\)$ and $\qsran=\(2/3, 2/3\)$. We can check that when $\thetabar= \(0/5, 0.5\)$ and $\qbar=\(0.5, 0.5\) \in \EQ(\thetabar)$, $[\thetabar] \subseteq \Sequiv(\qbar)=\{\s_1, \s_2\}$. Thus, $\(\thetabar, \qbar\)$ is another fixed point. Moreover, any $\theta \neq \thetasran$ must include $\s_2$ in the support set; but $\qbar= \(0.5, 0.5\)$ is the only strategy profile for which $\s_1$ and $\s_2$ are payoff-equivalent. Thus, there does not exist any other fixed points apart from $\(\thetasran, \qsran\)$ and $\(\thetabar, \qbar\)$, i.e. $\FP=\left\{(\thetasran, \qsran), (\thetabar, \qbar)\right\}$.



Now consider an initial state $\thetazero=\(0.1, 0.9\)$ and $\q^1=\(0.25, 0.25\)$. In each step $\t$, players entirely adopt the equilibrium strategy profile $\gwe(\thetatone)$ based on the updated belief, i.e. $\ait=1$ for all $\i$ and all $\t$. We can check that $\gwe(\thetatone)$ satisfies \textbf{(A1)}, and the stepsizes satisfy \textbf{(A2)}. Fig. \ref{fig:cournot_eq_belief_true} - \ref{fig:cournot_eq_strategy_true} demonstrate the sequence of beliefs and strategies in a realization of the learning dynamics that converges to the complete information fixed point $\(\thetasran, \qsran\)$. Fig. \ref{fig:cournot_eq_belief_fixed} -- \ref{fig:cournot_eq_strategy_fixed} illustrate the states of another realization that converges to the other fixed point $\(\thetabar, \qbar\)$. 
\begin{figure}[ht]
\centering
    \begin{subfigure}[b]{0.4\textwidth}
        \includegraphics[width=\textwidth]{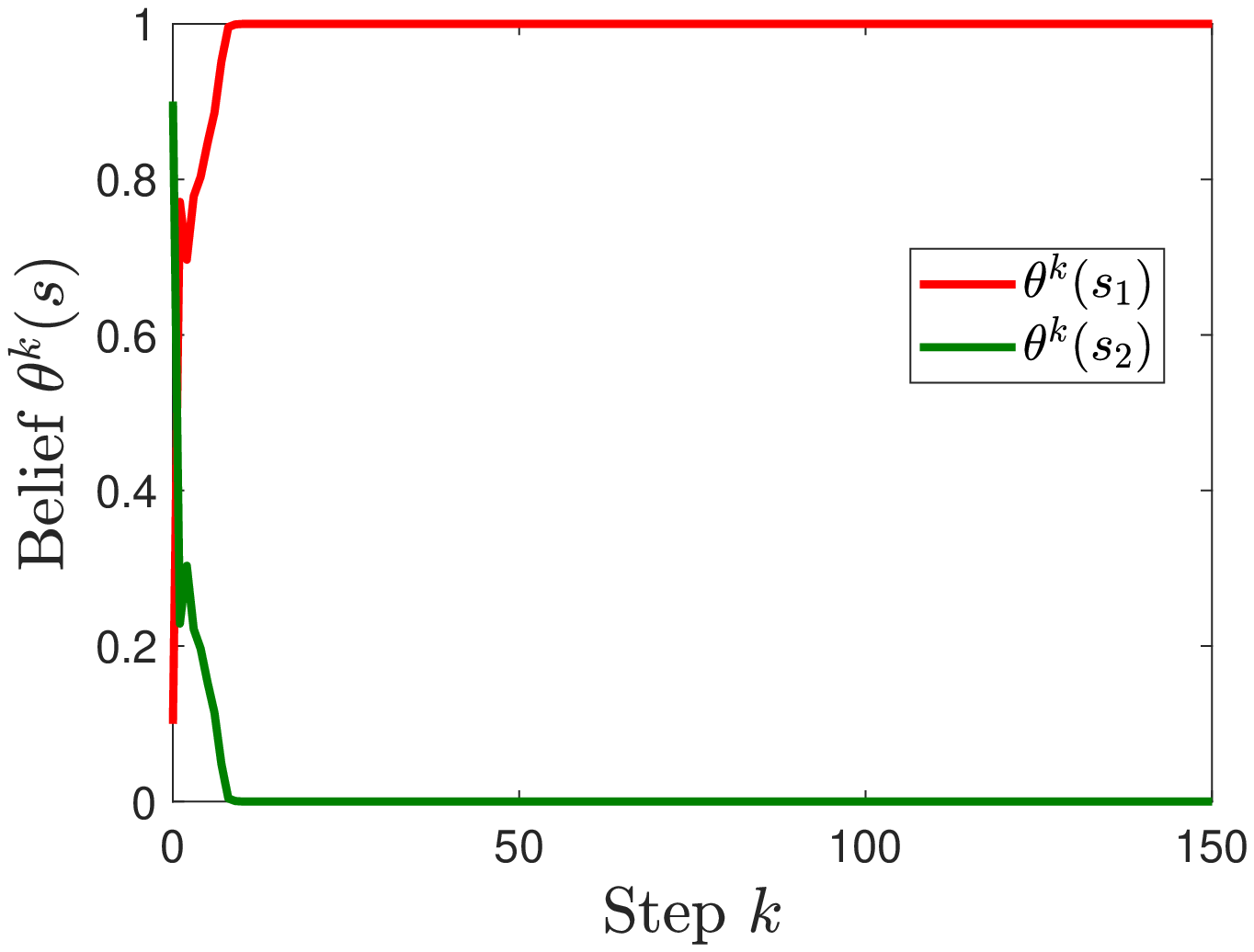}
        \caption{}
        \label{fig:cournot_eq_belief_true}
    \end{subfigure}
~
	\begin{subfigure}[b]{0.4\textwidth}
        \includegraphics[width=\textwidth]{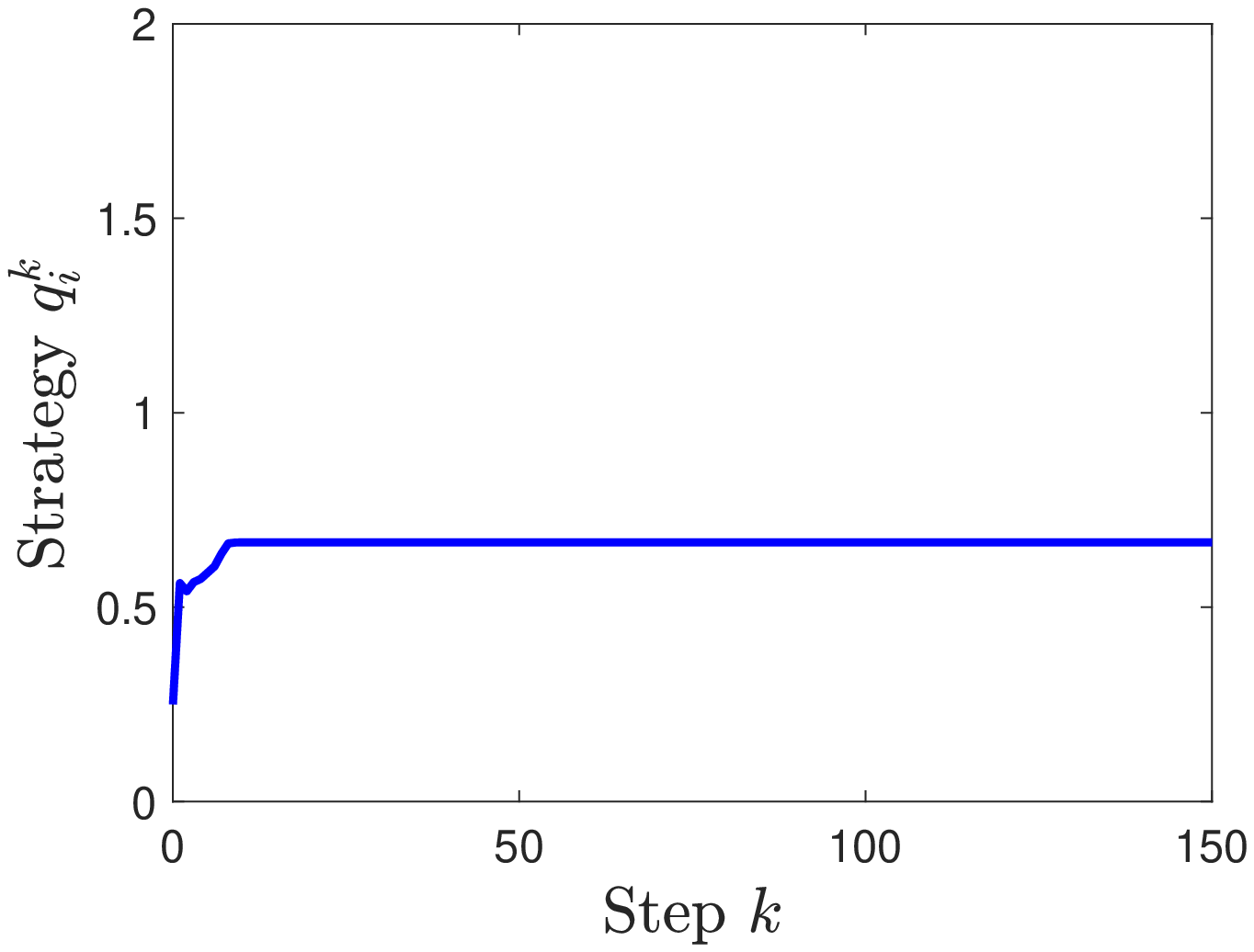}
        \caption{}
        \label{fig:cournot_eq_strategy_true}
    \end{subfigure}\\
 \begin{subfigure}[b]{0.4\textwidth}
        \includegraphics[width=\textwidth]{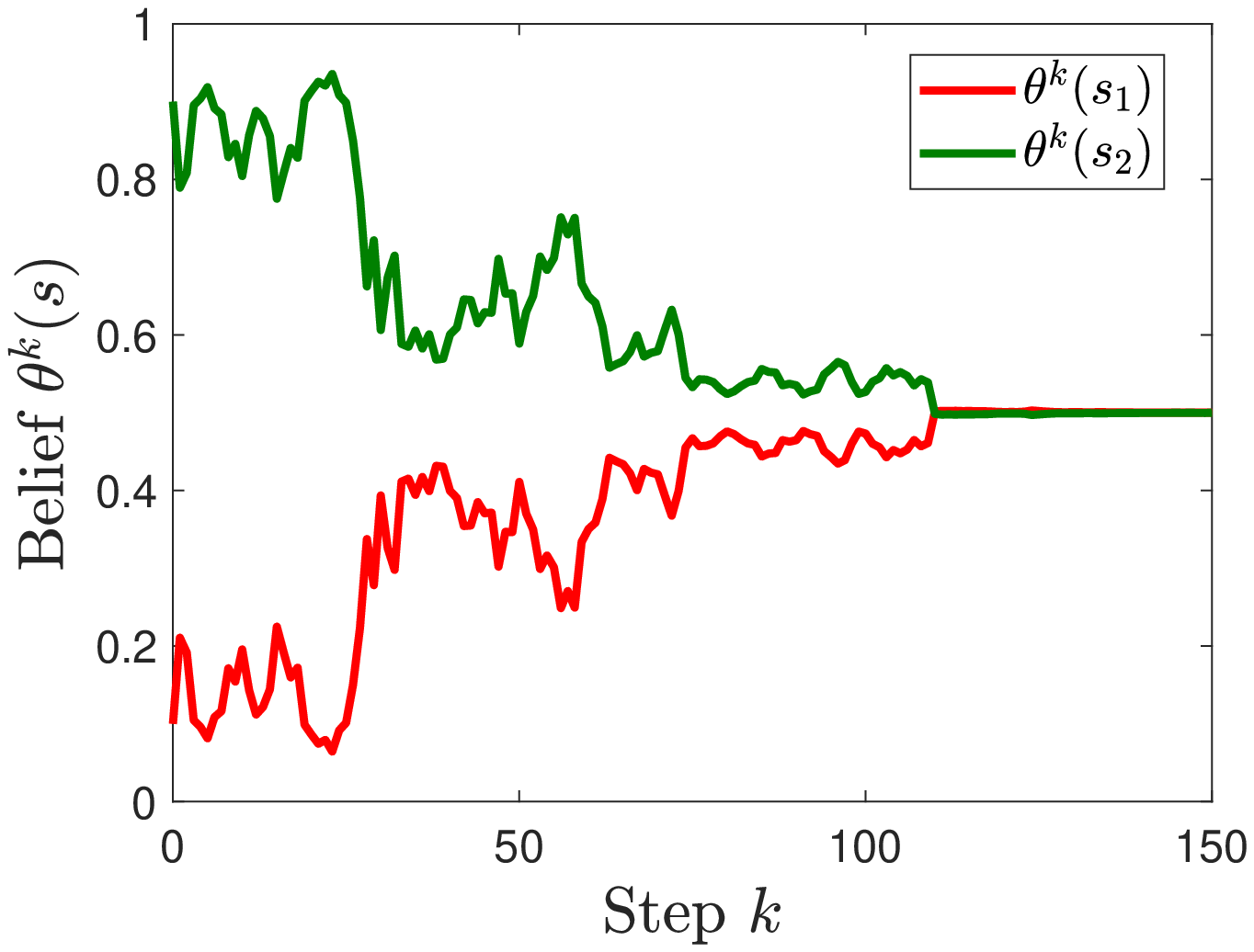}
        \caption{}
        \label{fig:cournot_eq_belief_fixed}
    \end{subfigure}
~
	\begin{subfigure}[b]{0.4\textwidth}
        \includegraphics[width=\textwidth]{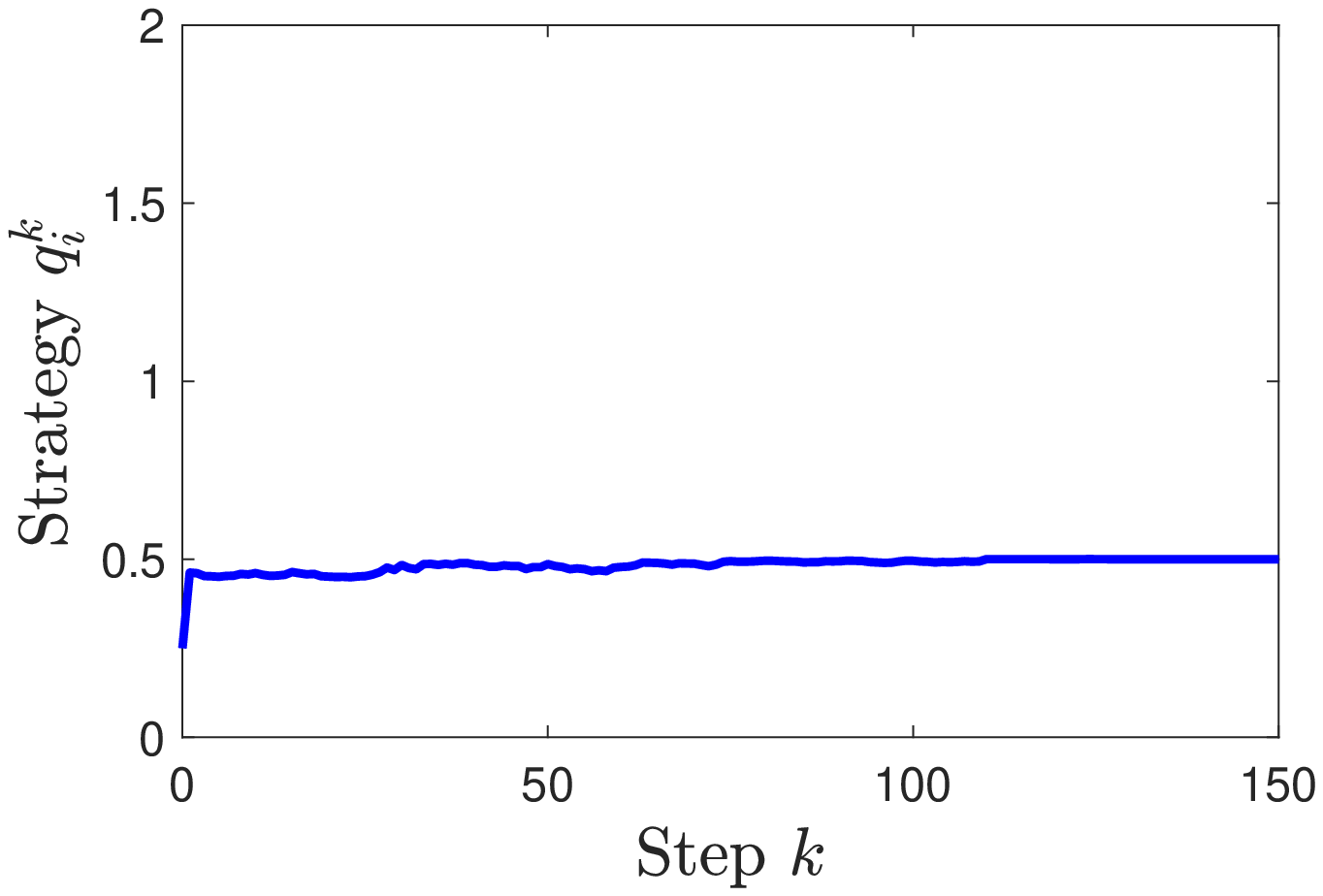}
        \caption{}
        \label{fig:cournot_eq_strategy_fixed}
    \end{subfigure}
    \caption{Beliefs and strategies in learning dynamics with equilibrium strategies in Cournot game: (a) - (b) Convergence to $\(\thetasran, \qsran\)$; (c) - (d) Convergence to $\(\thetabar, \qbar\)$.}
\label{fig:convergence_eq}
\end{figure}

 
}
  \end{example}

\begin{example}[Coordination with safe margin]\label{ex:coordinate}
{\normalfont Two players $\I = \{1, 2\}$ coordinate their strategies in a game. In each step $\t$, player 1's strategy is $\qt_1 \in [0, 2]$ and player 2's strategy is $\qt_2 \in [1, 4]$. The set of strategy profile is $\Q = [0, 2] \times [1, 4]$. Both players pay a cost if the difference between the two strategies $|\qt_1- \qt_2|$ exceeds a safe margin $\s$, which is an unknown parameter and belongs to the set $\S = \{\s_1=0, \s_2 =0.5, \s_3=1.5\}$. The true parameter is $\sran=\s_3$. Additionally, player 1 prefers to choose small $\qt_1$ while player 2 prefers high $\qt_2$. The player payoffs are as follows: 
\begin{align*}
    \ct_1 &= -2 \(\max\(|\qt_1 - \qt_2|, \s\) - \s\)^2- \qt_1 +\epsilon_1, \\
    \ct_2 &= -2 \(\max\(|\qt_1 - \qt_2|, \s\) - \s\)^2+ \qt_2 +\epsilon_2,
\end{align*}
where $\epsilon_1$ and $\epsilon_2$ are the noise terms with distribution $N(0,2)$. The information system updates the belief $\thetat$ based on $\qt$ and $\ct$ as in \eqref{eq:update_belief}. 

For each $\theta \in \Delta(\S)$, the equilibrium set $\EQ(\theta)$ is as follows: 
\begin{align}\label{eq:EQ_description}
    \EQ(\theta) = \left\{\begin{array}{ll}
    \left\{(\q_1, \q_2) \in \Q\left\vert\q_2-\q_1 = \frac{\theta(\s_2) + 3 \theta(\s_3)}{2}+\frac{1}{4}\right.\right\},&  \text{if $\theta(\s_2) + 3 \theta(\s_3) > 2.5$,} \\
    \left\{(\q_1, \q_2) \in \Q\left\vert\q_2-\q_1 = \frac{2 \theta(\s_2) + 1}{4 (\theta(\s_1)+\theta(\s_2))}\right.\right\},&  \text{if $\theta(\s_1)< 0.5$ and $\theta(\s_2) + 3 \theta(\s_3) \leq 2.5$,}\\
    \left\{(\q_1, \q_2)\in \Q\left\vert\q_2-\q_1 = \frac{1}{4\theta(\s_1)}\right.\right\},&  \text{if $\theta(\s_1) \geq 0.5$}.
    \end{array}\right.
\end{align}
 
We now characterize the fixed point set $\FP$. For any $\theta \in \Delta(\S)$, we have the following three cases from \eqref{eq:EQ_description}: (1) If $\theta(\s_2) + 3 \theta(\s_3) > 2.5$, then any $\q \in \EQ(\theta)$ satisfies $\q_2-\q_1 >1.5$. Since $\Sequiv(\q)=\{\s_3\}$ for any such $\q$, the only belief that can be a fixed point belief is $\thetasran$; (2) If $\theta(\s_1)< 0.5$ and $\theta(\s_2) + 3 \theta(\s_3) \leq 2.5$, then $\q_2-\q_1 \in (0.5, 1.5]$ for any $\q \in \EQ(\theta)$ and $\Sequiv(\q)=\{\s_3\}$. Again, we obtain that only $\thetasran$ is possible to be a fixed point belief. However, $\thetasran$ does not satisfy the assumption that $\theta(\s_2) + 3 \theta(\s_3) \leq 2.5$. Therefore, no fixed point exists in this case; (3) If $\theta(\s_1) \geq 0.5$, then $\q_2-\q_1 \leq 0.5$ for any $\q \in \EQ(\theta)$. In this case, $\Sequiv(\q)=\{\s_2, \s_3\}$. However, since $\theta(\s_1) \geq 0.5$, $\s_1$ must be in the support set of the fixed point belief. Hence, we know that no fixed point exists in this case. We can thus conclude the fixed point set of the coordination game is given by:
\[\FP = \{\(\thetasran, \EQ(\thetasran)\)\}=\left\{\(\thetasran, \qsran\)\left\vert
    \begin{array}{ll}
    \thetasran=\(0,0,1\), \text{ and}\\
    \qsran_1\in [0,2], \text{ and } \qsran_2 = \qsran_1+ 7/4
    \end{array}\right.\right\}.\]


Consider the learning dynamics with initial belief $\thetazero=(1/3, 1/3, 1/3)$ and the initial strategy $\q^1=\(1, 2\)$. The stepsizes are $a^\t_1=1$ and $a^\t_2=0$ for odd $\t$ and $a^\t_1=0$ and $a^\t_2=1$ for even $\t$, i.e. player 1 (resp. player 2) chooses the updated equilibrium strategy in odd (resp. even) steps, and does not update the strategy in even (resp. odd) steps. The strategy update uses the equilibrium $\gwe(\thetat) \in \EQ(\thetat)$ with $\gwe_1(\thetat)=0$. We can check that \textbf{(A1)} -- \textbf{(A2)} are satisfied. 
Fig. \ref{fig:safemargin_eq} shows that the states of the learning dynamics converge to a complete information fixed point $\thetasran=(0, 0, 1)$ and $\qsran=(0,7/4)$. 

\begin{figure}[ht]
\centering

	\begin{subfigure}[b]{0.4\textwidth}
        \includegraphics[width=\textwidth]{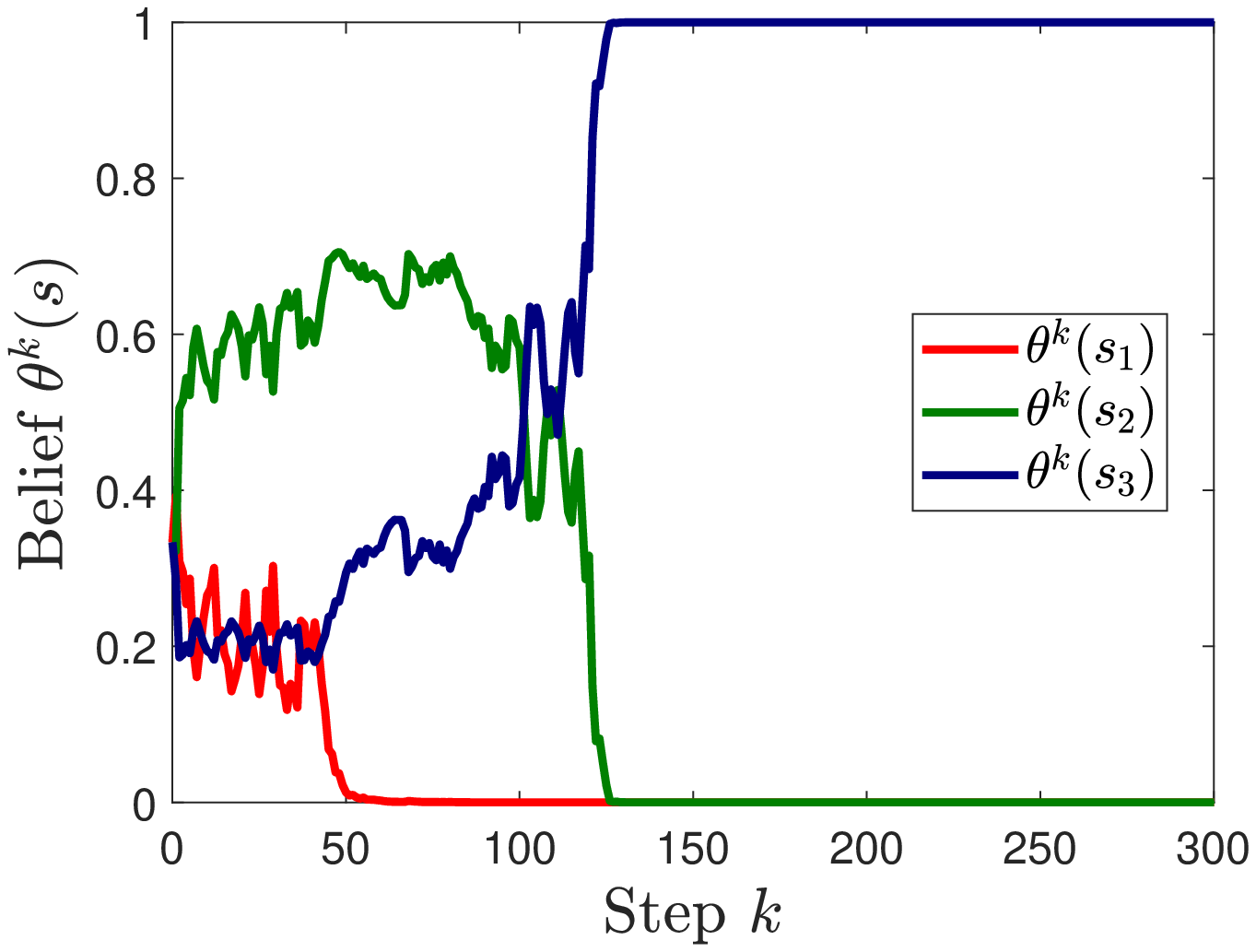}
        \caption{}
        \label{fig:safemargin_eq_belief}
    \end{subfigure}
    ~
    \begin{subfigure}[b]{0.4\textwidth}
        \includegraphics[width=\textwidth]{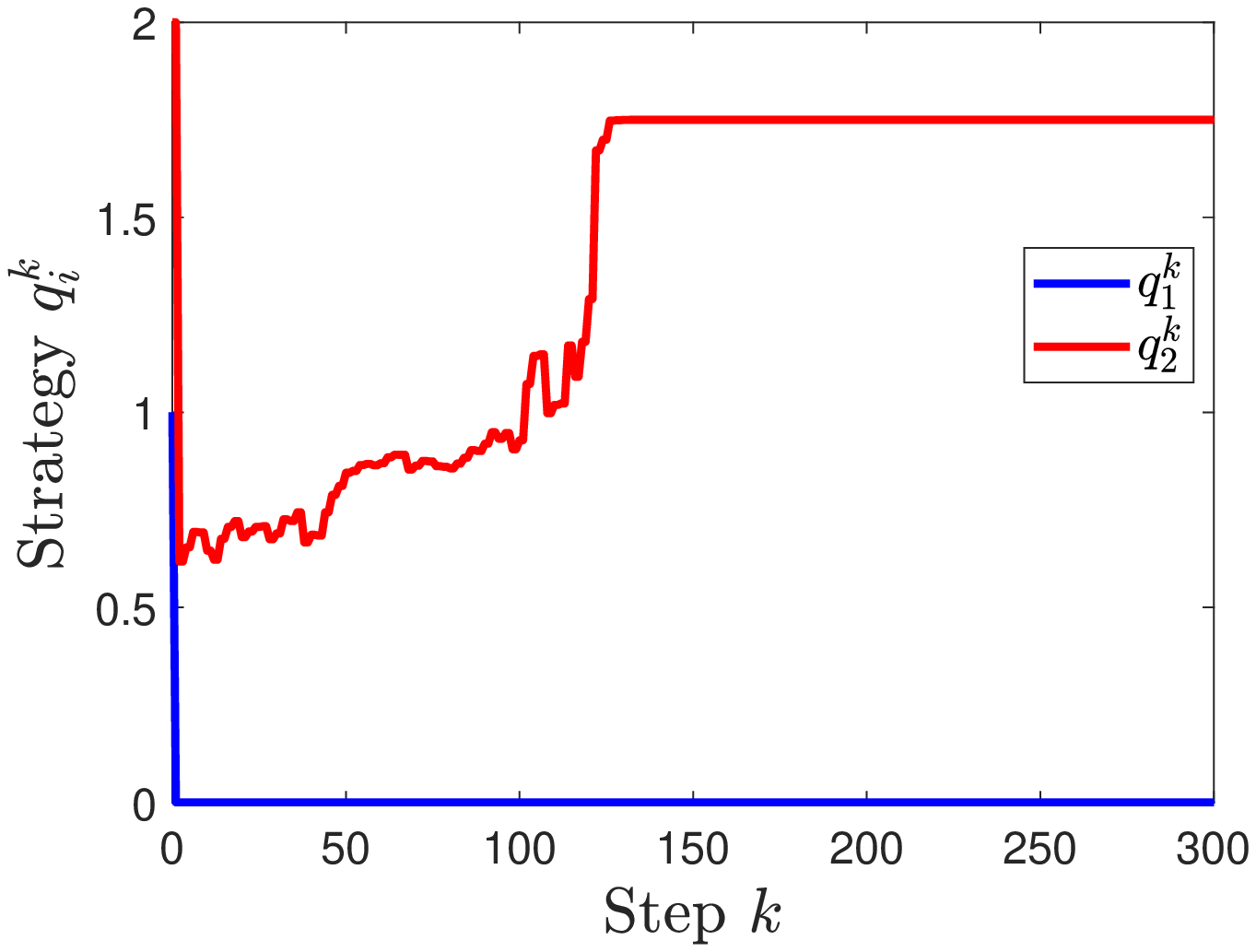}
        \caption{}
        \label{fig:safemargin_eq_strategy}
    \end{subfigure}
    \caption{Beliefs and strategies in learning dynamics with equilibrium strategies converge to a complete information fixed point in coordination game with safe margin.}
\label{fig:safemargin_eq}
\end{figure}
}
\end{example}

\begin{example}[Public good investment]\label{ex:supermodular}
{\normalfont Two players simultaneously invest in a public good project. In each step $\t$, the strategy $\qit \geq 0$ is the non-negative level of investment of player $\i$. Given the strategy profile $\qt = \(\qt_1, \qt_2\)$, the return of a unit investment in public good is randomly realized: $r^\t=\alphas +\qt_1+ \qt_2 + \epsilon^\s$, where $\s \in \S$ is the unknown parameter and $\epsilon^\s$ is the noise term with zero mean. The unknown parameter $\s$ takes value in the set $\S=\{l, m, h\}$. If $\s = l$, the mean and the variance of unit investment return are low: $\alpha^{l}=0$ and $\epsilon^{l} \sim N(0, 3)$. If $\s=m$, the mean and the variance of unit investment return are medium: $\alpha^{m}=1$ and $\epsilon^{m} \sim N(0, 5)$. If $\s=h$, the mean and the variance of unit investment return are high: $\alpha^{h}=2$ and $\epsilon^{h} \sim N(0, 10)$. The true parameter is $\sran=m$. The cost of investment for each player is $3 \(\qit\)^2$. Therefore, the payoff of each player $\i \in \I$ is $\ct_i= \qit(\alphas +\qt_1+ \qt_2+\epsilon^\s) - 3 \(\qit\)^2 = \qit(\alphas -2 \qit+ \qt_{\mi} + \epsilon^\s)$ for all $\s \in \S$. 

In each step $\t$, the information system updates belief $\thetat$ based on the total investment $\tilde{\q}^\t= \qt_1+\qt_2$ and the unit investment return $\tilde{\c}^\t=r^\t$. Analogous to Example \ref{ex:potential}, $\(\tilde{\q}^\t, \tilde{\c}^\t\)$ is a sufficient statistics of $\(\qt, \ct\)$, thus the belief update given $\(\tilde{\q}^\t, \tilde{\c}^\t\)$ is equivalent to that with $\(\qt, \ct\)$. 

For any $\theta \in \Delta(\S)$, the unique equilibrium strategy profile is $\gwe(\theta)= \(\frac{\alphabar(\theta)}{3}, \frac{\alphabar(\theta)}{3}\)$, where $\alphabar(\theta) = \sum_{\s \in \S} \theta(\s)\alpha^s$. Moreover, since $\Sequiv(\q)=\{\sran=m\}$ for any $\q \in \EQ(\theta)$, the unique fixed point is the complete information fixed point, i.e. $\FP = \{\(\thetasran, \qsran\)= \(\(0, 1, 0\), \(1/3, 1/3\)\)\}$.

Consider the learning dynamics starting with initial belief $\thetazero = \(0.5, 0.4, 0.1\)$ and strategy profile $\q^1=\(1, 0\)$. Player 1's stepsizes are $a^{\t}_1=1/\t$ for all $\t$. Player 2's stepsize is $\at_2=1/t$ in steps $\t=2t-1$, and $\at=1/2t$ for all $\t=2t$, where $t=1, 2, \dots$. The unique equilibrium strategy profile and the stepsizes satisfy \textbf{(A1)} and \textbf{(A2)}, respectively. Fig. \ref{fig:public_eq_belief_true} -- \ref{fig:public_eq_strategy_true} illustrate that the sequence of beliefs and strategies converge to the unique complete information fixed point.
\begin{figure}[htp]
\centering
    \begin{subfigure}[b]{0.4\textwidth}
        \includegraphics[width=\textwidth]{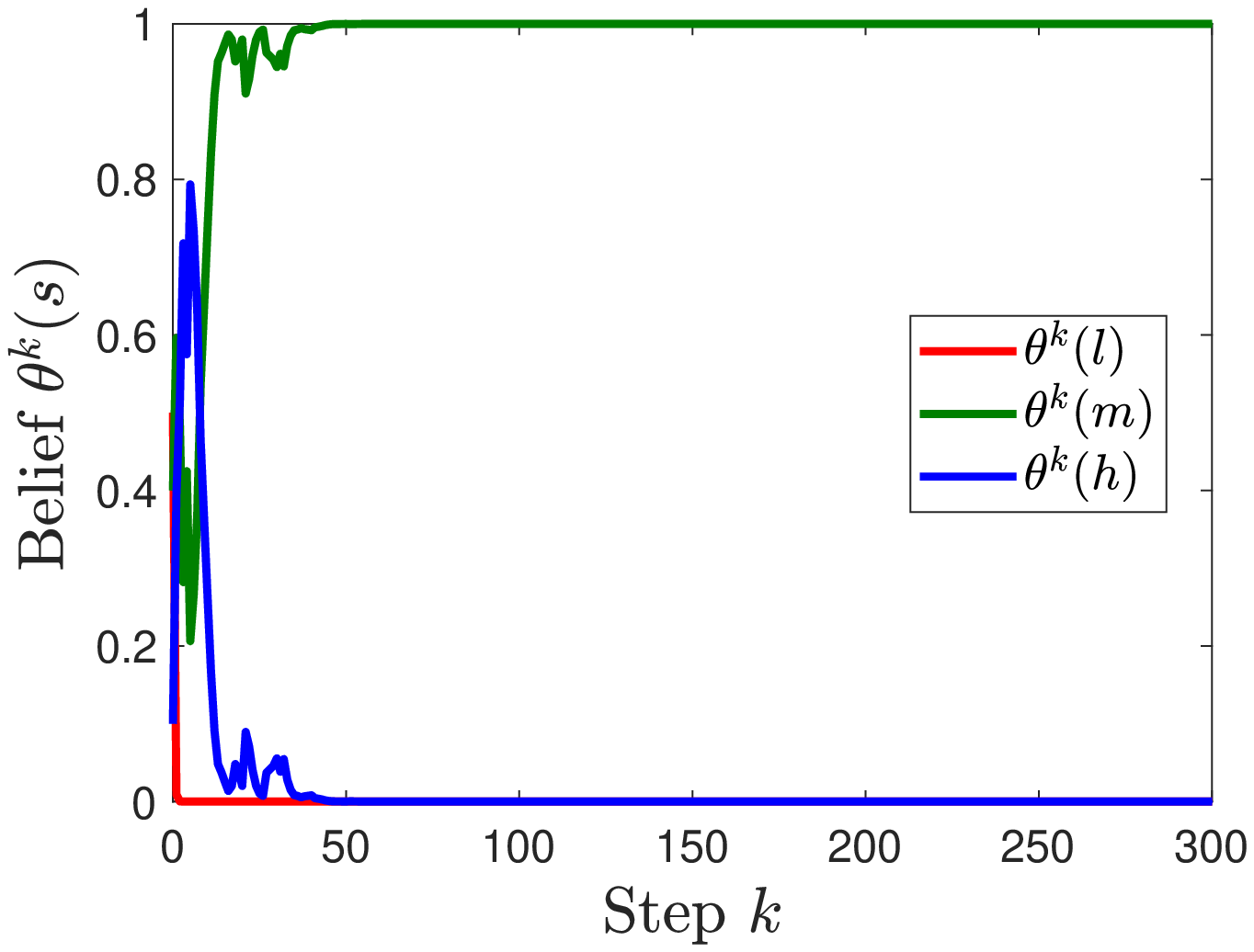}
        \caption{}
        \label{fig:public_eq_belief_true}
    \end{subfigure}
~
	\begin{subfigure}[b]{0.4\textwidth}
        \includegraphics[width=\textwidth]{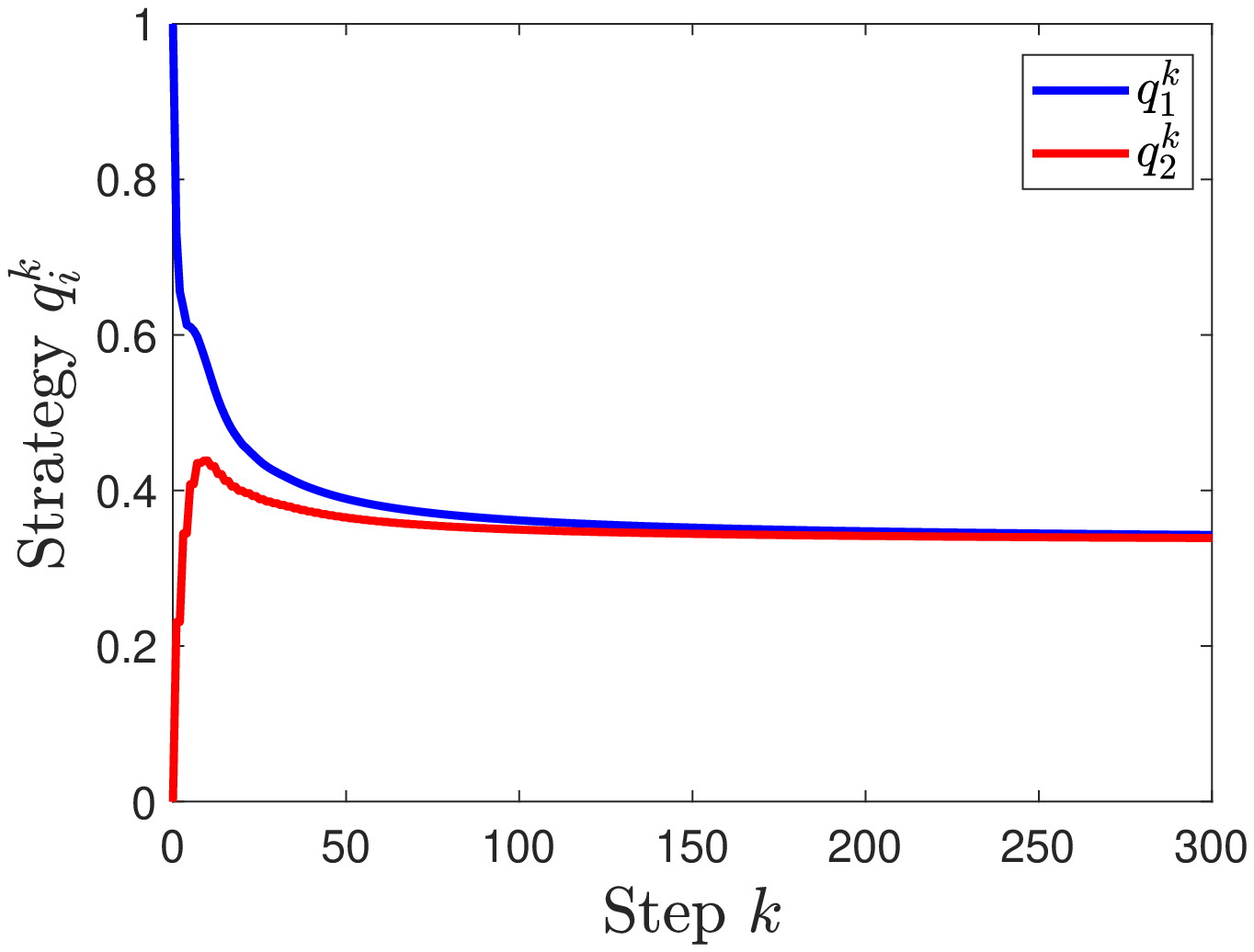}
        \caption{}
        \label{fig:public_eq_strategy_true}
    \end{subfigure}
    \caption{Beliefs and strategies in learning dynamics with equilibrium strategies converge to the complete information fixed point in the public good game.} 
\label{fig:learning_public_eq}
\end{figure}
}
    
    \end{example}

\subsection{Stability}\label{subsec:stability_eq}
In this section, we analyze both local and global stability properties of fixed point belief $\thetabar$ and the associated equilibrium set $\EQ(\thetabar)$. We first introduce the definitions of local and global stability. To begin with, for any $\epsilon>0$, an $\epsilon$-neighborhood of belief $\thetabar$ is defined as $N_{\epsilon}(\thetabar) \deleq \left\{\theta| \|\theta - \thetabar\|< \epsilon \right\}$. For any $\delta>0$, we define the $\delta$-neighborhood of equilibrium set as $N_{\delta}(\EQ(\thetabar)) \deleq  \left\{ \q| \mathrm{dist}\(\q, \EQ(\thetabar)\)<\delta\right\}$, where $\mathrm{dist}\(\q, \EQ(\thetabar)\) = \min_{\q'\in \EQ(\thetabar)}\|\q-\q'\|$ is the Euclidean distance between $\q$ and the set $\EQ(\thetabar)$. %
\begin{definition}[Local stability]\label{def:local}
A fixed point belief $\thetabar \in \Delta(\S)$ and the associated equilibrium set $\EQ(\thetabar)$ is \emph{locally stable} if for any $\gamma \in (0,1)$ and any $\thetaepfin, \loadepfin>0$, there exist $\thetaep, \loadep>0$ such that for learning dynamics that starts with $\thetazero\in \neighinitheta(\thetabar)$ and $\q^1 \in \neighiniload(\EQ(\thetabar))$,
\begin{center}$\lim_{\t \to \infty} \pro\(\thetat \in \neighinftheta(\thetabar), ~ \qt\in \neighinfload(\EQ(\thetabar))\)>\gamma$.
\end{center}
\end{definition}

This definition requires that when the learning starts with an initial state that is sufficiently close to a fixed point belief $\thetabar$ and the associated equilibrium set $\EQ(\thetabar)$, then the beliefs (resp. strategies) in the learning dynamics remain close to $\thetabar$ (resp. $\EQ(\thetabar)$) with high probability. In other words, when $(\bar\theta,\EQ(\bar\theta))$ is locally stable, the learning dynamics is robust to small perturbations around the belief $\thetabar$ and strategies in $\EQ(\bar\theta)$. On the other hand, if $\(\thetabar, \EQ(\thetabar)\)$ is locally unstable, then the state of learning dynamics can leave the neighborhood of $\thetabar$ and $\EQ(\thetabar)$ with a positive probability even when the initial state $\(\thetazero, \q^1\)$ is arbitrarily close to $\(\thetabar, \EQ(\thetabar)\)$.

\begin{definition}[Global stability]\label{def:global}
A fixed point belief $\thetabar \in \Delta(\S)$ and the associated equilibrium set $\EQ(\thetabar)$ is \emph{globally stable} if for any initial state $\(\thetazero, \q^1\)$, the beliefs of the learning dynamics $\(\thetat\)_{\t=1}^{\infty}$ converge to $\thetabar$ and the strategies $\(\qt\)_{\t=1}^{\infty}$ converge to $\EQ(\thetabar)$ with probability 1. 
\end{definition}


Note that these stability notions are not defined for individual fixed points, but rather for the
tuple $\(\bar\theta, \EQ(\bar\theta)\)$, i.e. the set of fixed points with an identical belief $\bar\theta$. This becomes important when the game $G$ has multiple equilibria; i.e., $\EQ(\theta)$ is not a singleton set for some belief $\theta\in\Delta(S)$. Our stability notions do not hinge on the choice of a particular equilibrium in the strategy updates, i.e. a fixed point that is locally or globally stable when the learning dynamics evolve under a given equilibrium $g$ retains this property under a different equilibrium $\tilde{g}$.

The following result provides sufficient conditions for local stability of the learning dynamics with equilibrium strategy updates: 
\begin{theorem}\label{theorem:stability_eq}
A fixed point belief $\bar\theta\in\Delta(S)$ and the associated equilibrium set $EQ(\bar\theta)$ is locally stable under the learning dynamics \eqref{eq:update_belief}
and \eqref{eq:eq_update} if Assumptions \textbf{(A1)} -- \textbf{(A2)} are satisfied and the following conditions hold: (a) $EQ(\theta)$ is upper-hemicontinuous in $\theta$; and
(b) $\exists \delta>0$ such that $[\bar\theta]\subseteq \Sequiv(q)$ for all
$q\in N_\delta(\EQ(\bar\theta))$. 
\end{theorem}

In Theorem \ref{theorem:stability_eq}, condition {\em (a)} ensures that when the belief is locally perturbed in the neighborhood of $\thetabar$, the equilibrium of perturbed belief remains close to the fixed point equilibrium set $\EQ(\thetabar)$; thus the updated strategy given the perturbed belief in \eqref{eq:eq_update} also remains close to the set $\EQ(\thetabar)$. Condition {\em (b)} further ensures that the set of payoff-equivalent parameters do not change under local perturbations of $\q$, so that the belief update \eqref{eq:update_belief} given any strategy in the small neighborhood of $\EQ(\thetabar)$ keeps the beliefs of all parameter in $[\thetabar]$ close to their probabilities in $\thetabar$, and excludes the parameters that are not in $[\thetabar]$. One can verify this condition by checking whether or not the KL-divergence  $D_{KL}\(\phi^{\s'}(\c|\q)||\phi^s(\c|\q)\)$ between any two parameters $\s , \s' \in [\thetabar]$ changes with $\q$ in the neighborhood of $\EQ(\thetabar)$. These two sufficient conditions together ensure that, with high probability, the beliefs and strategies in learning with equilibrium strategies remain in a local neighborhood of the fixed point $\thetabar$ and $\EQ(\thetabar)$.

To prove that $\(\thetabar, \EQ(\thetabar)\)$ is locally stable, for any $\thetaepfin, \loadepfin>0$ and any $\gamma \in (0, 1)$, we need to find positive numbers $\thetaep, \loadep>0$ such that if the initial state satisfies $\thetazero \in N_{\thetaep}(\thetabar)$ and $\q^1 \in N_{\loadep}(\EQ(\thetabar))$, then $\lim_{\t \to \infty} \pro\(\thetat \in \neighinftheta(\thetabar), ~ \qt\in \neighinfload(\EQ(\thetabar))\)>\gamma$ (Definition \ref{def:local}). The Lemmas \ref{lemma:constrained_set} -- \ref{lemma:other_parameters} characterize such $\thetaep$ and $\loadep$. 

\begin{lemma}\label{lemma:constrained_set}
For any $\thetaepfin, \loadepfin>0$, there exists $\epsilonhat \in (0, \thetaepfin)$ and $\deltahat \in (0, \loadepfin)$ that satisfies (i) $[\thetabar] \subseteq \Sequiv(\q)$ for any $\q \in N_{\deltahat}(\EQ(\thetabar))$; (ii) $\EQ(\theta) \subseteq N_{\deltahat}(\EQ(\thetabar))$ for any $\theta \in N_{\epsilonhat}(\thetabar)$. Additionally, $\lim_{\t \to \infty} \pro\(\thetat \in \neighinftheta(\thetabar),\right.$ $ \left. \qt \in  \neighinfload(\EQ(\thetabar))\)\geq \pro\(\thetat \in  N_{\epsilonhat}(\thetabar), ~ \qt \in N_{\deltahat}(\EQ(\thetabar)),  \forall \t\)$. 
\end{lemma}

\noindent\textbf{\emph{Proof of Lemma \ref{lemma:constrained_set}.}} {\em(i)} We set $\deltahat = \min \{\delta, \loadepfin\}$, where $\delta$ is taken from condition (b) in Theorem \ref{theorem:stability_eq}. Since $[\thetabar] \subseteq \Sequiv(\q)$ for any $\q \in N_{\delta}(\EQ(\thetabar))$ and $N_{\deltahat} \(\EQ(\thetabar)\) \subseteq N_{\delta}(\EQ(\thetabar))$, we know that $[\thetabar] \subseteq \Sequiv(\q)$ for all $\q \in N_{\deltahat}\(\EQ(\thetabar)\)$. 

{\em (ii)} Since $\EQ(\theta)$ is upper-hemicontinuous in $\theta$ and $\deltahat>0$, there exists $\epsilon>0$ such that if $\|\theta-\thetabar\|<\epsilon$, then $\EQ(\theta) \subseteq N_{\deltahat}\(\EQ(\thetabar)\)$. By setting $\epsilonhat = \min\{\epsilon, \thetaepfin\}$, we know that $\EQ(\theta) \subseteq N_{\deltahat}\(\EQ(\thetabar)\)$ for any $\theta \in N_{\epsilonhat}(\thetabar)$. Additionally, if $\thetat \in N_{\thetaepfin}(\thetabar)$ and $\qt \in N_{\loadepfin}\(\EQ(\thetabar)\)$ for all $\t$, then we must have $\lim_{\t \to \infty} \thetat \in N_{\thetaepfin}(\thetabar)$ and $\lim_{\t \to \infty} \qt \in N_{\loadepfin}\(\EQ(\thetabar)\)$. Therefore, we have $\lim_{\t \to \infty} \pro\(\thetat \in \neighinftheta(\thetabar),\right.$ $\left. \qt \in \neighinfload(\EQ(\thetabar))\) \geq \pro\(\thetat \in N_{\thetaepfin}(\thetabar), ~ \qt \in N_{\loadepfin}(\EQ(\thetabar)), ~ \forall \t\)$. Finally, since $\epsilonhat \leq \thetaepfin$ and $\deltahat \leq \loadepfin$, we conclude that $\pro\(\thetat \in N_{\thetaepfin}(\thetabar), ~ \qt \in N_{\loadepfin}(\EQ(\thetabar)), ~ \forall \t\) \geq \pro\(\thetat \in N_{\epsilonhat}(\thetabar), ~ \qt \in N_{\deltahat}(\EQ(\thetabar)), ~ \forall \t\)$.

 \QEDA

\vspace{0.2cm}



Since the sub-neighborhoods $N_{\epsilonhat}(\thetabar) \subseteq N_{\thetaepfin}(\thetabar)$ and $N_{\deltahat}\(\EQ(\thetabar)\) \subseteq N_{\loadepfin}\(\EQ(\thetabar)\)$, it is sufficient to characterize $\thetaep$ and $\loadep$ such that $\thetat \in N_{\epsilonhat}(\thetabar)$ and $\qt \in N_{\deltahat}\(\EQ(\thetabar)\)$ for all $\t$ with probability higher than $\gamma$. From Property {\em (ii)} in Lemma \ref{lemma:constrained_set}, we know that if $\thetat \in N_{\epsilonhat}(\thetabar)$ for all $\t$, then the equilibrium strategy profile $\gwe(\thetat)$ must be in the neighborhood $N_{\deltahat}\(\EQ(\thetabar)\)$ for all $\t$. From the strategy update \eqref{eq:eq_update}, we further know that the initial strategy $\q^1 \in N_{\deltahat}\(\EQ(\thetabar)\)$ guarantees that $\qt \in N_{\deltahat}\(\EQ(\thetabar)\)$ for all $\t$. Therefore, what remains is to find the neighborhood of the initial belief $\theta^1$ such that $\thetat \in N_{\epsilonhat}(\thetabar)$ for all $\t$ with probability higher than $\gamma$.

Now to show that $\thetat \in N_{\epsilonhat}\(\thetabar\)$ for all $\t$, we need to show that $|\thetat(\s)-\thetabar(\s)|<\frac{\epsilonhat}{|\S|}$ for all $\s \in \S$ and all $\t$. We separately analyze the beliefs of all $\s \in \S\setminus [\thetabar]$ (i.e. the set of parameters with zero probability in $\thetabar$) in Lemma \ref{lemma:stopping_time}, and that of $\s \in [\thetabar]$ in Lemma \ref{lemma:other_parameters}. To proceed, we need the following thresholds:
\begin{subequations}
\begin{align}
    \rhoone &\deleq \min_{\s \in [\thetabar]} \left\{\frac{(1-\gamma) \thetabar(\s)\epsilonhat}{(1-\gamma +|\Sbar|)(|\S\setminus [\thetabar]|+1)|\S|+(1-\gamma) \epsilonhat}\right\}, \label{eq:rhoone} \\
    \rhotwo &\deleq\frac{\epsilonhat}{(|\Sbar|+1)|\S|}, \label{eq:rhotwo}\\
    \rhothree &\deleq \min_{\s \in [\thetabar]}\left\{\frac{\epsilonhat - |\Sbar| |\S|\rhotwo \thetabar(\s)}{|\S|-|\Sbar||\S| \rhotwo},~  \frac{\epsilonhat}{|\S|+ |\Sbar|\(\thetabar(\s)|\S|+ \epsilonhat\)}, ~ \thetabar(\s) \right\}. \label{eq:rho_three}
\end{align}
\end{subequations}

Lemma \ref{lemma:stopping_time} provides a condition on the initial belief $\thetazero$ under which the belief of parameters in $\Sbar$ is less than the threshold $\rhotwo$ for all steps with probability higher than $\gamma$. Note that $\thetat(\s) \leq \rhotwo$ ensures $|\thetat(\s)-\thetabar(\s)|< \frac{\epsilonhat}{|\S|}$ since $\thetabar(\s)=0$ for all $\s \in \Sbar$ and $\rhotwo < \frac{\epsilonhat}{|\S|}$. (The threshold $\rhotwo$ is specifically constructed to bound the beliefs of the remaining parameters $[\thetabar]$ as shown in Lemma \ref{lemma:other_parameters}.)


\begin{lemma}\label{lemma:stopping_time}
For any $\gamma \in (0, 1)$, if the initial belief satisfies
\begin{subequations}
\begin{align}
    &\thetazero(\s)< \rhoone, \quad \forall \s \in \Sbar, \label{eq:epsilonlow}\\
    &\thetabar(\s) - \rhoone < \thetazero(\s) < \thetabar(\s)+\rhoone, \quad \forall \s \in [\thetabar],\label{eq:epsilonlow_positive}
\end{align}
\end{subequations}
then $\pro\(\thetat(\s)\leq \rhotwo, ~\forall \s \in \Sbar, ~ \forall \t\)>\gamma$. 
\end{lemma}

We first discuss the main idea behind the proof of this lemma: If the initial belief of a parameter $\s \in \Sbar$ is smaller than $\rhoone$ but higher than $\rhotwo$ in some step $\t$, then the belief sequence $\(\theta^j(\s)\)_{j=1}^{\t}$ must complete at least one upcrossing of the interval $[\rhoone, \rhotwo]$ before step $\t$, i.e. the belief of $\s$ increases from below $\rhoone$ to above $\rhotwo$. Therefore, the event that $\thetat(\s)\leq \rhotwo$ for all $\t$ is equivalent to the event that belief $\thetat(\s)$ never upcrosses the interval $[\rhoone, \rhotwo]$. Additionally, by bounding the initial belief of parameters $\s \in [\thetabar]$ as in \eqref{eq:epsilonlow_positive}, we construct another interval $\left[\rhoone/\(\thetabar(\sran)- \rhoone\), \rhotwo\right]$ such that the number of upcrossings with respect to this interval completed by the sequence of belief ratios $\(\frac{\thetat(\s)}{\thetat(\sran)}\)_{\t=1}^{\infty}$ is no less than the number of upcrossings with respect to interval $[\rhoone, \rhotwo]$ completed by $\(\thetat(\s)\)_{\t=1}^{\infty}$. Recall that the belief ratios $\(\frac{\thetat(\s)}{\thetat(\sran)}\)_{\t=1}^{\infty}$ form a martingale process (Lemma \ref{lemma:theta_eq}). By applying Doob's upcrossing inequality, we obtain an upper bound on the expected number of upcrossings completed by the belief ratio of each parameter $\s \in \Sbar$, which is also an upper bound on the expected number of upcrossings made by the belief of $\s$. Using Markov inequality and the upper bound of the expected number of upcrossings, we show that with probability higher than $\gamma$, no belief $\thetat(\s)$ of any parameter $\s \in \Sbar$ can ever complete a single upcrossing with respect to the interval $[\rhoone, \rhotwo]$ given by \eqref{eq:rhoone} -- \eqref{eq:rhotwo}. Hence, $\thetat(\s)$ remains lower than the threshold $\rhotwo$ for all $\s \in \Sbar$ and all $\t$ with probability higher than $\gamma$. We now present the formal proof:

\vspace{0.2cm}
\noindent \textbf{\emph{Proof of Lemma \ref{lemma:stopping_time}.}} First, note that $0< \rhoone < \rhotwo< \frac{\epsilonhat}{|\S|}$. For any $\s \in \Sbar$ and any $\t>1$, we denote $U^{\t}(\s)$ the number of upcrossings of the interval $[\rhoone, \rhotwo]$ that the belief $\theta^j(\s)$ completes by step $\t$. That is, $U^{\t}(\s)$ is the maximum number of intervals $\([\underline{\t}_{i}, \overline{\t}_{i}]\)_{\i=1}^{U^{\t}(\s)}$ with $1 \leq \underline{\t}_{1} < \overline{\t}_{1} < \underline{\t}_{2} <  \overline{\t}_{2}< \cdots< \underline{\t}_{U^{\t}(\s)}< \overline{\t}_{U^{\t}(\s)} \leq \t$, such that $\theta^{\underline{\t}_{i}}(\s)<\rhoone< \rhotwo <\theta^{\overline{\t}_{i}}(\s)$ for $i=1, \dots U^{\t}(\s)$. Since the beliefs $\(\theta^\j(\s)\)_{\j=1}^{\t}$ are updated based on randomly realized payoffs $\(\c^j\)_{j=1}^{\t}$ as in \eqref{eq:update_belief}, $U^{\t}(\s)$ is also a random variable. For any $\t>1$, $U^{\t}(\s)\geq1$ if and only if $\thetazero(\s)<\rhoone$ and there exists a step $j \leq \t$ such that $\theta^j(\s)>\rhotwo$. Equivalently, $\lim_{\t \to \infty} U^\t(\s)\geq1$ if and only if $\thetazero(\s)<\rhoone$ and there exists a step $\t>1$ such that $\thetat(\s)>\rhotwo$. Therefore, if $\thetazero(\s)< \rhoone$ for all $\s \in \Sbar$, then:
\begin{align}
    &\pro\(\thetat(\s)\leq\rhotwo, ~ \forall \s \in \Sbar, ~\forall \t\) = 1- \pro\(\exists \s \in \Sbar \text{ and } \t, ~s.t. ~\thetat(\s)>\rhotwo\) \notag\\
    \geq & 1- \sum_{\s \in \Sbar} \pro\(\exists \t, ~s.t. ~~\thetat(\s)>\rhotwo\) = 1- \sum_{\s \in 
    \Sbar}\lim_{\t \to \infty} \pro\(U^\t(\s)\geq 1\). \label{eq:up_one}
\end{align}

Next, we define $\alpha \deleq \thetabar(\sran)-\rhoone$. Since $0<\rhoone < \min_{\s \in [\thetabar]} \{\thetabar(\s)\}$ and $\sran$ is in the support set, we have $\alpha \in (0, \thetabar(\sran))$. If $\thetazero(\s)<\rhoone$ for all $\s \in \Sbar$ and $\thetazero(\sran) > \thetabar(\sran)-\rhoone = \alpha$, then $\frac{\thetazero(\s)}{\thetazero(\sran)}< \frac{\rhoone}{\alpha}$ for all $\s \in \Sbar$. Additionally, for any step $\t$ and any $\s \in \Sbar$, if $\thetat(\s)>\rhotwo$, then $\frac{\thetat(\s)}{\thetat(\sran)} \geq \rhotwo$ because $\thetat(\sran)\leq 1$. Hence, whenever $\thetat(\s)$ completes an upcrossing of the interval $\left[\rhoone, \rhotwo\right]$, $\frac{\thetat(\s)}{\thetat(\sran)}$ must also have completed an upcrosssing of the interval $\left[\frac{\rhoone}{\alpha}, \rhotwo\right]$. From \eqref{eq:rhoone} -- \eqref{eq:rhotwo}, we can check that $\frac{\rhoone}{\alpha} < \rhotwo$ so that the interval $\left[\frac{\rhoone}{\alpha}, \rhotwo\right]$ is valid. We denote $\Uhat^\t(\s)$ as the number of upcrossings of the sequence $\(\frac{\theta^j(\s)}{\theta^j(\sran)}\)_{j=1}^{\t}$ with respect to the interval $\left[\frac{\rhoone}{\alpha}, \rhotwo\right]$ until step $\t$. Then, $U^\t(\s) \leq \Uhat^\t(\s)$ for all $\t$. Therefore, we can write:
\begin{align}\label{eq:up_two}
    \pro\(U^\t(\s)\geq 1\) \leq \pro\(\Uhat^\t(\s)\geq 1\) \leq \mathbb{E}\left[\Uhat^\t(\s)\right], 
\end{align}
where the last inequality is due to Makov inequality. 

From the proof of Lemma \ref{lemma:theta_eq}, we know that the sequence $\(\frac{\thetat(\s)}{\thetat(\sran)}\)_{\t=1}^{\infty}$ is a martingale. Therefore, we can apply the Doob's upcrossing inequality as follows: 
\begin{align}\label{eq:up_three}
    \mathbb{E}\left[\Uhat^\t(\s)\right] \leq \frac{\mathbb{E}\left[\max \{\frac{\rhoone}{\alpha}-\frac{\thetat(\s)}{\thetat(\sran)}, 0\}\right]}{\rhotwo- \frac{\rhoone}{\alpha}} \leq \frac{\frac{\rhoone}{\alpha}}{\rhotwo- \frac{\rhoone}{\alpha}}, \quad \forall \t.
\end{align}
From \eqref{eq:up_one} -- \eqref{eq:up_three} and \eqref{eq:rhoone} -- \eqref{eq:rhotwo}, we can conclude that: 
\begin{align*}
    \pro\(\thetat(\s)\leq\rhotwo, ~ \forall \s \in \Sbar, ~ \forall \t\) \geq 1- \frac{\frac{\rhoone}{\alpha} |\Sbar|}{\rhotwo- \frac{\rhoone}{\alpha}}= 1- \frac{\frac{\rhoone}{\thetabar(\sran)-\rhoone} |\Sbar|}{\rhotwo- \frac{\rhoone}{\thetabar(\sran)-\rhoone}}> \gamma.
\end{align*} \QEDA

Finally, Lemma \ref{lemma:other_parameters} provides conditions on the initial belief and strategy, under which if the beliefs of parameters in $\Sbar$ are bounded by $\rhotwo$, then with probability 1, we have $|\thetat(\s)-\thetabar(\s)|<\frac{\epsilonhat}{|\S|}$ for all $\s \in [\thetabar]$ and $\qt \in N_{\deltahat}\(\EQ(\thetabar)\)$ for all $\t$.

\begin{lemma}\label{lemma:other_parameters}
If $|\thetazero(\s)-\thetabar(\s)|<\rhothree$ for all $\s \in [\thetabar]$ and $\q^1 \in N_{\deltahat}\(\EQ(\thetabar)\)$, then 
\begin{align}\label{eq:ensure}
\pro\(\left.\begin{array}{l}
|\thetat(\s)-\thetabar(\s)|<\frac{\epsilonhat}{|\S|}, ~\forall \s \in [\thetabar], ~\forall \t\\
\text{and } \qt \in N_{\deltahat}\(\EQ(\thetabar)\), ~\forall \t
\end{array}\right\vert
\thetat(\s)<\rhotwo, ~\forall \s \in \Sbar, ~\forall \t\)=1.
\end{align}
\end{lemma}

Lemma \ref{lemma:other_parameters} builds on Lemmas \ref{lemma:constrained_set} and \ref{lemma:stopping_time}, and it is proved by mathematical induction. 


\vspace{0.2cm}
\noindent \emph{\textbf{Proof of Lemma \ref{lemma:other_parameters}.}}
Recall from Lemma \ref{lemma:constrained_set}, $[\thetabar] \subseteq \Sequiv(\q^1)$ if $\q^1 \in N_{\deltahat}\(\EQ(\thetabar)\)$. Hence, $\phibar^\s(\c^1|\q^1) = \phibar^{\sran}(\c^1|\q^1)$ for any $\s \in [\thetabar]$ and any realized payoff $\c^1$. Therefore, 
\begin{align}\label{eq:two_one}
    \frac{\theta^{2}(\s)}{\theta^{2}(\sran)}=\frac{\theta^{1}(\s)}{\theta^{1}(\sran)}\frac{\phibar^{\s}(\c^1|\q^1)}{\phibar^{\sran}(\c^1|\q^1)}= \frac{\theta^{1}(\s)}{\theta^{1}(\sran)}, \quad w.p.~1, \quad \forall \s \in [\thetabar].
\end{align}
This implies that $\frac{\sum_{\s \in [\thetabar]} \theta^{2}(\s)}{\theta^{2}(\sran)}=\frac{\sum_{\s \in [\thetabar]} \thetazero(\s)}{\thetazero(\sran)}$, and for all $\s \in [\thetabar]$:
\begin{align*}
\frac{\theta^{2}(\s)}{\sum_{\s \in [\thetabar]} \theta^{2}(\s)}=\frac{\theta^{2}(\s)}{\theta^{2}(\sran)} \frac{\theta^{2}(\sran)}{\sum_{\s \in [\thetabar]} \theta^{2}(\s)}=\frac{\thetazero(\s)}{\thetazero(\sran)} \frac{\thetazero(\sran)}{\sum_{\s \in [\thetabar]} \thetazero(\s)}=\frac{\thetazero(\s)}{\sum_{\s \in[\thetabar]} \thetazero(\s)}.
\end{align*}
Thus, we have
\begin{align*}
    \frac{\theta^{2}(\s)}{\thetazero(\s)}=\frac{\sum_{\s \in [\thetabar]} \theta^{2}(\s)}{\sum_{\s \in[\thetabar]} \theta^1(\s)}, \quad w.p.~1,  \quad \forall \s \in [\thetabar].\end{align*}

Since $\sum_{\s \in [\thetabar]} \thetazero(\s) \leq 1$, if $\theta^{2}(\s)<\rhotwo$ for all $\s \in \Sbar$, then we have $\frac{\theta^{2}(\s)}{\thetazero(\s)}> 1-|\Sbar|\rhotwo$. Additionally, since $\sum_{\s \in [\thetabar]} \theta^{2}(\s)<1$ and $\thetazero(\s)< \rhothree$ for all $\s \in [\thetabar]$, we have $\frac{\theta^{2}(\s)}{\thetazero(\s)}<\frac{1}{1-|\Sbar| \rhothree}$. Since by \eqref{eq:rho_three}, $\rhothree \leq  \thetabar(\s)$ for all $\s \in \Sbar$, any $\thetazero(\s) \in \(\thetabar(\s)-\rhothree, \thetabar(\s)+ \rhothree\)$ is a non-negative number for all $\s \in [\thetabar]$. Therefore, we have the following bounds: 
\begin{align}\label{eq:refer_end}
\(\thetabar(\s)-\rhothree\) \(1-|\Sbar|\rhotwo\)<\theta^{2}(\s) <\frac{\thetabar(\s)+ \rhothree}{1-|\Sbar| \rhothree}.
\end{align}
Since
\begin{align}\label{eq:up}
    \rhothree \stackrel{\eqref{eq:rho_three}}{\leq} \frac{\epsilonhat - |\Sbar| |\S|\rhotwo \thetabar(\s)}{|\S|-|\Sbar||\S| \rhotwo}, \quad \forall \s \in [\thetabar],
\end{align}
we can check that $\(\thetabar(\s)-\rhothree\) \(1-|\Sbar|\rhotwo\) \geq \thetabar(\s) -\frac{\epsilonhat}{|\S|}$ for all $\s \in [\thetabar]$. To ensure the right-hand-side of \eqref{eq:up} is positive, we need to have $\rhotwo < \frac{\epsilonhat}{|\Sbar| |\S| \thetabar(\s)}$ for all $\s \in [\thetabar]$, which is satisfied by \eqref{eq:rhotwo}.  Also, since $\rhothree \stackrel{\eqref{eq:rho_three}}{\leq} \frac{\epsilonhat}{|\S|+ |\Sbar|\(\thetabar(\s)|\S|+ \epsilonhat\)}$ for all $\s \in [\thetabar]$, we have $\frac{\thetabar(\s)+ \rhothree}{1-|\Sbar| \rhothree}< \thetabar(\s)+\frac{\epsilonhat}{|\S|}$ for all $\s \in [\thetabar]$. Therefore, we can conclude that $\theta^{2}(\s) \in \(\thetabar(\s)-\frac{\epsilonhat}{|\S|}, \thetabar(\s)+\frac{\epsilonhat}{|\S|}\)$ for all $\s \in [\thetabar]$. Additionally, if $\theta^2(\s) \leq \rhotwo < \frac{\epsilonhat}{|\S|}$ for all $\s \in \S \setminus [\thetabar]$, then $\theta^2 \in N_{\epsilonhat}\(\thetabar\)$. From Lemma \ref{lemma:constrained_set}, we know that $\gwe(\theta^2) \in N_{\deltahat}\(\EQ(\thetabar)\)$. Since $\q^1 \in N_{\deltahat}\(\EQ(\thetabar)\)$, the updated strategy $\q^2$ given by \eqref{eq:eq_update} must also be in the neighborhood $N_{\deltahat}\(\EQ(\thetabar)\)$.  

We now use mathematical induction to prove that the belief of any $\s \in [\thetabar]$ satisfies $\theta^{\t}(\s) \in \(\thetabar(\s)-\frac{\epsilonhat}{|\S|}, \thetabar(\s)+\frac{\epsilonhat}{|\S|}\)$ for steps $\t>2$. If in steps $j=1, \dots, \t$, $|\theta^j(\s)-\thetabar(\s)|<\frac{\epsilonhat}{|\S|}$ for all $\s \in [\thetabar]$ and $\theta^j(\s)<\rhotwo<\frac{\epsilonhat}{|\S|}$ for all $\s \in \Sbar$, then Lemma \ref{lemma:constrained_set} ensures that $g(\theta^j) \in \EQ(\theta^{j}) \subseteq N_{\deltahat}\(\EQ(\thetabar)\)$ for all $j=1, \dots, \t$. Since $\qj$ is a linear combination of $\q^1$ and $\(\gwe(\theta^i)\)_{i=2}^{j}$, if $\q^1 \in N_{\deltahat}\(\EQ(\thetabar)\)$, then $\qj \in N_{\deltahat}\(\EQ(\thetabar)\)$ for all $j=1, \dots, \t$.

From Lemma \ref{lemma:constrained_set}, we know that $[\thetabar] \subseteq \Sequiv(\qj)$ for all $j=1, \dots, \t$. Therefore, for any $\s \in [\thetabar]$ and any $j=1, \dots, \t$, $\phibar^\s(\cj|\loadj)=\phibar^{\sran}(\cj|\loadj)$ with probability 1. Then, by iteratively applying \eqref{eq:two_one}, we have $\frac{\theta^{\t+1}(\s)}{\thetazero(\s)}=\frac{\sum_{\s \in [\thetabar]} \theta^{\t+1}(\s)}{\sum_{\s \in[\thetabar]} \thetazero(\s)}$ for all $\s \in [\thetabar]$ with probability 1. Analogous to $\t=2$, we can prove that if $|\thetazero(\s)-\thetabar(\s)|<\rhothree$ for all $\s \in [\thetabar]$, then $|\thetatone(\s)-\thetabar(\s)|<\frac{\epsilonhat}{|\S|}$ for all $\s \in [\thetabar]$. From the principle of mathematical induction, we conclude that in all steps $\t$,  $|\thetat(\s)-\thetabar(\s)|< \frac{\epsilonhat}{|\S|}$ for all $\s \in [\thetabar]$, and $\qt \in N_{\deltahat}\(\EQ(\thetabar)\)$ for all $\t$. Therefore, we have proved \eqref{eq:ensure}. \QEDA

\vspace{0.2cm}

Finally, we are ready to prove Theorem \ref{theorem:stability_eq}. 

\vspace{0.2cm}
\noindent\textbf{\emph{Proof of Theorem \ref{theorem:stability_eq}.}}
We combine Lemmas \ref{lemma:constrained_set} -- \ref{lemma:other_parameters}. For any $\gamma \in (0, 1)$, and any $\thetaepfin, \loadepfin >0$, consider $\loadep \deleq  \min\{\delta, \loadepfin\}$ as in Lemma \ref{lemma:constrained_set} and $\thetaep \deleq \min\{\rhoone, \rhothree\}$ given by \eqref{eq:rhoone} -- \eqref{eq:rho_three}. If $\thetazero \in N_{\epsilon^1}(\thetabar)$, then $|\thetazero(\s)-\thetabar(\s)|<\thetaep$ for all $\s \in \S$. 
Recall from Lemma \ref{lemma:constrained_set}, $\lim_{\t \to \infty} \pro\(\thetat\in N_{\thetaepfin}(\theta), ~ \qt \in \neighinfload(\EQ(\thetabar)) \) \geq  \pro\(\thetat \in N_{\epsilonhat}(\thetabar), ~ \qt \in N_{\deltahat}\(\EQ(\thetabar)\), ~ \forall \t\)$. Since $\rhotwo \leq \epsilonhat/|\S|$, we further have: 
\begin{align*}
    &\pro\(\thetat \in N_{\epsilonhat}(\thetabar), ~ \qt \in N_{\deltahat}\(\EQ(\thetabar)\), ~ \forall \t\) \geq \pro\(\begin{array}{l}
|\thetat(\s)-\thetabar(\s)|<\frac{\epsilonhat}{|\S|}, ~ \forall \s \in [\thetabar],~ \forall \t \text{ and}\\
\thetat < \rhotwo,~ \forall \s \in \Sbar,~ \qt \in N_{\deltahat}\(\EQ(\thetabar)\), ~\forall \t
\end{array}\)\\
&= \pro\(\thetat(\s)<\rhotwo, \forall \s \in \Sbar, \forall \t\) \cdot \pro\(\left.\begin{array}{l}
|\thetat(\s)-\thetabar(\s)|<\frac{\epsilonhat}{|\S|}, \forall \s \in [\thetabar], \forall \t\\
\text{and }\qt \in N_{\deltahat}\(\EQ(\thetabar)\), ~\forall \t
\end{array}\right\vert
\begin{array}{l}
\thetat(\s)<\rhotwo.\\
\forall \s \in \Sbar, \forall \t\end{array}\)
\end{align*}
For any $\thetazero \in N_{\epsilon^1}(\thetabar)$ and any $\q^1 \in N_{\loadep}\(\EQ(\thetabar)\)$, we know from Lemmas \ref{lemma:stopping_time} -- \ref{lemma:other_parameters} that: 
\begin{align*}
&\pro\(\thetat(\s)<\rhotwo, ~\forall \s \in \Sbar, ~\forall \t\)>\gamma, \text{ and } \\
&\pro\(\left.\begin{array}{l}
|\thetat(\s)-\thetabar(\s)|<\frac{\epsilonhat}{|\S|}, ~ \forall \s \in [\thetabar], ~\forall \t\\
\text{and }\qt \in N_{\deltahat}\(\EQ(\thetabar)\), ~\forall \t
\end{array}\right\vert
\thetat(\s)<\rhotwo, ~\forall \s \in \Sbar, ~\forall \t\) =1
\end{align*}
Therefore, for any $\thetazero \in N_{\epsilon^1}(\thetabar)$ and any $\q^1 \in N_{\loadep}\(\EQ(\thetabar)\)$, the states of learning dynamics satisfy $\lim_{\t \to \infty} \pro\(\thetat\in N_{\thetaepfin}(\theta), ~ \qt \in  N_{\loadepfin}\(\EQ(\thetabar)\) \)> \gamma$. Thus, $\(\thetabar, \qbar\)$ is locally stable under conditions (a) and (b).
\QEDA

We can check that any complete information fixed point $\(\thetasran, \qsran\)$ such that $\qsran \in \EQ(\thetasran)$ trivially satisfies condition {\em (b)} in Theorem \ref{theorem:stability_eq} since $[\thetasran]=\{\sran\} \subseteq \Sequiv(\q)$ for any $\q \in \Q$. Therefore, condition {\em (a)} in Theorem \ref{theorem:stability_eq} is sufficient to guarantee local stability of complete information fixed points: 
\begin{corollary}\label{cor:complete_stability}
If $\EQ(\theta)$ is upper-hemicontinuous in $\theta$, then any complete information fixed points $\(\thetasran, \EQ(\thetasran)\)$ is locally stable under the learning dynamics \eqref{eq:update_belief} -- \eqref{eq:eq_update}.  
\end{corollary}

Finally, we show that the learning dynamics has a globally stable fixed point if and only if all fixed points have complete information of the unknown parameter. 
\begin{proposition}\label{prop:global}
For learning dynamics \eqref{eq:update_belief} -- \eqref{eq:eq_update}, there exists globally stable fixed points if and only if $\Omega = \left\{\(\thetasran, \EQ(\thetasran)\)\right\}$. Then, all fixed points in $\(\thetasran, \EQ(\thetasran)\)$ are globally stable.
\end{proposition}

\noindent\emph{\textbf{Proof of Proposition \ref{prop:global}.}} We first show that there exist globally stable fixed points if and only if $\Omega = \left\{\(\thetasran, \EQ(\thetasran)\)\right\}$. If $\FP= \{\(\thetasran, \EQ(\thetasran)\)\}$, then the states converge to the set $\FP$ with probability 1 for any initial state; hence $\(\thetasran, \EQ(\thetasran)\)$ is globally stable. On the other hand, if the set $\FP$ contained other fixed points $\(\thetabar, \qbar\) \notin \(\thetasran, \EQ(\thetasran)\)$, then the states of the learning dynamics starting from $\(\thetabar, \qbar\)$ (resp. $\(\thetasran, \qsran\)$) would remain at $\(\thetabar, \qbar\)$ (resp. $\(\thetasran, \qsran\)$) with probability 1, and hence no fixed point would be globally stable. Moreover, when the condition is satisfied, all fixed points in $\(\thetasran, \EQ(\thetasran)\)$ are globally stable. 

\QEDA


\begin{example}[Cournot competition continued]\label{ex:stability_cournot}
{\normalfont Recall from Example \ref{ex:potential}, the fixed point set in the Cournot game is $\FP=\left\{\(\thetasran, \qsran\)=\(\(1, 0\), \(2/3, 2/3\)\), \(\thetabar, \qbar\)= \(\(0.5, 0.5\), \(0.5, 0.5\)\)\right\}$. The unique equilibrium strategy profile $\gwe(\theta)$ is continuous in $\theta$; thus $\EQ(\theta)$ is upper-hemicontinuous in $\theta$. From Corollary \ref{cor:complete_stability}, the complete information fixed point $\(\thetasran, \qsran\)$ is locally stable. Now, consider the second fixed point $\(\thetabar, \qbar\)$: Note that $\qbar$ is the only strategy profile in $\Q$ for which $\s_1$ is payoff equivalent to $\s_2$, because the payoff functions are affine in $\q$. Therefore, condition {\em (ii)} in Theorem \ref{theorem:stability_eq} is violated in that there does not exist a neighborhood of $\qbar$ in which $\s^2$ remains to be payoff equivalent to $\s^1$. Thus, the sufficient condition of local stability is not satisfied by the fixed point $\(\thetabar, \qbar\)$.

Moreover, since the complete information fixed point is not the unique fixed point, no fixed point is globally stable. Indeed, as shown in Fig. \ref{fig:convergence_eq}, the states of the learning dynamics can converge to either one of the two fixed points. 
} 
\end{example}

\begin{example}[Coordination with safe margin continued]\label{ex:coordinate_stability}
{\normalfont Recall from Example \ref{ex:coordinate}, the fixed point set only contains complete information fixed points, i.e. $\FP=\{\(\thetasran, \EQ(\thetasran)\)\}$. Therefore, we know from Proposition \ref{prop:global} that $\(\thetasran, \EQ(\thetasran)\)$ is globally stable. }
\end{example}

\begin{example}[Public good investment continued]\label{ex:stability_public}
{\normalfont Recall from Example \ref{ex:supermodular}, the unique fixed point of the public good investment game is the complete information fixed point $\(\thetasran, \qsran\)= \(\(0, 1, 0\), \(1/2, 1/3\)\)$. From Proposition \ref{prop:global}, the complete information fixed point is globally stable}
\end{example}


\section{Learning with Best-response Strategies}\label{sec:br}
In Sec. \ref{subsec:results_br}, we derive convergence and stability results for learning with best-response strategies \eqref{eq:update_belief} -- \eqref{eq:eq_update} under certain assumptions of stepsizes and best response correspondence. In Sec. \ref{subsec:two_class}, we show that these assumptions hold in two classes of games -- potential games and dominance solvable games. 

\subsection{Convergence and Stability Analysis}\label{subsec:results_br}


Since belief is updated as in \eqref{eq:update_belief}, analogous to Lemma \ref{lemma:theta_eq}, the beliefs converge to a fixed point belief $\thetabar$ with probability 1. However, in contrast to \eqref{eq:eq_update}, the best response strategy profile $\gbr(\thetatone, \qt)=\(\gibr(\thetatone, \qt_{\mi})\)_{\i \in \I}$ in \eqref{eq:br_update} depends on the updated belief $\thetatone$ as well as the current strategy profile $\qt$. Hence, the strategy profile in each step can no longer be expressed as a linear combination of the initial strategy and continuous functions of beliefs. Therefore, the approach of proving the convergence of strategies in Lemma \ref{lemma:q_eq} does not apply to learning with best response strategies. We now develop a new approach to analyze the convergence of $\qt$ for \eqref{eq:br_update}.

Based on Lemma \ref{lemma:theta_eq}, we consider any sequence of beliefs $\(\thetat\)_{\t=1}^{\infty}$ that converges to a fixed point belief $\thetabar$. In each step $\t$ with the belief $\theta^{\t+1}$, for any best response strategy $\gibr(\thetatone, \qt_{\mi}) \in \BRi\(\theta^{\t+1}, \qt_{\mi}\)$ in the strategy update \eqref{eq:br_update}, we can find another strategy $\tilde{h}_\i(\thetabar, \qt_{\mi})$ in the best response correspondence $\BRi(\thetabar, \qt_{\mi})$ with respect to the fixed point belief $\thetabar$ such that the distance between $\tilde{h}_\i(\thetabar, \qt_{\mi})$ and $\gibr(\thetatone, \qt_{\mi})$ attains a minimum, i.e. $\tilde{h}_\i(\thetabar, \qt_{\mi})= \argmin_{\qi \in \BRi(\thetabar, \qt_{\mi})} $ $|\qi-\gibr(\thetatone, \qt_{\mi})|$. We define $\xi_i^{\t}\deleq \gibr(\thetatone, \qt_{\mi})-\tilde{h}_\i(\thetabar, \qt_{\mi})$. Then, the strategy update \eqref{eq:br_update} in each step $\t$ can be re-written as follows: 
\begin{align}\label{eq:rewrite_dynamics}
    \q^{\t+1}_i  -\qit&=\ait \(\gibr(\thetatone, \qt_{\mi})-\qit\)= \ait\(\tilde{h}_\i(\thetabar, \qt_{\mi})-\qit+\xi_i^{\t}\).
\end{align}
We define the maximum stepsize in step $\t$ as $\bar{a}^\t=\max_{\i \in \I} \ait$, and assume that the ratio between the stepsizes of all players and the maximum stepsize are lower bounded by a positive number: 

\textbf{(A3)} $\frac{\ait}{\bar{a}^\t} \geq \nu>0$ for all $\i \in \I$ and all $\t$.

Then, since $\tilde{h}_\i(\thetabar, \qt_{\mi}) \in \BRi(\thetabar, \qt_{\mi})$ for all $\i \in \I$ and all $\t$, we can write \eqref{eq:rewrite_dynamics} as a discrete-time asynchronous best response dynamics for the game $\G$ with the fixed point belief $\thetabar$ and residual terms $\xi^{\t}= \(\xi^{\t}_i\)_{i \in \I}$: 
\begin{align}\label{eq:discrete}
    \q^{\t+1} -\qt \in \bar{a}^\t \mathcal{A}\cdot \(\BR\(\thetabar, \qt\) - 
    \qt + \xi^{\t}\),
\end{align}
where $\mathcal{A} \cdot\(\BR\(\thetabar, \qt\)-\qt+ \xi^{\t}\) \deleq \left\{A \cdot \(\tilde{h}(\thetabar, \qt)-\qt+ \xi^{\t}\) \left\vert A \in \mathcal{A}, ~ \tilde{h}(\thetabar, \qt) \in \BR(\thetabar, \qt) \right. \right\}$, and
\begin{align}
    \mathcal{A} \deleq \left\{A \in \mathbb{R}^{dim(\Q)\times dim(\Q)}\left\vert
     \begin{array}{l}
     A=diag(\underbrace{\alpha_1 \dots \alpha_1}_{dim\(\Q_1\)}, \dots, \underbrace{\alpha_i \dots \alpha_i}_{dim\(\Qi\)} , \dots, \underbrace{\alpha_I \dots \alpha_I}_{dim\(Q_I\)}).\\
     \alpha_i \in [\nu, 1], \quad \forall \i \in \I
    \end{array}\right.\right\}.\label{subeq:Lambda}
\end{align}
The diagonal matrix $A \in \mathcal{A}$ in \eqref{subeq:Lambda} captures the asynchronous nature of players' strategy updates, with $\alpha_i$ denoting the ratio between the stepsize of player $\i \in \I$ and the maximum stepsize in step $\t$. From Assumption \textbf{(A3)}, we know that each diagonal value $\alpha_i$ is lower bounded by $\nu$ for all $\i \in \I$. If the step sizes of all players are identical (i.e. $\ait=\at_{j}$ for all $i, j \in \I$ and all $\t$), then the best response dynamics is synchronous and the matrix $A$ is an identity matrix.

We apply the theory of stochastic approximation to show the convergence of strategies in \eqref{eq:discrete}; see \cite{borkar1998asynchronous}, \cite{benaim2005stochastic}, \cite{benaim2006stochastic}, and \cite{perkins2013asynchronous}. The theory allows us to analyze the asymptotic properties of the strategy sequence $\(\qt\)_{\t=1}^{\infty}$ by approximating the discrete-time dynamics \eqref{eq:discrete} with a continuous time best-response differential inclusion given by: 
\begin{align}\label{eq:differential_inclusion}
    \frac{d \qdiff(\tau)}{d \tau} \in \mathcal{A} \cdot \(\BR\(\thetabar, \qdiff\)-\qdiff\)
\end{align}
where $\mathcal{A} \cdot\(\BR\(\thetabar, \qdiff\)-\qdiff\) \deleq \left\{A \cdot \(\tilde{\gbr}(\thetabar, \qdiff)-\qdiff\) \left\vert A \in \mathcal{A}, ~ \tilde{\gbr}(\thetabar, \qdiff) \in \BR(\thetabar, \qdiff) \right. \right\}$ and $\mathcal{A}$ is defined as in \eqref{subeq:Lambda}.
A solution of \eqref{eq:differential_inclusion} with the initial strategy profile $\q^1 \in \Q$ is an absolutely continuous function $\tilde{\q}(\tau): [1, \infty) \to \Q$ such that $\tilde{\q}(1)=\q^1$, and $\tilde{\q}(\tau)$ satisfies \eqref{eq:differential_inclusion} for almost all $\tau \geq 1$. We adopt the assumptions from \cite{perkins2013asynchronous} for asynchronous best response dynamics:

\textbf{(A4)} For each $\t$, the maximum stepsizes $\(\atbar\)_{\t=1}^{\infty}$ satisfy the following conditions: 
\begin{equation*}
\begin{split}
    \sum_{\t=1}^{\infty} \atbar &=\infty, \quad 
    \sum_{\t=1}^{\infty} \(\atbar\)^2 < \infty, \quad 
    \atbar \geq \abar^{\t+1}, \quad \forall \t, \\
    \sup_{\t} \frac{\abar^{\lfloor x\t \rfloor}}{\atbar} &< \infty, \quad \forall x \in (0, 1], 
\end{split}
\end{equation*}
where $\lfloor x\t \rfloor$ is the largest integer that is smaller than $x\t$.

\textbf{(A5)} For any $\theta \in \Delta(\S)$ and any $\qmi \in \Q_{\mi}$, the best response correspondence $\BRi(\theta, \qmi)$ is a convex and compact set in $\Q$. Additionally, $\BRi(\theta, \qmi)$ is upper-hemicontinuous in both $\theta$ and $\qmi$ for all $\i \in \I$.

\textbf{(A6)} For any $\theta \in \Delta(\S)$ and any $\q^1 \in \Q$, any solution $\qdiff(\tau)$ of \eqref{eq:differential_inclusion} such that $\qdiff(1)=\q^1$ satisfies $\lim_{\tau \to \infty} \mathrm{dist}\(\qdiff(\tau), \EQ(\theta)\)=0$.

Assumption \textbf{(A4)} is a standard requirement on step sizes in stochastic approximation \cite{borkar2009stochastic}. Assumption \textbf{(A5)} ensures that the solutions of the differential inclusion \eqref{eq:differential_inclusion} exist given any initial strategy $\q^1 \in \Q$. Assumption \textbf{(A6)} requires that given any initial strategy $\q^1 \in \Q$, the continuous-time best response dynamics converges to the equilibrium set of the game $\G$ for any constant belief $\theta$. 

Not all games satisfy \textbf{(A5)} -- \textbf{(A6)}. For example, best response dynamics is cyclic in the well-known generalized rock-paper-scissors game with complete information so \textbf{(A6)} is not satisfied by this game, see \cite{shapley1964some}. In Sec. \ref{subsec:two_class}, we demonstrate that \textbf{(A5)} -- \textbf{(A6)} are guaranteed in two classes of games -- potential games and dominance solvable games.

Based on \textbf{(A3)} -- \textbf{(A6)}, we have the following:

\begin{lemma}[\cite{perkins2013asynchronous}]\label{lemma:asynchronous}
Under assumptions \textbf{(A3)} -- \textbf{(A6)}, if $\xi^{\t}=\(\xi_i^\t\)_{\i \in \I}$ is bounded for all $\t$ and $\lim_{\t \to \infty} \xi^{\t}=0$, then the sequence $\(\qt\)_{\t=1}^{\infty}$ given by \eqref{eq:rewrite_dynamics} converges to the equilibrium set $\EQ(\thetabar)$ for any $\thetabar \in \Delta(\S)$. 
\end{lemma}

Based on \textbf{(A5)}, we can show that for any sequence of beliefs $\(\thetat\)_{\t=1}^{\infty}$ that converge to $\thetabar$, the sequence $\(\xi^{\t}\)$ indeed converges to zero. 
\begin{lemma}\label{lemma:residual_zero}
Under Assumption \textbf{(A5)}, for any belief sequence $\(\thetat\)_{\t=1}^{\infty}$ such that $\lim_{\t \to \infty} \thetat = \thetabar$, $\xi^{\t}=\(\xi_i^\t\)_{\i \in \I}$ is bounded for all $\t$ and $\lim_{\t \to \infty} \xi^{\t}=0$. 
\end{lemma}

The proof of Lemma \ref{lemma:residual_zero} is included in Appendix \ref{apx:proof}. From Lemmas \ref{lemma:asynchronous} and \ref{lemma:residual_zero}, we can conclude that as the beliefs $\(\thetat\)_{\t=1}^{\infty}$ converge to a fixed point belief vector $\thetabar$, the strategies $\(\qt\)_{\t =1}^{\infty}$ also converge to the equilibrium set $\EQ(\thetabar)$ corresponding to fixed point belief $\thetabar$. 
\begin{lemma}\label{lemma:q_br}
Under Assumptions \textbf{(A3)} -- \textbf{(A6)}, $\lim_{\t \to \infty} d \(\qt, \EQ(\thetabar)\)=0$ with probability 1. 
\end{lemma}

Now, recall from Lemma \ref{lemma:consistency} in Sec. \ref{subsec:convergence_eq} that in learning with equilibrium strategies, any fixed point belief $\thetabar$ identifies the true parameter $\sran$ in the payoff-equivalent parameter set given the corresponding fixed point strategy $\qbar$. In learning with best response strategies, if the game with fixed point belief $\thetabar$ has multiple equilibria, then the strategy profiles $\(\qt\)_{\t=1}^{\infty}$ converge to the equilibrium set $\EQ(\thetabar)$, but not necessarily converge to a single fixed point strategy. Thus, we need the notion of the limit set of $\(\qt\)_{\t=1}^{\infty}$:
\begin{definition}[Limit set]\label{def:limit_point}
Set $\limitq \subseteq \Q$ is the limit set of the sequence $\(\qt\)_{\t=1}^{\infty}$ if for any $\qbar \in \limitq$, there exists a subsequence of strategy profiles $\(\qtk\)_{j=1}^{\infty}$ such that $\lim_{j \to \infty} \qtk=\qbar$. 
\end{definition}
That is, limit set $\Qbar$ is the set of strategies that are limit points of converging subsequences in $\(\qt\)_{\t=1}^{\infty}$. The limit set must be nonempty since the feasible strategy set $\Q$ is bounded. From Lemma \ref{lemma:q_br}, we know that $\limitq \subseteq \EQ(\thetabar)$.  

Analogous to Definition \ref{def:payoff_equivalence}, we define payoff-equivalent parameters on a set of strategies:
\begin{definition}\label{def:set_payoff}
The set of parameters that are payoff-equivalent to $\sran$ on a set $\widehat{\Q} \subseteq \Q$ is:
\begin{align*}
  \Sequiv(\widehat{Q})\deleq \left\{\S|D_{KL} \left(\phibar^{\sran}(\cbar|\q)||\phibar^{\s}(\cbar|\q)\right)=0,  ~ \forall \q \in \widehat{\Q}\right\}. 
\end{align*}
\end{definition}

We now show that in learning with best response strategies, $\thetabar$ only assigns positive probability on parameters that are payoff-equivalent to $\sran$ on the set $\limitq$. That is, the fixed point belief $\thetabar$ consistently estimates the payoff distribution of the strategies in the limit set $\limitq$. 
\begin{lemma}\label{lemma:consistency_br}
$[\thetabar] \subseteq \Sequiv(\limitq)$ with probability 1, where $\limitq \subseteq \EQ(\thetabar)$ is the limit set of the strategy sequence $\(\qt\)_{\t=1}^{\infty}$. 
\end{lemma}
The proof of Lemma \ref{lemma:consistency_br} is as follows: For any $\qbar \in \limitq$, from Definition \ref{def:limit_point}, there must exist a subsequence of strategies in $\(\qt\)_{\t=1}^{\infty}$ that converges to $\qbar$. Then, we use the same approach developed in Lemma \ref{lemma:consistency} to show that the belief ratio $\(\frac{\thetat(\s)}{\thetat(\sran)}\)_{\t=1}^{\infty}$ converges to 0 for any $\s \in \S \setminus \Sequiv(\qbar)$. Since this argument holds for any $\qbar \in \Qbar$, we know that $\thetabar(\s)$ is positive only if $\s$ is payoff-equivalent to $\sran$ for all $\qbar \in \Qbar$. The proof of this lemma is in Appendix \ref{apx:proof}.

Based on Lemmas \ref{lemma:theta_eq}, \ref{lemma:q_br} and \ref{lemma:consistency_br}, we obtain the following convergence result: 
\begin{theorem}\label{prop:asynchronous}
For any initial state~$(\theta^1, \q^1) \in \Delta(\S) \times \Q$, under assumptions \textbf{(A3)} -- \textbf{(A6)}, the sequence of states $(\thetat, \qt)_{\t=1}^{\infty}$ generated by the learning dynamics~\eqref{eq:update_belief} and \eqref{eq:br_update} satisfy $\lim_{\t \to \infty} \thetat =\thetabar$ and $\lim_{\t \to \infty} \mathrm{dist}\(\qt, \EQ(\thetabar)\)=0$ with probability 1. 

Moreover, the belief $\thetabar$ satisfies $[\thetabar] \subseteq \Sequiv(\limitq)$ with probability 1, where $\limitq \subseteq \EQ(\thetabar)$ is the limit set of the strategy sequence $\(\qt\)_{\t=1}^{\infty}$. 
\end{theorem}

Analogous to Theorem \ref{theorem:convergence_eq}, this result ensures that in learning with best response strategies, the convergent belief $\thetabar$ accurately estimates the payoff distribution given players' strategies, and players eventually play equilibrium strategies in game $\G$ with belief $\thetabar$. 

We now provide a set of sufficient conditions that guarantee local stability of $\(\thetabar, \EQ(\thetabar)\)$ for learning with best response strategies. The global stability property is identical to that of the learning dynamics with equilibrium strategies, as stated in Proposition \ref{prop:global}.  

\begin{theorem}\label{theorem:stability_br}
A fixed point belief $\bar\theta\in\Delta(S)$ and the corresponding equilibrium set $EQ(\bar\theta)$ is locally stable under the learning dynamics \eqref{eq:update_belief}
and \eqref{eq:br_update} if Assumptions \textbf{(A3)} -- \textbf{(A6)} are satisfied, and the following conditions hold: (a) $EQ(\theta)$ is upper-hemicontinuous in $\theta$;
(b) $\exists \delta>0$ and $\epsilon>0$ such that $[\bar\theta]\subseteq \Sequiv(q)$ and $\BR(\theta, \q) \subseteq N_{\loaddelta}(\EQ(\thetabar))$ for any $q\in N_\delta(EQ(\bar\theta))$ and any $\theta \in N_{\epsilon}\(\thetabar\)$
\end{theorem}
Condition {\em (a)} in Theorem \ref{theorem:stability_br} is the same as that in Theorem \ref{theorem:stability_eq}. This condition ensures that if the convergent belief is close to $\thetabar$, then the convergent strategy will also be close to $\EQ(\thetabar)$. Condition {\em (b)} in Theorem \ref{theorem:stability_br} is more restrictive than in Theorem \ref{theorem:stability_eq} -- it not only ensures that all parameters in $[\thetabar]$ are payoff-equivalent (given any strategy) in a local neighborhood of $\EQ(\thetabar)$, but also requires that the best response strategy profile $\gbr(\theta, \q)$ remains close to the set $\EQ(\thetabar)$ when $\q$ and $\theta$ are locally perturbed in a small neighborhood of $\EQ(\thetabar)$ and $\thetabar$, respectively. These two conditions together guarantee local stability. The proof builds on Theorem \ref{theorem:stability_br}, and is included in Appendix \ref{apx:proof}.

\subsection{Potential Games and Dominance Solvable Games}\label{subsec:two_class}

We now show that Assumptions \textbf{(A5)} -- \textbf{(A6)} are guaranteed in two classes of games -- potential games and dominance solvable games. 

\vspace{0.2cm}
\noindent\textbf{[Potential games]}
A game $G_P$ with parameter $\s \in \S$ is a potential game if there exists a potential function $\Potentials(\q): \Q \to \mathbb{R}$ such that 
\begin{align*}
\Potentials(\qi, \qmi) - \Potentials(\qi', \qmi) = \usi(\qi, \qmi) - \usi(\qi', \qmi), \quad \forall \qi, \qi' \in \Qi, \quad \forall \qmi \in \Q_{\mi}, \quad \forall \i \in \I.
\end{align*}
We assume that the potential function $\Potentials(\q)$ is continuous and differentiable, and $\Potentials(\qi, \qmi)$ is concave in each player's strategy $\qi$ for all $\i \in \I$ and all $\s \in \S$. 

The next proposition shows that for any $\theta \in \Delta(\S)$, the equilibrium set $\EQ(\theta)$ in game $G_P$ is upper-hemicontinuous in $\theta$, and the best-response strategies satisfy \textbf{(A5)} -- \textbf{(A6)}. 
\begin{proposition}\label{prop:potential}
In game $G_P$, 
\begin{enumerate}
\item The equilibrium set $\EQ(\theta)$ is upper-hemicontinuous in $
\theta$ for all $\theta \in \Delta(\S)$. In addition, if $\Psi^s(\q)$ is strictly concave in $\q$ for all $\s \in \S$, then the equilibrium of $G_P$ is unique and is continuous in $\theta$.  
\item For each $\i \in \I$, the best-response correspondence $\BRi(\theta, \q)$ is a closed convex set in $\Qi$, and is upper-hemicontinuous in both $\theta$ and $\q$.
\item For any $\theta \in \Delta(\S)$ and any $\q^1 \in \Q$, any solution $\qdiff(\tau)$ of \eqref{eq:differential_inclusion} such that $\qdiff(1)=\q^1$ satisfies $\lim_{\tau \to \infty} \mathrm{dist}\(\qdiff(\tau), \EQ(\theta)\)=0$.
\end{enumerate}
\end{proposition}

\noindent\textbf{\emph{Proof of Proposition \ref{prop:potential}.}}
We prove \emph{1-3} in sequence: 

1. We can show that the function $\mathbb{E}_{\theta}[\Psi^s(\q)]$ is a potential function of game $G_P$ with belief $\theta$: 
\begin{align*}
&\mathbb{E}_{\theta}[\Psi^s(\qi, \qmi)] - \mathbb{E}_{\theta}[\Psi^s(\qi', \qmi)]= \sum_{\s \in \S} \theta(\s)\(\usi(\qi, \qmi) - \usi(\qi', \qmi)\)\\
&= \mathbb{E}_{\theta}[\usi(\qi, \qmi)] - \mathbb{E}_{\theta}[\usi(\qi', \qmi)], \quad \forall \i \in \I, \quad \forall \qi, \qi' \in \Qi, \quad \forall \qmi \in \Qmi.
\end{align*}
Therefore, the equilibrium set can be solved as $\EQ(\theta)=\argmax_{\q \in \Q} \mathbb{E}_{\theta}[\Psi^s(\q)]$. Note that $\mathbb{E}_{\theta}[\Psi^s(\q)]$ is a continuous function in $\q$ and $\theta$, and the set $\Q$ is a closed convex set. From Berge's maximum theorem, we obtain that the equilibrium set is upper-hemicontinuous in $\theta$. If the function $\Psi^s(\q)$ is strictly concave in $\q$ for all $\s \in \S$, then $\mathbb{E}_{\theta}[\Psi^s(\q)]$ is also strictly concave in $\q$. Therefore, the optimal solution set $\argmax_{\q \in \Q} \mathbb{E}_{\theta}[\Psi^s(\q)]$ is a singleton set, and the unique equilibrium strategy profile $\gwe(\theta)$ must be continuous in $\theta$.

2. For any player $\i \in \I$ and any other players' strategy $\qmi \in \Qmi$, the set of best response strategies is solved as $\BR_i(\theta, \qmi)=\argmax_{\qi \in \Qi} \mathbb{E}_{\theta}[\Psi^s(\qi, \qmi)]$. Again from the Berge's maximum theorem, since the potential function $\mathbb{E}_{\theta}[\Psi^s(\qi, \qmi)]$ is continuous in $\theta$, $\qi$ and $\qmi$, we know that $\BR_i(\theta, \qmi)$ must be upper-hemicontinuous in $\theta$ and $\qmi$. Since the function $\mathbb{E}_{\theta}[\Psi^s(\q)]$ is concave in $\qi$ and $\Qi$ is a convex bounded set, we know that the set of optimal solutions $\BR_i(\theta, \qmi)$ must also be a convex set.

3. For any $\theta$ and any solution $\tilde{\q}(\tau)$ of the differential inclusion \eqref{eq:differential_inclusion}, we consider the expected value of the potential function $\mathbb{E}_{\theta}[\Psi^s(\tilde{q}(\tau))]$. Since $\Psi^s(\tilde{q}(\tau))$ is differentiable in $\tilde{q}$, we can compute the derivative of $\mathbb{E}_{\theta}[\Psi^s(\tilde{q}(\tau))]$ with respect to $\tau$:
\begin{align*}
&\frac{d \mathbb{E}_{\theta}[\Psi^s(\tilde{q}(\tau))]}{d \tau} = \sum_{\s \in \S} \theta(\s) \sum_{\i \in \I} \nabla_{\qdiff_i(\tau)}\mathbb{E}_{\theta}[\Psi^s(\tilde{q}(\tau))] \cdot \frac{d \qdiff_i(\tau)}{d \tau}\\
\stackrel{\eqref{eq:differential_inclusion}}{\in}&\sum_{\s \in \S} \theta(\s) \( \sum_{\i \in \I} \nabla_{\qdiff_i(\tau)}\mathbb{E}_{\theta}[\Psi^s(\tilde{q}(\tau))] \cdot  \alpha_i \(\BRi(\theta, \qdiff_{\mi}(\tau))- \qdiff_i(\tau)\)\),
\end{align*}
where $\alpha_i \in [\nu, 1]$ and $\epsilon>0$. 
Since $\mathbb{E}_{\theta}[\Psi^s(\tilde{q}(\tau))]$ is concave in $\qdiff_i(\tau)$, for any best response strategy $\gibr(\theta, \qdiff_{\mi}(\tau)) \in \BRi(\theta, \qdiff_{\mi}(\tau))$, we have: 
\begin{align*}
\nabla_{\qdiff_i(\tau)}\mathbb{E}_{\theta}[\Psi^s(\tilde{q}(\tau))] \(\gibr(\theta, \qdiff_{\mi}(\tau))- \qdiff_i(\tau)\)\geq \mathbb{E}_{\theta}[\Potentials\(\gibr(\theta, \qdiff_{\mi}(\tau)), \qdiff_{\mi}(\tau)\)] - \mathbb{E}_{\theta}[\Potentials(\qdiff_i(\tau), \qdiff_{\mi}(\tau))].
\end{align*}
Since for all $\i \in \I$, $\gibr(\theta, \qdiff_{\mi}(\tau)) \in \argmax_{\qdiff_i(\tau) \in \Qi} \mathbb{E}_{\theta}[\Psi^s(\qdiff_i(\tau), \qdiff_{\mi}(\tau))]$ and $\alpha_i>0$, we have:
\begin{align*}
 \frac{d \mathbb{E}_{\theta}[\Psi^s(\tilde{q}(\tau))]}{d \tau}  \left\{
\begin{array}{ll}
>0, & \quad \text{If $\qdiff(\tau) \notin \EQ(\theta)$}, \\
=0, & \quad \text{If $\qdiff(\tau) \in \EQ(\theta)$. }
\end{array}
\right.
\end{align*} 
That is, the value of the potential function $\mathbb{E}_{\theta}[\Psi^s(\tilde{q}(\tau))]$ strictly increases in $\tau$ except for $\tilde{q}(\tau)$ in the equilibrium set $\EQ(\theta)$. Since the value of the potential function is finite, we must have $\lim_{\tau \to \infty} \mathrm{dist}\( \tilde{q}(\tau), \EQ(\theta)\)=0$ for all $\theta \in \Delta(\S)$.
\QEDA

\vspace{0.2cm}

\noindent\textbf{[Dominance-solvable games with diminishing returns]} A set of players $\I$ play a game $G_D$, where strategy of each player $\i \in \I$ is a scalar $\qi$ in a closed interval $\Qi \subseteq \mathbb{R}$. Given any strategy profile $\q \in \Q$, the utility function $\usi(\q)$ of each player $\i \in \I$ depends on an unknown parameter $\s \in \S$. In $\G_D$, each player's utility marginally decreases in their own strategy in that $\usi(\qi, \qmi)$ is concave in the strategy $\qi$ for all $\s \in \S$ and all $\qmi \in \Qmi$. 

For any $\theta \in \Delta(\S)$, the game $G_D$ with belief $\theta$ is a dominance solvable game in that each player has a unique rationalizable strategy -- the strategy that survives the process of iterated deletion of strictly dominated strategies. This rationalizable strategy is the unique equilibrium strategy $\gwe_i(\theta)$ of each player $i \in I$. Thus, the equilibrium set $\EQ(\theta) = \left\{\(\gwe_i(\theta)\)_{\i \in \I}\right\}$ is a singleton set.

The next proposition shows that in game $G_D$ with any belief $\theta \in \Delta(\S)$, the unique equilibrium strategy profile $\gwe(\theta)$ is continuous in $\theta$, and the best-response strategies satisfy \textbf{(A5)} -- \textbf{(A6)}.
\begin{proposition}\label{prop:supermodular}
In game $G_D$,
\begin{enumerate} 
\item The unique equilibrium strategy profile $\gwe(\theta)$ is continuous in $\theta$. 
\item For any player $\i \in \I$ and any strategy $\q \in \Q$, the best-response correspondence $\BRi(\theta, \qmi) = [\hmin_i(\theta, \qmi), \hmax_i(\theta, \qmi)]$ is an interval in $\Qi$ and is upper-hemicontinuous in $\theta$ and $\qmi$.
\item  For any $\theta \in \Delta(\S)$ and any $\q^1 \in \Q$, any solution $\qdiff(\tau)$ of \eqref{eq:differential_inclusion} such that $\qdiff(1)=\q^1$ satisfies $\lim_{\tau \to \infty} \qdiff(\tau) = \gwe(\theta)$. 
\end{enumerate}
\end{proposition}

\noindent\textbf{\emph{Proof of Proposition \ref{prop:supermodular}.}}

We first prove \emph{2}. For any $\i \in \I$ and any $\theta$, $\BRi(\theta, \q_{\mi}) = \argmax_{\qi \in \Qi} \mathbb{E}_{\theta}[u_i^s(\qi, \qmi)]$. Since $u_i^s(\qi, \qmi)$ is concave in $\qi$ for all $\s \in \S$, the function $\mathbb{E}_{\theta}[u_i^s(\qi, \qmi)]$ is also concave in $\qi$. Therefore, the set $\BRi(\theta, \qmi) $ must be an interval with the minimum best response strategy $\hmin_i(\theta, \q)$ and the maximum best-response strategy $\hmax_i(\theta, \q)$. Moreover, from Berge's maximum theorem, since $\mathbb{E}_{\theta}[u_i^s(\qi, \qmi)]$ is continuous in $\q$ and $\theta$, $\BRi(\theta, \qmi) $ is upper-hemicontinuous in $\qmi$ and $\theta$. 

Next, we prove \emph{1}. Consider the process of iterated deletion of strictly dominated strategy of game $G_D$ with belief $\theta$: In the first round, we delete the set of strategies of player $\i$ that are not a best response strategy for any $\qmi \in \Qmi$ in game $G_D$ with belief $\theta$. We denote the remaining strategies of each player $\i \in \I$ as $R^1_i(\theta)$. Then, any strategy $\qi \in R_i^1(\theta)$ must be a best response of at least one $\qmi \in \Qmi$, i.e. $R_i^1(\theta) = \cup_{\qmi \in \Qmi} \BRi(\theta, \qmi)$. From \emph{2}, we know that $\BRi(\theta, \qmi)$ is an interval in the set $\Qi$ for any $\qmi \in \Qmi$, and $\BRi(\theta, \qmi)$ is upper-hemicontinuous in both $\theta$ and $\qmi$. Therefore, the strategy set $R_i^1(\theta)$ is an interval in $\Qi$, and $R_i^1(\theta)$ is upper-hemicontinuous in $\theta$. We denote the set of player $\i$'s strategies that are not deleted after round $n \geq 1$ as $R^{n}_i(\theta)$. In step $n+1$, we delete the set of player $\i$'s strategies that are not a best response strategy for any $\qmi \in \prod_{j \in \I \setminus \{\i\}} R_i^{n}(\theta)$, and the set of remaining strategies is $R_i^{n+1}(\theta)$. Then, any strategy $\qi \in R_i^n(\theta)$ must be a best response of some opponents' strategy profile in $R_{\mi}^{n-1}(\theta) \deleq \prod_{j \in \I \setminus \{i\}} R_{j}^{n-1}(\theta)$, i.e. $R_i^n(\theta) = \cup_{\qmi \in R_{\mi}^{n-1}(\theta)} \BRi(\theta, \qmi)$.
Therefore, the strategy set $R_i^n(\theta)$ must be an interval in $\Qi$, and $R_i^n(\theta)$ is upper-hemicontinuous in $\theta$ for each round $n$. 

The process of iterated deletion of dominated strategies stops when no strategy can be deleted for any player. Since the game $G_D$ is a dominance solvable game, the rationalizable strategies $\R(\theta) = \prod_{\i \in \I} \R_i(\theta)$ is a singleton set that contains the unique equilibrium strategy profile $\gwe(\theta)$. Therefore, we can conclude that the rationalizable strategy set $R(\theta) = \{\gwe(\theta)\}$ is upper-hemicontinuous in $\theta$, which implies that the unique equilibrium strategy profile $\gwe(\theta)$ is continuous in $\theta$.

Finally, we prove \emph{3}: To begin with, we can show that any solution $\qdiff(\tau)$ of the differential inclusion \eqref{eq:differential_inclusion} satisfies that for any $\i \in \I$, $\lim_{\t \to \infty} \mathrm{dist}\(\qdiff_i(\tau), R^1_i(\theta)\) =0$, where $ R^1_i(\theta)$ is the set of player $\i$'s strategies that survive the first round of deletion of strictly dominated strategies. For any $\i \in \I$, we consider the function $v^1_i(\tau)= \mathrm{dist}\(\qdiff_i(\tau), R^1_i(\theta)\) ^2$. From \emph{1}, we know that the set $R^1_i(\theta)$ is an interval in $\Q_i$. Therefore, we can write $R^1_i(\theta) = [\Rlow_i^1, \Rup_i^1]$, where $\Rlow_i^1$ (resp. $\Rup_i^1$) is the minimum (resp. maximum) strategy in the set $R^1_i(\theta)$. Therefore, 
\begin{align*}
v^1_i(\tau)= \left\{
\begin{array}{ll}
\(\qdiff_i(\tau) - \Rlow^1_i\)^2, & \quad \text{if $\qdiff_i(\tau) <\Rlow^1_i$},\\
0, & \quad \text{if $\qdiff_i(\tau) \in [\Rlow^1_i, \Rup^1_i]$},\\
\(\qdiff_i(\tau) - \Rup^1_i\)^2, & \quad \text{if $\qdiff_i(\tau) > \Rup^1_i$}. 
\end{array}
\right.
\end{align*}
Then, 
\begin{align*}
&\frac{d v_i^1(\tau)}{d \tau} = \frac{d v_i^1(\tau)}{d \qdiff_i(\tau)}\frac{d \qdiff_i(\tau)}{d \tau}\\
\stackrel{\eqref{eq:differential_inclusion}}{\in}&\left\{
\begin{array}{ll}
2 \alpha_i\(\qdiff_i(\tau) - \Rlow^1_i\) \(\BRi(\theta, \qdiff(\tau))-\qdiff_i(\tau)\), &  \quad \text{if $\qdiff_i(\tau) <\Rlow^1_i$},\\
0, & \quad \text{if $\qdiff_i(\tau) \in [\Rlow^1_i, \Rup^1_i]$},\\
2 \alpha_i\(\qdiff_i(\tau) - \Rup^1_i\)  \(\BRi(\theta, \qdiff(\tau))-\qdiff_i(\tau)\), &  \quad \text{if $\qdiff_i(\tau) >\Rup^1_i$},
\end{array}
\right.
\end{align*}
where $\alpha_i \in [\nu, 1]$. Since $\BRi(\theta, \qdiff(\tau)) \in [\Rlow^1_i, \Rup^1_i]$, we have 
\begin{align*}
\frac{d v^1_i(\tau)}{d \tau}  \leq -2 \alpha_i \mathrm{dist}\(\qdiff_i(\tau), R^1_i(\theta)\) ^2 = - 2 \alpha_i v^1_i(\tau) \left\{
\begin{array}{ll}
<0, &  \quad \text{if $\qdiff_i(\tau) <\Rlow^1_i$},\\
=0, & \quad \text{if $\qdiff_i(\tau) \in [\Rlow^1_i, \Rup^1_i]$},\\
<0, &  \quad \text{if $\qdiff_i(\tau) >\Rup^1_i$}.
\end{array}
\right.
\end{align*}
That is, as $\tau$ increases, the value of the function $v^1_i(\tau)$ strictly decreases so long as $\qdiff_i(\tau) \neq R_i^1(\theta)$. Since $v^1_i(\tau) \geq 0$, we must have $\lim_{\t \to \infty}  v^1_i(\tau) =0$ for all $\i \in \I$, i.e. $\lim_{\t \to \infty} \mathrm{dist}\(\qdiff_i(\tau), R^1_i(\theta)\) =0$ for all $\i \in \I$. 

Suppose that in $n$-th round, $\lim_{\t \to \infty} \mathrm{dist}\(\qdiff_i(\tau), R^n_i(\theta)\) =0$ for all $\i \in \I$. Following the same procedure as $n=1$, we can show that the set $R^{n+1}_{\i}$ is an interval $[\Rlow^{n+1}_i, \Rup^{n+1}_i]$, where $\Rlow^{n+1}_i$ (resp. $ \Rup^{n+1}_i$) is the smallest (resp. largest) strategy of player $\i$ after the $n+1$-th iteration. Then, by defining $v^{n+1}_i(\t)= \mathrm{dist}\(\qdiff_i(\tau), R^{n+1}_i(\theta)\) ^2$ for all $\i \in \I$, we can show that $\lim_{\t \to \infty}  v^{n+1}_i(\tau) =0$ and hence  $\lim_{\tau \to \infty} \mathrm{dist}\(\qdiff_i(\tau), R^{n+1}_i(\theta)\) =0$ for all $\i \in \I$. From mathematical induction, we can conclude that $\lim_{\tau \to \infty} \mathrm{dist}\(\qdiff_i(\tau), R_i^n(\theta)\) =0$ for all $\i \in \I$ and all $n\geq 1$. Therefore, $\lim_{\tau \to \infty} \mathrm{dist}\(\qdiff_i(\tau), R_i(\theta)\) =0$. Since $G_D$ is dominance-solvable, we must have $\lim_{\tau \to \infty} \qdiff_i(\tau)= \gwe_i(\theta)$. That is, any solution $\qdiff(\tau)$ of the differential inclusion \eqref{eq:differential_inclusion} converges to the unique equilibrium strategy profile $\gwe(\theta)$. 
\QEDA 

We demonstrate the convergence and stability properties of learning with best response strategies in the three examples introduced in Sec. \ref{sec:eq}. In particular, the Cournot game in Example \ref{ex:potential} and the coordination game in Example \ref{ex:coordinate} are potential games, and the public investment game in Example \ref{ex:supermodular} is a dominance solvable game.
\begin{example}[Cournot competition continued]\label{ex:potential_br}
{\normalfont The Cournot game with parameter $\s \in \S$ introduced in Example \ref{ex:potential} admits a potential function $\Potentials(\q)=  \alpha^\s\(q_1+q_2\) - \beta^\s\(\(\q_1\)^2+ \(\q_2\)^2\) - \beta^\s \q_1 \q_2$, which is concave in $\qi$ for any $\s \in \S$ and any $\i \in \I$. 
We consider stepsizes $a_1^\t=a_2^\t=\frac{1}{\t}$ for all $\t$, which satisfy \textbf{(A3)} -- \textbf{(A4)}. For any $\theta$ and any $\qmi \in \Qmi$, player $\i$ has a unique best-response strategy $\gibr(\theta, \qmi)=\frac{\alphabar(\theta)-\betabar(\theta) \q_{-\i}}{2\betabar(\theta)}$, where $\alphabar(\theta)= \sum_{\s \in \S} \theta(\s) \alphas$ and $\betabar(\theta) = \sum_{\s \in \S}\theta(\s) \betas$. From Proposition \ref{prop:potential}, we know that the best response strategy satisfies \textbf{(A5)} -- \textbf{(A6)}.  

Recall from Example \ref{ex:potential}, the fixed point set of the game is $\FP =\left\{\(\thetasran, \qsran\)=\(\(1, 0\), \(2/3, 2/3\)\),\right. $ $\left. \(\thetabar, \qbar\)= \(\(0.5, 0.5\), \(0.5, 0.5\)\)\right\}$. Fig. \ref{fig:cournot_br_belief_true} - \ref{fig:cournot_br_strategy_true} demonstrate the states of a realization that converges to the complete information fixed point $\(\thetasran, \qsran\)$, and Fig. \ref{fig:cournot_br_belief_fixed} -- \ref{fig:cournot_br_strategy_fixed} illustrate the states of another realization that converges to the other fixed point $\(\thetabar, \loadbar\)$. 

 \begin{figure}[ht]
\centering
    \begin{subfigure}[b]{0.4\textwidth}
        \includegraphics[width=\textwidth]{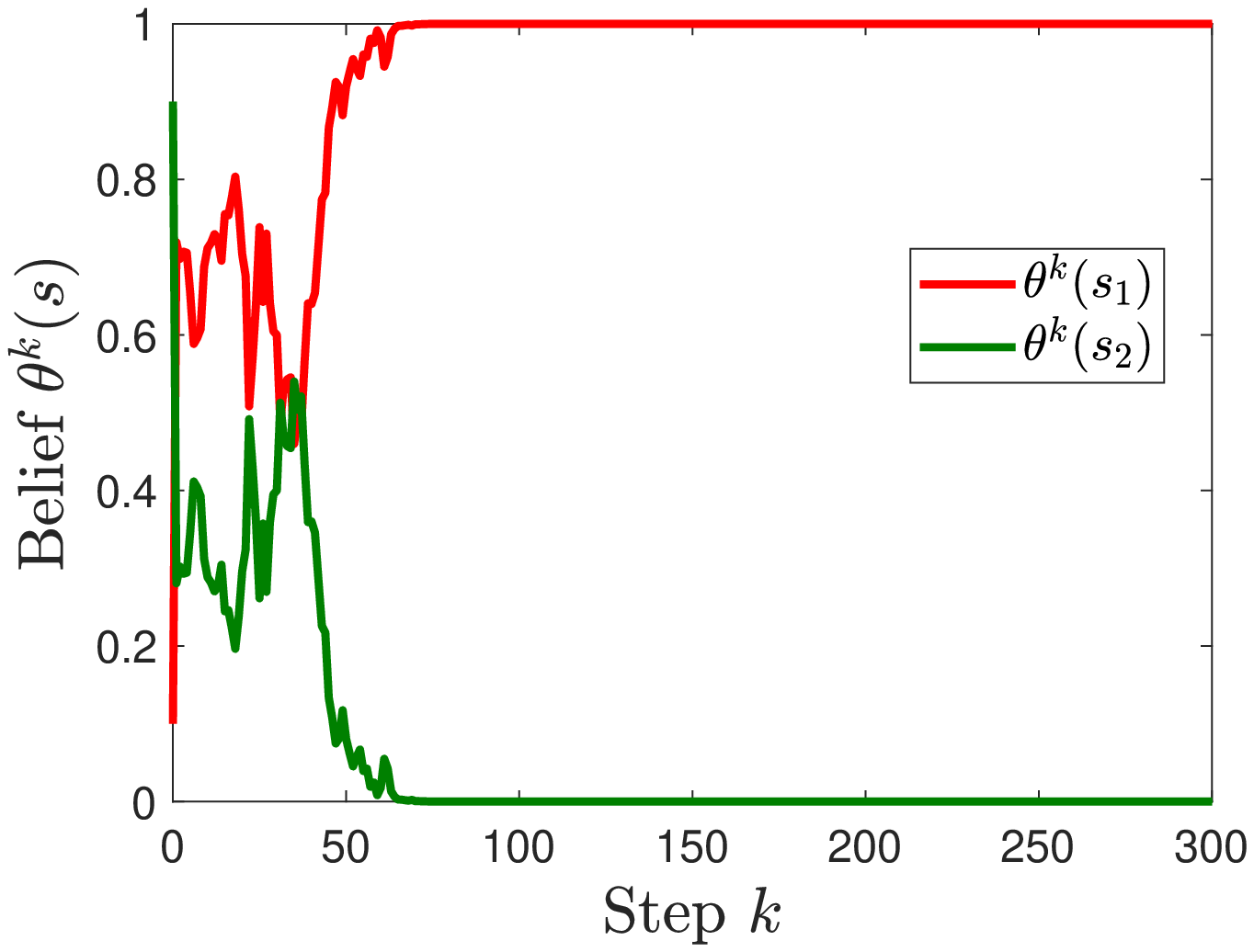}
        \caption{}
        \label{fig:cournot_br_belief_true}
    \end{subfigure}
~
	\begin{subfigure}[b]{0.4\textwidth}
        \includegraphics[width=\textwidth]{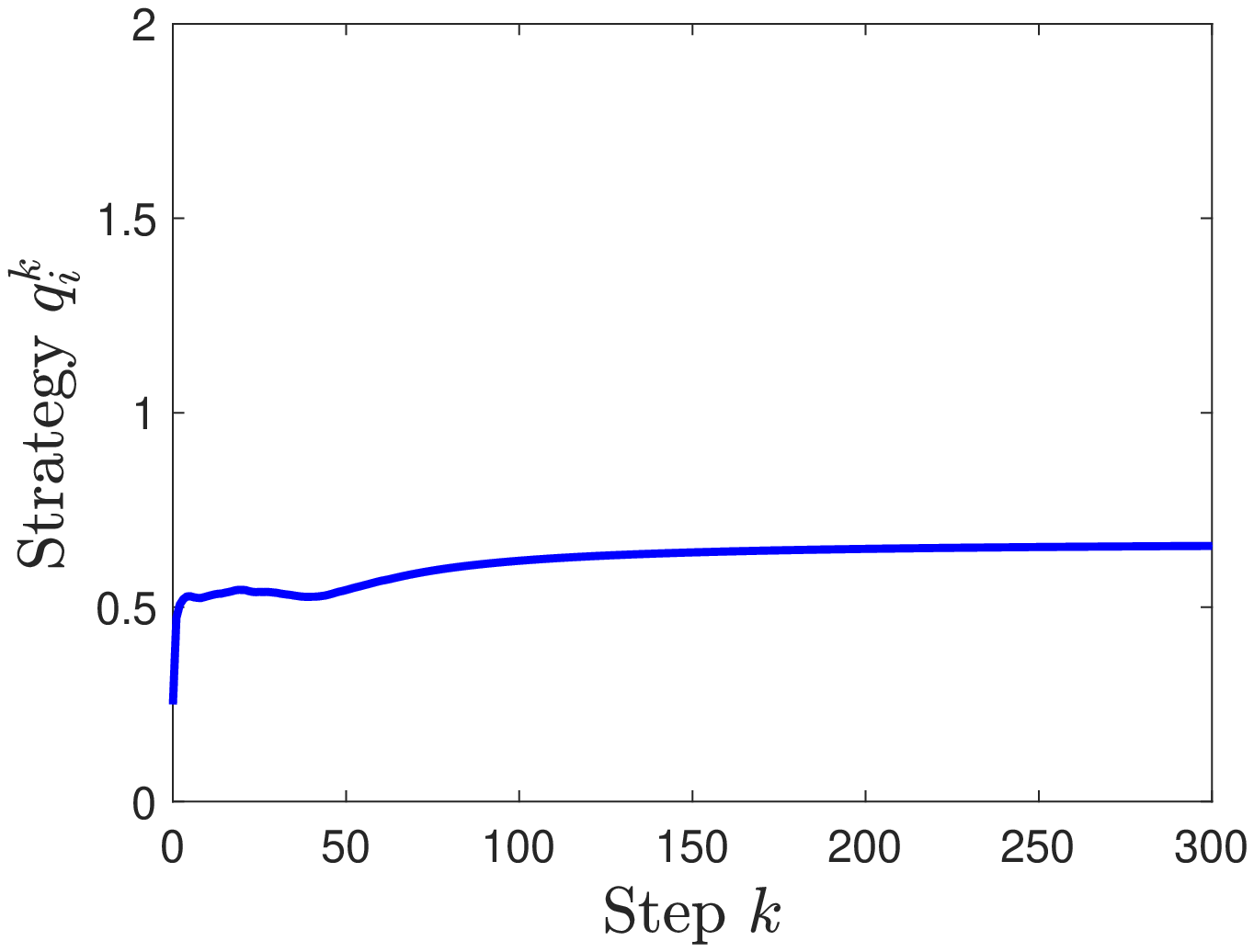}
        \caption{}
        \label{fig:cournot_br_strategy_true}
    \end{subfigure}\\
 \begin{subfigure}[b]{0.4\textwidth}
        \includegraphics[width=\textwidth]{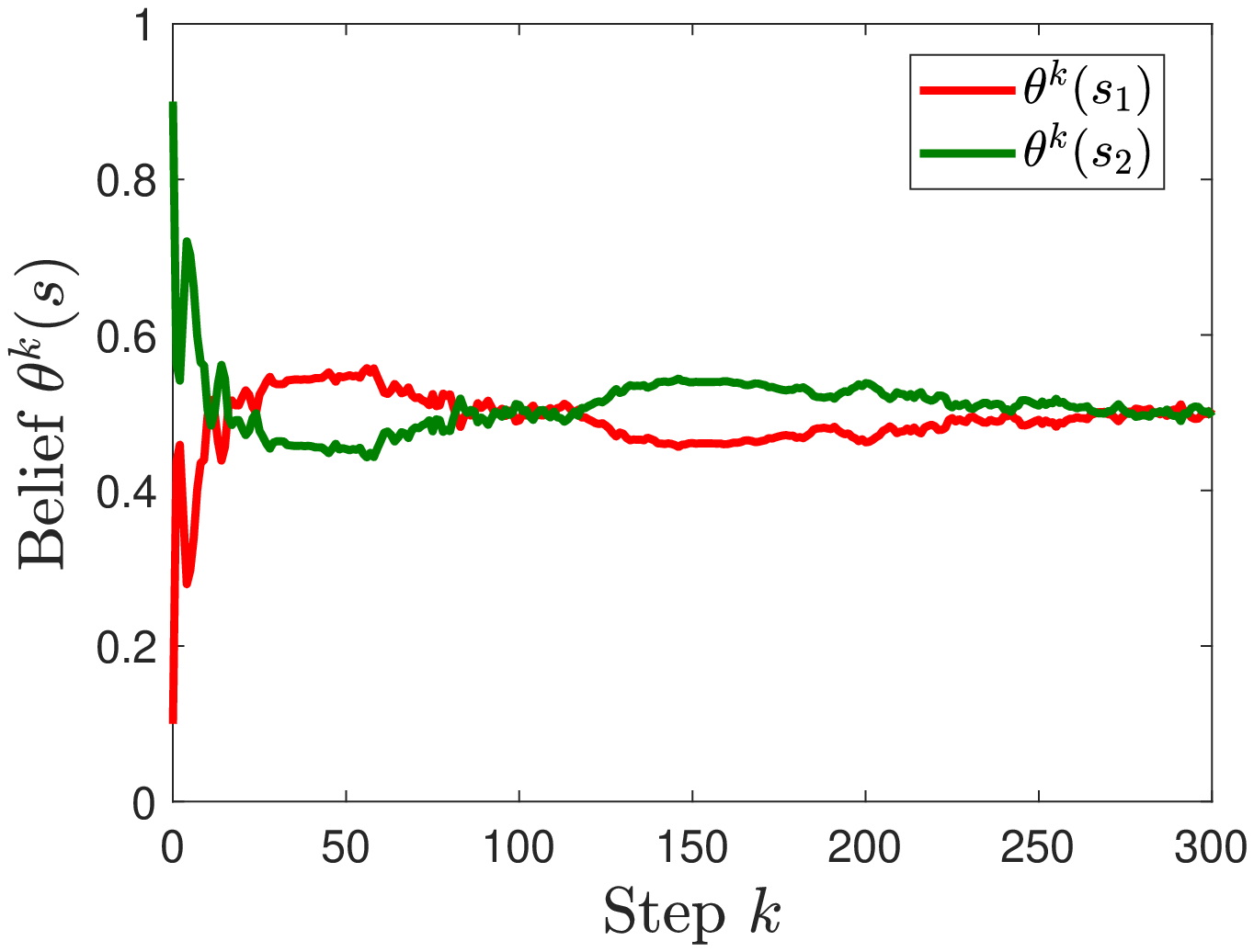}
        \caption{}
        \label{fig:cournot_br_belief_fixed}
    \end{subfigure}
~
	\begin{subfigure}[b]{0.4\textwidth}
        \includegraphics[width=\textwidth]{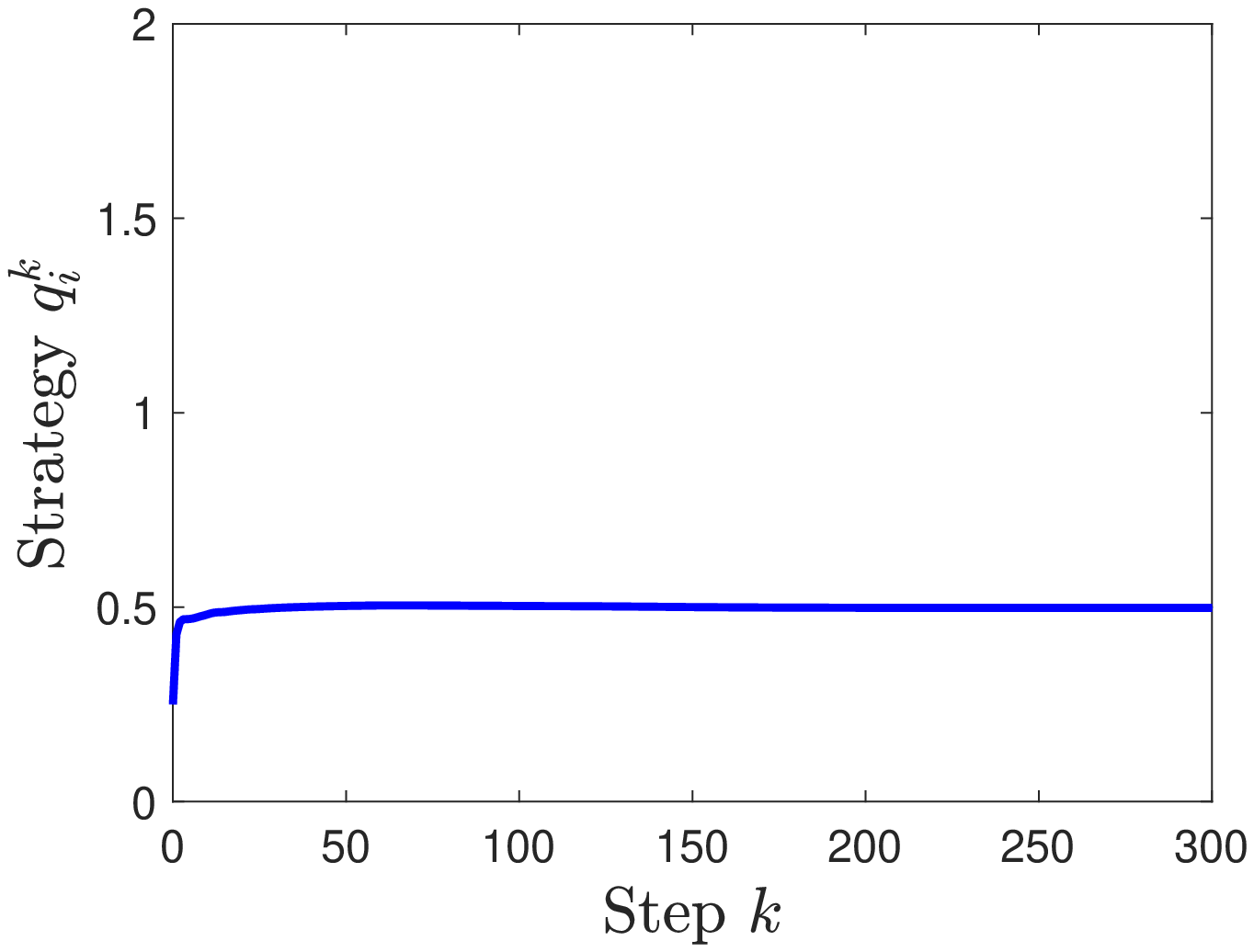}
        \caption{}
        \label{fig:cournot_br_strategy_fixed}
    \end{subfigure}
    \caption{Beliefs and strategies in learning with best response strategies in Cournot game: (a) - (b) Complete information fixed point $\(\thetasran, \qsran\)$; (c) - (d) The other fixed point $\(\thetabar, \qbar\)$.}
\label{fig:convergence_br}
\end{figure}

Recall from Example \ref{ex:stability_cournot}, since the complete information fixed point is not the unique fixed point, there is no globally stable fixed point in learning with best response strategies. The complete information fixed point is locally stable in learning with best response strategy because both conditions in Theorem \ref{theorem:stability_br} are satisfied.

Moreover, we can show that condition (b) in Theorem \ref{theorem:stability_br} is not satisfied by the fixed point $\(\thetabar, \qbar\) = \(\(0.5, 0.5\), \(0.5, 0.5\)\)$. This is because for any $\delta>0$ and any $\q \in N_{\delta}\(\EQ(\thetabar)\) \setminus \{\qbar\}$, $\Sequiv(\q) = \{\sran=\s_1\}$. Since $[\thetabar]=\{\s_1, \s_2\}$, there does not exist $\epsilon, \delta>0$ that satisfy condition (b) in Theorem \ref{theorem:stability_br}. 
}
\end{example}

\begin{example}[Coordination with safe margins continued]\label{ex:coordinate_br}
{\normalfont The coordination game introduced in Example \ref{ex:coordinate} admits a potential function $\Psi^s(\q) = -2 \(\max\(|\q_1 - \q_2|, \s\) - \s\)^2- q_1+\q_2$ for any $\s \in \S$, which is concave in $\qi$ for each $\i \in \I$. From \eqref{eq:EQ_description}, we can write the best response correspondence $\BRi(\theta, \qmi)$ for any $\qmi \in \Q_{\mi}$, any $\i \in \I$ and any $\theta \in \Delta(\S)$: 
\begin{align*}
    \BR_1(\theta, \q_2) &=\left\{ \begin{array}{ll}
    \max\{0, \q_2- \frac{\theta(\s_2) + 3 \theta(\s_3)}{2}-\frac{1}{4}\},& ~ \text{if $\theta(\s_2) + 3 \theta(\s_3) > \frac{5}{2}$,}\\
    \max\{0, \q_2- \frac{2 \theta(\s_2) + 1}{4 (\theta(\s_1)+\theta(\s_2))}\},& ~ \text{if $\theta(\s_1)< \frac{1}{2}$, and $\theta(\s_2) + 3 \theta(\s_3) \leq \frac{5}{2}$,}\\
    \{\q_2- \frac{1}{4\theta(\s_1)}\},& ~ \text{if $\theta(\s_1) \geq \frac{1}{2}$},
    \end{array}\right.\\
    \BR_2(\theta, \q_1) &=\left\{ \begin{array}{ll}
    \{\q_1+ \frac{\theta(\s_2) + 3 \theta(\s_3)}{2}+\frac{1}{4}\}, & ~ \text{if $\theta(\s_2) + 3 \theta(\s_3) > \frac{5}{2}$,}\\
    \{\q_1+ \frac{2 \theta(\s_2) + 1}{4 (\theta(\s_1)+\theta(\s_2))}\},& ~ \text{if $\theta(\s_1)< \frac{1}{2}$, and $\theta(\s_2) + 3 \theta(\s_3) \leq \frac{5}{2}$,}\\
    \{\q_1+ \frac{1}{4\theta(\s_1)}\},& ~ \text{if $\theta(\s_1) \geq \frac{1}{2}$}.
    \end{array}\right.
\end{align*}

The stepsize of player 1 (resp. player 2) is $a_1^\t=1/\t$ (resp. $a_2^\t=1/2\k$) when $\t$ is odd and $a_1^\t=1/2\t$ (resp. $a_2^\t=1/\t$) when $\t$ is even. This sequence of stepsizes satisfies \textbf{(A3)} -- \textbf{(A4)}. From Proposition \ref{prop:potential}, the best response correspondence $\BR(\thetatone, \qt)$ satisfies \textbf{(A5)} -- \textbf{(A6)}.

Fig. \ref{fig:safemargin_br_belief} - \ref{fig:safemargin_br_strategy} demonstrate that the states of learning with best response strategies converge to a complete information fixed point $\(\thetasran=\(0, 0,1\), \qsran=\(0.85, 2.6\)\)$.
 \begin{figure}[ht]
\centering
 \begin{subfigure}[b]{0.4\textwidth}
        \includegraphics[width=\textwidth]{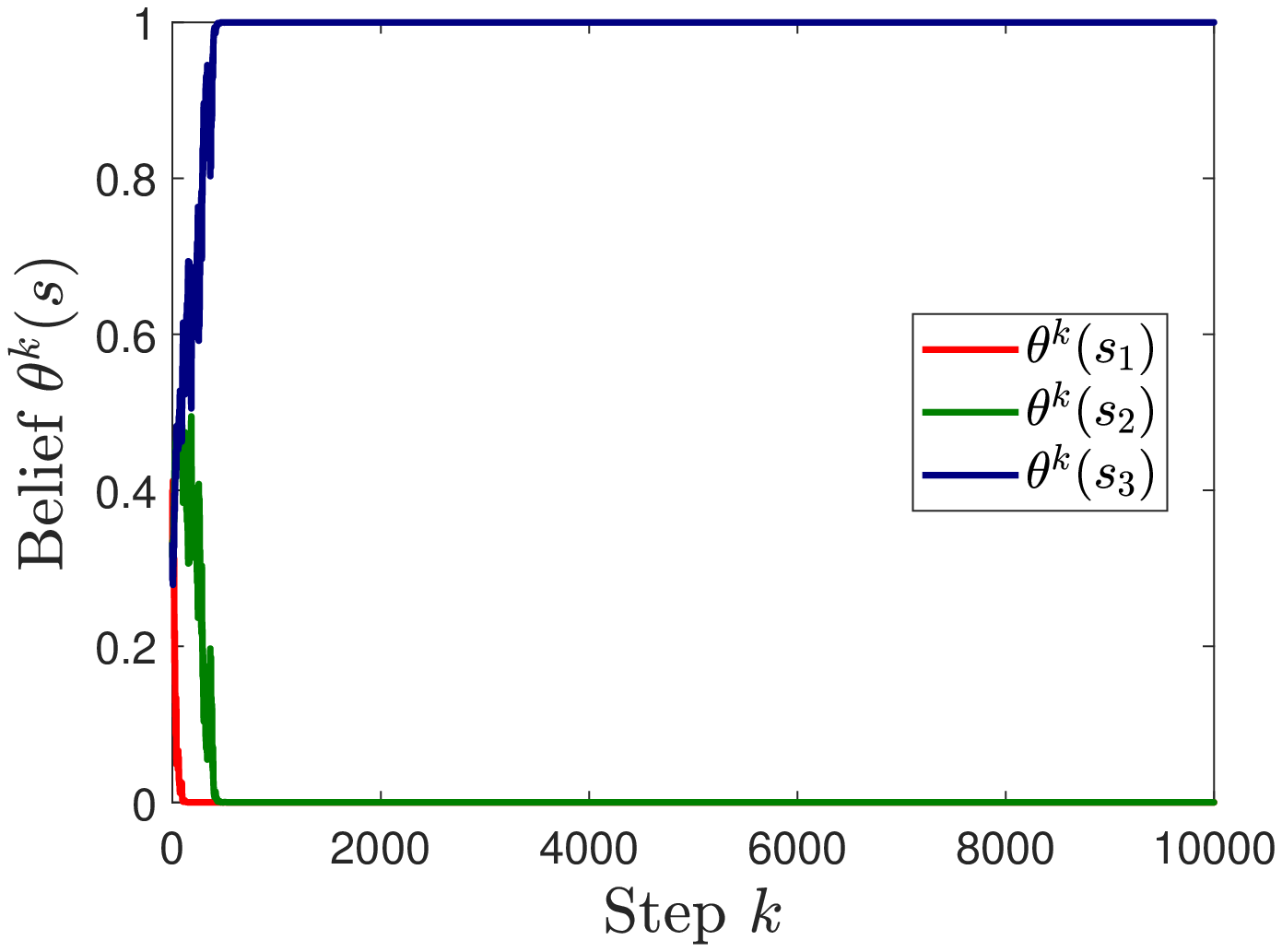}
        \caption{}
        \label{fig:safemargin_br_belief}
    \end{subfigure}
~
	\begin{subfigure}[b]{0.4\textwidth}
        \includegraphics[width=\textwidth]{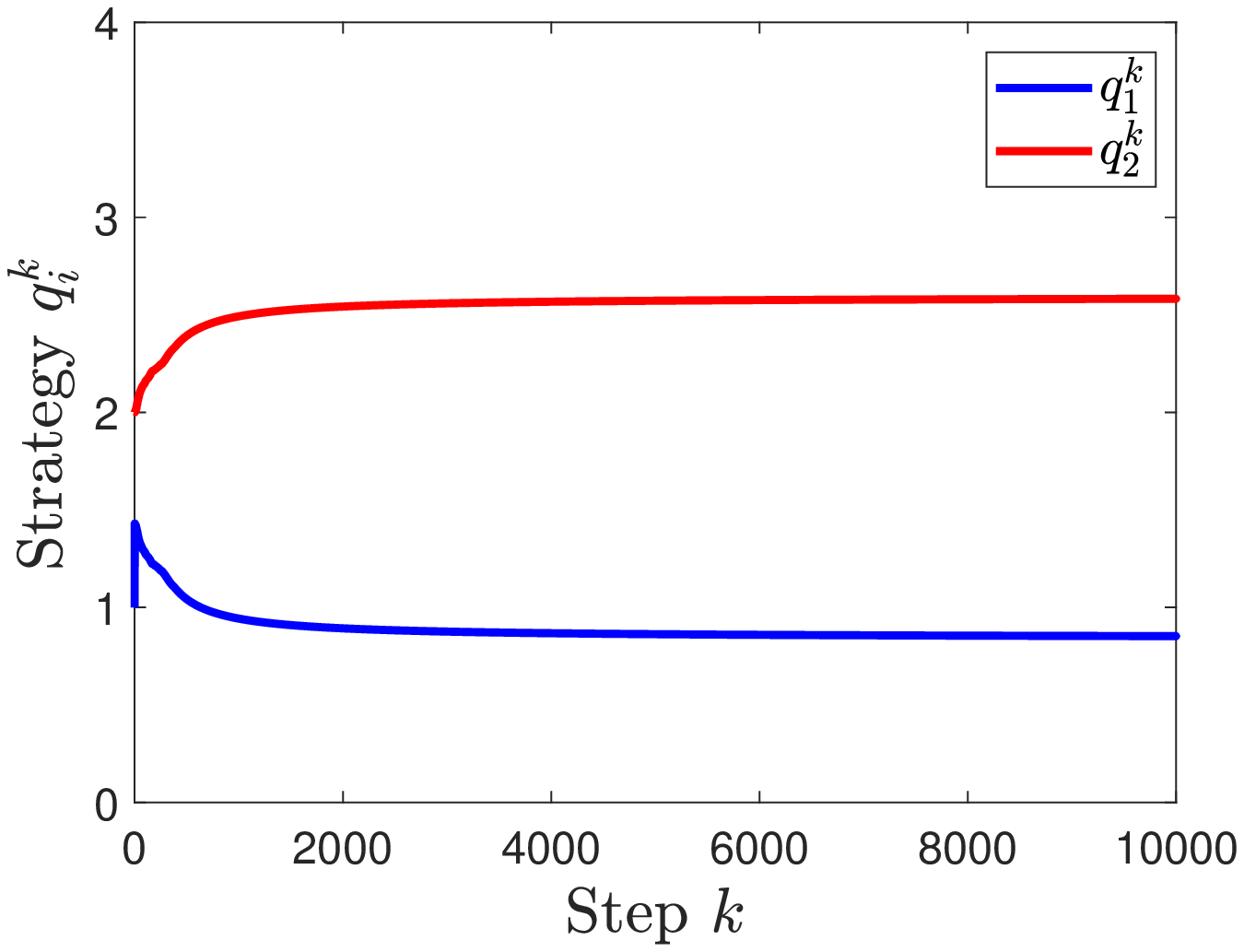}
        \caption{}
        \label{fig:safemargin_br_strategy}
    \end{subfigure}
    \caption{Beliefs and strategies in learning with best response strategies converge to complete information fixed point of coordination game with safe margin.}
\label{fig:convergence_br_safemargin}
\end{figure}
}
\end{example}

\begin{example}[Public good investment continued]\label{ex:supermodular_br}
{\normalfont We can check that the public good investment game introduced in Example \ref{ex:supermodular} is dominance solvable, and each player $\i$'s payoff is concave in their $\qi$. We consider the same stepsizes $\(\ait\)_{\t =1}^{\infty}$ as in Example \ref{ex:supermodular}. This sequence of stepsizes satisfies \textbf{(A3)} -- \textbf{(A4)}. Additionally, given any belief $\theta \in \Delta(\S)$ and any $\qmi\in \Qmi$, each player $\i \in \I$ has a unique best response strategy $\gibr(\theta, \q_{\mi})=\frac{\alphabar(\theta) + \q_{-\i}}{4}$. Since this game is dominance solvable, we know from Proposition \ref{prop:supermodular} that \textbf{(A5)} -- \textbf{(A6)} are also satisfied. Fig. \ref{fig:public_br_belief_true} -- \ref{fig:public_br_strategy_true} illustrate that the states of learning with best-response strategies converge to the unique complete information fixed point. Analogous to Example \ref{ex:supermodular_br}, this unique fixed point is globally stable.
\begin{figure}[htp]
\centering
        \begin{subfigure}[b]{0.4\textwidth}
        \includegraphics[width=\textwidth]{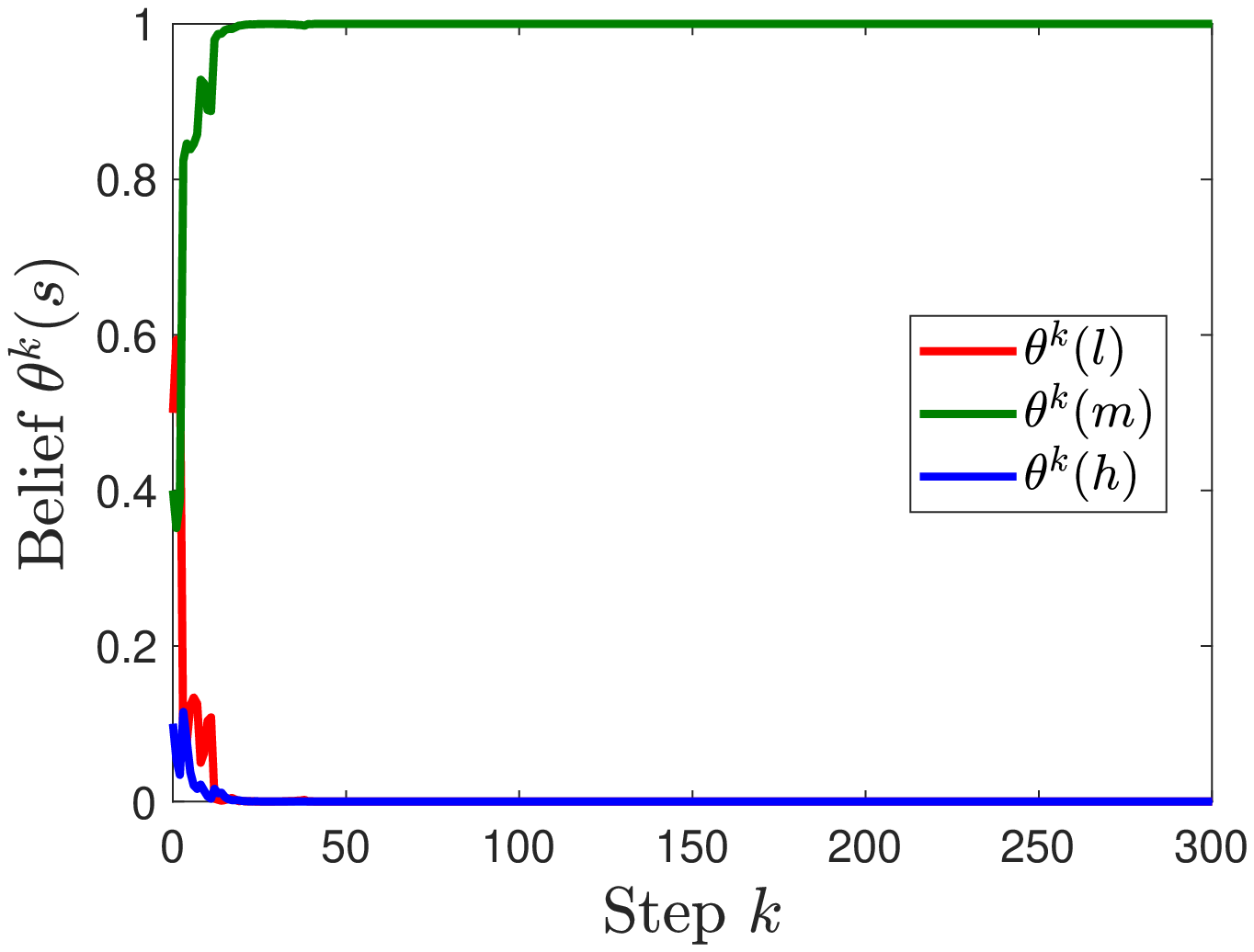}
        \caption{}
        \label{fig:public_br_belief_true}
    \end{subfigure}
~
	\begin{subfigure}[b]{0.4\textwidth}
        \includegraphics[width=\textwidth]{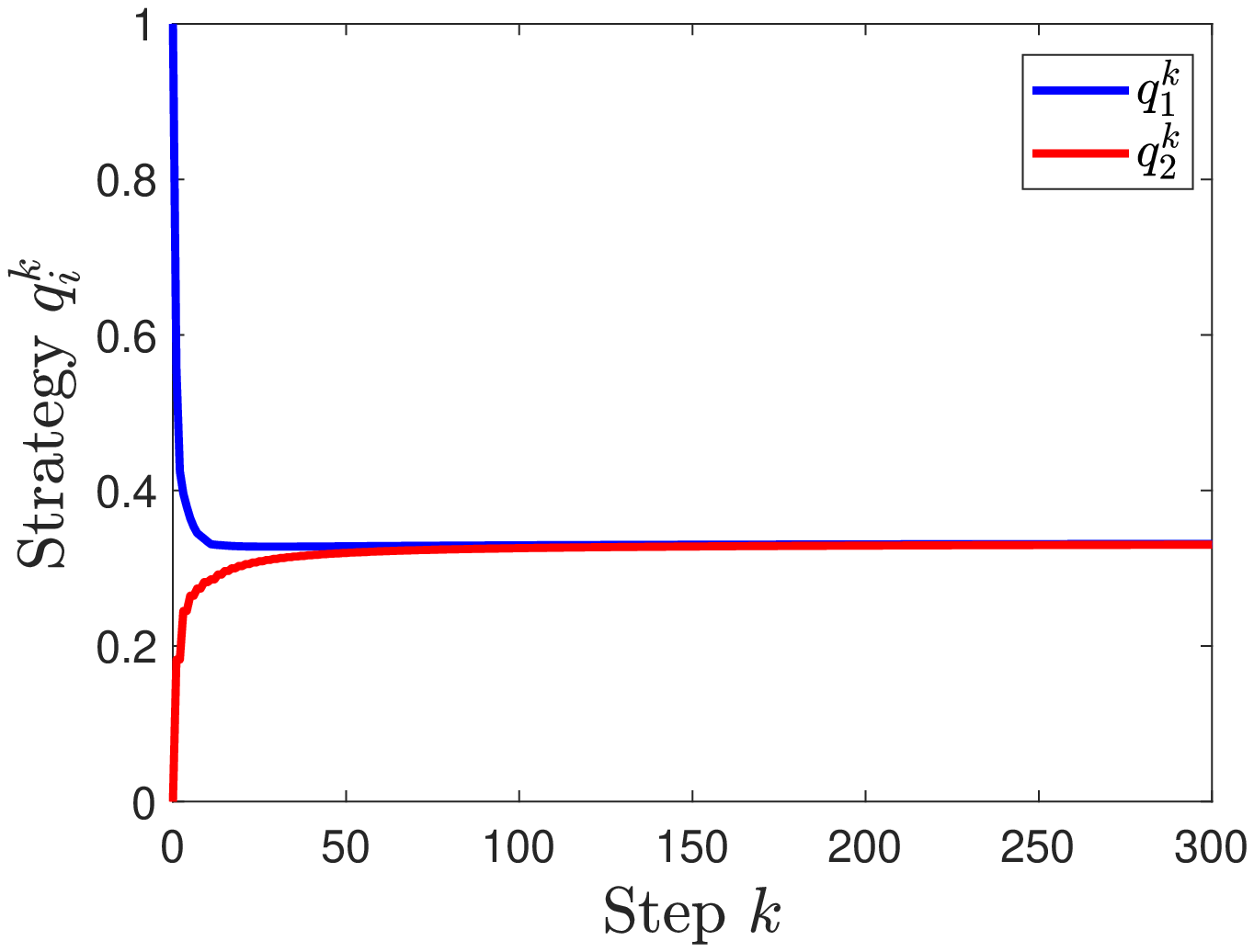}
        \caption{}
        \label{fig:public_br_strategy_true}
    \end{subfigure}
    \caption{Beliefs and strategies in learning with best response strategies converge to the complete information fixed point in public good game.} 
\label{fig:learning_public_br}
\end{figure}}

\end{example}

\section{Convergence to Complete Information Equilibrium}\label{sec:discussion}
We present a sufficient and necessary condition under which all fixed points are complete information fixed points. We also derive a set of sufficient conditions on fixed points and the average payoff functions of the game, which ensure that the strategy played in the fixed point is equivalent to a complete information equilibrium although the fixed point belief may not have complete information of the true parameter. Finally, we discuss how players can learn the true parameter and find the complete information equilibrium when such conditions are not satisfied.

Recall from \eqref{eq:FP}, if all fixed points in $\FP$ are complete information fixed points, then any $\theta \in \Delta(\S)$ other than the complete information belief $\thetasran$ cannot be a fixed point belief. Therefore, from Theorem \ref{theorem:convergence_eq}, all fixed points being complete information fixed points is equivalent to the condition that the support set of any $\theta \in \Delta(\S)\setminus \{\thetasran\}$ contains at least one parameter that is not payoff-equivalent to the true parameter $\sran$ for any equilibrium strategy profile corresponding to $\theta$. 
\begin{corollary}\label{cor:complete_belief}
    The fixed point set $\FP \equiv \{\(\thetasran, \qsran\)|\qsran \in \EQ(\thetasran)\}$ if and only if $[\theta] \setminus \Sequiv\(\q\)$ is non-empty for any $\theta \in \Delta(\S) \setminus \{\thetasran\}$ and any $\q \in \EQ(\theta)$. 
\end{corollary}

From Proposition \ref{prop:complete}, we obtain that the condition in Corollary \ref{cor:complete_belief} is also the sufficient and necessary condition of the existence of globally stable fixed points. Therefore, under this condition, the learning dynamics recovers the complete information environment with probability 1. 

On the other hand, when the condition in Corollary \ref{cor:complete_belief} is not satisfied, there must exist at least one fixed point $\(\thetabar, \qbar\) \in \FP$ such that $\thetabar \neq \thetasran$. In this case, the true parameter $\sran$ is not fully identified by $\thetabar$, and $\qbar$ may or may not be a complete information equilibrium. Next, we present a condition under which the fixed point strategy is a complete information equilibrium even when $\thetabar \neq \thetasran$:
\begin{proposition}\label{prop:complete}
Given fixed point belief $\thetabar$, $\EQ(\thetabar)=\EQ(\thetasran)$ if the payoff function $\usi(\qi, \qmi)$ is concave in $\qi$ for all $\i \in \I$ and all $\s \in [\thetabar]$ and there exists a positive number $\loaddelta >0$ such that $[\thetabar] \subseteq \Sequiv(\q)$ for any $\q \in N_{\delta}\(\EQ(\thetabar)\)$. 
\end{proposition}

We can verify the condition that $\exists \loaddelta >0$ such that $[\thetabar] \subseteq \Sequiv(\q)$ for all $\q \in N_{\delta}\(\EQ(\thetabar)\)$ by analyzing whether or not the KL-divergence $D_{KL}\(\phi^s(\c|\q)||\phi^{\s'}(\c|\q)\)$ between each pair of parameters $\s, \s' \in [\thetabar]$ remains zero under perturbations of $\q \in \EQ(\thetabar)$. 

Proposition \ref{prop:complete} is intuitive: For any fixed point belief $\thetabar$ and any fixed point strategy $\qbar \in \EQ(\thetabar)$, since $\qbar_i$ is a best response strategy of $\qbar_{\mi}$, $\qbar_i$ is a local maximizer of the expected payoff function $\mathbb{E}_{\thetabar}[u_i^s(\qi, \qbar_{\mi})]$. From the condition in Proposition \ref{prop:complete}, we know that the value of the expected payoff function is identical to that with the true parameter $\sran$ for $\qi$ in a small neighborhood of $\qbar_i$. Therefore, $\qbar_i$ must also be a local maximizer of the payoff function with the true parameter $u_i^{\sran}(\qi, \qbar_{\mi})$. Since the payoffs are concave functions of the strategies for any $\s \in [\thetabar]$, $\qbar_i$ is a global maximizer of $u_i^{\sran}(\qi, \qbar_{\mi})$. Thus, any $\qbar \in \EQ(\thetabar)$ is an equilibrium of the game with complete information of $\sran$. We include the formal proof in Appendix \ref{apx:proof}. 


We introduce a modified coordination game that satisfies the condition in Proposition \ref{prop:complete}. 

\begin{example}[Coordination with increasing penalty]\label{ex:concave}
{\normalfont We consider a modified version of the coordination game in Example \ref{ex:coordinate}, in which player 1 chooses strategy $\q_1 \in [0, 2]$ and player 2 chooses strategy $\q_2 \in [1,4]$. Players pay a cost that increases with the difference between their strategies $|\q_1- \q_2|$. In particular, when $|\q_1- \q_2|<1$, the cost is $\(\q_1- \q_2\)^2$. If $|\q_1- \q_2|>1$, then the cost is $\(1+\s\(|\q_1-\q_2|-1\)\)^2$, where each unit of strategy difference that exceeds 1 is penalized by $\s$. The unknown parameter $\s$ is in the set $\S = \{\s_1=2, \s_2 =4\}$, and the true parameter is $\sran=\s_1$. Similar to Example \ref{ex:coordinate}, player 1 prefers to choose small $\q_1$, and player 2 prefers high $q_2$. The payoff of each player is given as follows: 
\begin{align*}
    \c_1 &= \left\{
    \begin{array}{ll}
    -\(\q_1- \q_2\)^2- q_1 +\epsilon_1, & \quad \text{If $|\q_1-\q_2|\leq 1$} \\
    -\(1+\s\(|\q_1-\q_2|-1\)\)^2- q_1 +\epsilon_1, & \quad \text{If $|\q_1-\q_2|> 1$}
    \end{array}
    \right.\\
    \c_2 &= \left\{
    \begin{array}{ll}
    -\(\q_1- \q_2\)^2+q_2 +\epsilon_1, & \quad \text{If $|\q_1-\q_2|\leq 1$} \\
    -\(1+\s\(|\q_1-\q_2|-1\)\)^2+q_2 +\epsilon_1, & \quad \text{If $|\q_1-\q_2|> 1$,}
    \end{array}
    \right.
\end{align*}
where $\epsilon_1$ and $\epsilon_2$ are the noise terms with zero means. 

For any belief $\theta \in \Delta(\S)$, the set of equilibrium strategy $\EQ(\theta)$ is given by: 
\begin{align*}
\EQ(\theta)=\{\(\q_1, \q_2\) \in \Q\left\vert \(\q_2-\q_1\)=1/2\right.\}.
\end{align*}
We can check that the conditions in Proposition \ref{prop:complete} are satisfied in this example. Therefore, even though players do not have complete information of the unknown parameter, any fixed point strategy is a complete information equilibrium.}

\end{example}

If $\thetabar \neq \thetasran$ and the conditions in Proposition \ref{prop:complete} are not satisfied, then the fixed point strategy $\qbar$ may not be a complete information equilibrium. In general, players' payoffs $u^{\sran}_i(\qbar)$ given $\qbar$ may be higher or lower than $u^{\sran}_i(\qsran)$ with a complete information equilibrium $\qsran$. Recall from Example \ref{ex:potential}, the fixed point strategy profile $\qbar=\(0.5, 0.5\)$ has higher payoff for both players than the complete information strategy profile $\qsran=\(2/3, 2/3\)$. However, in many other games, players may want to learn the complete information equilibrium either because the payoffs under the complete information equilibrium are higher than the payoffs given by other fixed point strategies\footnote{In \cite{wu2019learning}, we proved that in routing games on series-parallel networks, average equilibrium costs of players are higher when they have less than perfect information of the network condition.}, or simply because they prefer to identify the true parameter.  

We now outline a procedure that players can use to distinguish the true parameter $\sran$ from other parameters in the set $[\thetabar]$ and learn the complete information equilibrium after the states of the learning dynamics have converged to a fixed point $\(\thetabar, \qbar\)$. Note that any two parameters $\s, \s' \in [\thetabar]$ must be payoff equivalent given the fixed point strategy $\qbar$. Therefore, to distinguish $\s$ from $\s'$ based on the realized payoffs, players need to take strategy profiles for which $\s$ and $\s'$ are not payoff equivalent. One simple example is when the average payoff functions are affine in $\q$, i.e. $\c_i=\alpha^s_i \q+\beta_i^s+\epsilon_i^s$ for all $\i \in \I$, where the vectors $\alpha^\s = \(\alpha^\s_\i\)_{\i\in I}$ and $\beta^\s= \(\beta_i^\s\)_{\i \in \I}$ depend on the unknown parameter $\s$. In this case, players can distinguish all $\s \in [\thetabar] \setminus \{\sran\}$ from the true parameter $\sran$ by repeatedly taking any strategy that is in a local neighborhood of $\qbar$. This is because two affine payoff functions can only have identical values for at most on one strategy profile. Hence $\qbar$ must be the only strategy for which $\s \in [\thetabar] \setminus \{\sran\}$ is payoff equivalent to $\sran$. Moreover, since players' payoff functions are continuous in their strategies, local perturbation of $\qbar$ only changes the players' average payoffs by a small number. 

On the other hand, when the payoffs are nonlinear functions, parameters in $[\thetabar]$ may remain to be payoff equivalent in the local neighborhood of $\qbar$, but result in different payoffs for strategies outside of the neighborhood. Then, local perturbations of $\qbar$ may not identify the true parameter, and players need to repeatedly play strategies in set $\widehat{\Q}^\s \deleq \{\q \in \Q| \s \not \in \Sequiv(\q)\}$ in order to distinguish each $\s \in [\thetabar]$ with the true parameter $\sran$. However, some players' payoffs given $\q \in \widehat{\Q}^\s$ may be significantly lower than their payoffs at the fixed point. Therefore, players need to be incentivized to take such strategies in order to learn the true parameter. The design of such mechanism is beyond the scope of this article.

\section{Learning with MAP and OLS Estimates}\label{sec:other_estimate}
In this section, we highlight that our convergence results in Sec. \ref{subsec:convergence_eq} and \ref{subsec:results_br} can be extended to learning dynamics, where the unknown parameter is in a continuous set, and the parameter is estimated by maximum a posteriori probability (MAP) or ordinary least squares (OLS). 

\vspace{0.2cm}

\noindent\textbf{(1) MAP estimator.} We now consider a continuous and bounded parameter set $\S$. The initial belief $\thetazero(\s)$ is a probability density function of $\s$ on the set $\S$, and $\thetazero(\s)>0$ for all $\s \in \S$. Since the unknown parameter $\s$ is continuous, Bayesian belief update in \eqref{eq:update_belief} is as follows: 
\begin{align*}
\theta^{\t+1}(\s)&= \frac{\thetat(\s)\phibar^\s(\ct|\qt)}{\int_{s \in \S} \thetat(\s) \phibar^{\s}(\ct|\qt) ds}, \quad \forall \s \in \S.
\end{align*}
 Instead of computing the full posterior belief in each step (which entails computing the continuous integration in the denominator of the Bayesian update), we consider learning with maximum a posteriori (MAP) estimator. In each step $\t$, the MAP estimate $\thetatmap$ is in set $\S$ that maximizes the posterior belief of the unknown parameter based on the history of strategies $\(\q^j\)_{j=1}^{\t-1}$ and the history of payoffs $\(\c^j\)_{j=1}^{\t-1}$:
\begin{align}\tag{$\theta_M$-update}\label{eq:map}
\thetatmap = \argmax_{\s \in \S} \thetat(\s) =\argmax_{\s \in \S} \thetazero(\s) \prod_{j=1}^{\t-1} \phi^s(\c^j|\q^j).
\end{align}
Note that if the initial belief $\thetazero$ is a uniform distribution of all parameters, then $\thetatmap$ maximizes the likelihood function based on the history of strategies and payoffs, i.e. $\thetatmap = \argmax_{\s \in \S}  \prod_{j=1}^{\t-1} \phi^s(\c^j|\q^j)$. In this case, $\thetatmap$ is also a maximum likelihood estimate (MLE). 
 
In learning with MAP estimates, the information system updates the MAP estimate in each state according to \eqref{eq:map}, and players update their strategies either with an equilibrium strategy profile $\gwe(\thetat_M) \in \EQ(\thetat_M)$ as in \eqref{eq:eq_update}, or with a best response strategy $\gbr(\thetat_M, \qt) \in \BR(\thetat_M, \qt)$ as in \eqref{eq:br_update}. The results of state convergence as in Theorems \ref{theorem:convergence_eq} and Theorem \ref{prop:asynchronous} can be directly extended to learning with MAP estimate: The following proposition shows that both the MAP estimate and the strategies converge with probability 1. In particular, MAP estimate converges to a payoff equivalent parameter $\thetabar_M$ given the fixed point strategy profile, and the strategies converge to the equilibrium set $\EQ(\thetabar_M)$ of game $\G$ with parameter $\thetabar_M$. 
\begin{proposition}\label{prop:map_convergence}
Under Assumptions \textbf{(A1)} -- \textbf{(A2)}, the sequence of states $\(\thetat_M, \qt\)_{\t=1}^{\infty}$ of learning dynamics \eqref{eq:map} -- \eqref{eq:eq_update} satisfy $\lim_{\t \to \infty} \thetat_M=\thetabar_M$ and $\lim_{\t \to \infty} \qt=\qbar$ with probability 1, where $\thetabar_M \in \Sequiv(\qbar)$ and $\qbar = \gwe(\thetabar_M) \in \EQ(\thetabar_M)$. 

Under Assumptions \textbf{(A3)} -- \textbf{(A6)}, the sequence of states $\(\thetat_M, \qt\)_{\t=1}^{\infty}$ of learning dynamics \eqref{eq:map} -- \eqref{eq:br_update} satisfy $\lim_{\t \to \infty} \thetat_M=\thetabar_M$ and $\lim_{\t \to \infty} \mathrm{dist}\(\qt, \EQ(\thetabar_M)\)=0$ with probability 1, where $\thetabar_M \in \Sequiv(\Qbar)$ and $\Qbar  \subseteq  \EQ(\thetabar_M)$ is the limit set of the strategy sequence $\(\qt\)_{\t=1}^{\infty}$. 
\end{proposition}

The proof of this proposition is included in Appendix \ref{apx:proof}. 

\vspace{0.4cm}
\noindent\textbf{(2) Linear payoff functions and OLS estimate.} Finally, we consider average payoff functions that are affine in the strategy profile, i.e. 
\begin{align}\label{eq:linear_utility}
    \c_i=\(\q, 1\) \s_i+\epsilon_i^\s, \quad \forall \i \in I,
\end{align} 
where $\s_i$ is the $|\qi|+1$-dimensional unknown parameter vector that affects the payoff of player $\i$, and the noise term $\epsilon_i^\s$ is realized from a normal distribution with zero mean and finite variance. The unknown parameter vector is $\s=\(\s_i\)_{\i \in \I}$.

From step $1$ to $\t$, player $\i$'s realized payoff $\(\c_i^j\)_{j=1}^{\t}$ can be written as a linear function of the strategies $\(\q^j\)_{j=1}^{\t}$ in the following matrix form: 
\begin{align*}
    \underbrace{\left(\begin{array}{l}
    \c_i^1\\
    \c_i^2\\
    \vdots\\
    \c_i^\t
    \end{array}\right)}_{Y_i^{\t}} = \underbrace{\left(\begin{array}{ll}
    \q^1, &1\\
    \q^2, &1\\
    \vdots &\vdots\\
    \q^\t,& 1
    \end{array}\right)}_{\tilde{\Q}^{\t}} \s_i+\left(\begin{array}{l}\epsilon_i^1\\
    \epsilon_i^2\\
    \vdots\\
    \epsilon_i^\t
    \end{array}
    \right).
\end{align*}
In each step $\t$, the information system computes an OLS estimator of the parameter, denoted as $\shat^\t=\(\shat_i^\t\)_{\i \in \I}$ where
\begin{align}\label{eq:OLS}\tag{$\shat$ - update}
    \shat_i^\t= \(\(\tilde{\Q}^\t\)' \tilde{\Q}^\t\)^{-1}\(\tilde{\Q}^\t\)'Y_i^{\t}, \quad \forall \i \in \I, \quad \forall \t.
\end{align}

In learning dynamics with OLS estimate and equilibrium strategies (resp. best response strategies), the information system updates the OLS estimator $\shat^\t=\(\shat^\t_i\)_{\i \in \I}$ as in \eqref{eq:OLS}, and players update their strategy profile given by \eqref{eq:eq_update} (resp. \eqref{eq:br_update}) with $\gwe(\shat^\t) \in \EQ(\shat^{\t+1})$ (resp. $\gbr(\shat^{\t+1}, \qt) \in \BR(\shat^{\t+1}, \qt)$) being an equilibrium strategy profile (resp. a best response strategy) in game $\G$ with the updated parameter estimate $\shat^{\t+1}$. 

In fact, the convergence of the OLS estimator $\shat^\t$ as well as the strategy $\qt$ can be viewed as a special case of learning with MAP estimator because of the following well-known result: In each step $\t$, the OLS estimator $\shat^\t$ is identical to the MLE estimator $\thetat_M$ when each player's payoff as in \eqref{eq:linear_utility} is an affine function of the strategy profile plus a noise term with Normal distribution, i.e. 
\begin{align*}
    \thetat_M =\argmax_{\s \in \S} \prod_{j=1}^{\t-1} \phi^{\s}(\c^j|\q^j) = \shat^\t.
\end{align*}

Therefore, we directly obtain the convergence result of learning with OLS estimator from Proposition \ref{prop:map_convergence}: 
\begin{corollary}\label{cor:OLS}
Under Assumptions \textbf{(A1)} -- \textbf{(A2)}, the sequence of states $\(\shat^\t, \qt\)_{\t=1}^{\infty}$ of learning dynamics \eqref{eq:OLS} -- \eqref{eq:eq_update} satisfy $\lim_{\t \to \infty} \shat^\t=\sbar \in \S$ and $\lim_{\t \to \infty} \qt=\qbar$ with probability 1, where $u^{\sbar}_i(\qbar) =u^{\sran}_i(\q)$ for all $\i \in \I$, and $\qbar = \gwe(\sbar) \in \EQ(\sbar)$. 

Under Assumptions \textbf{(A3)} -- \textbf{(A6)}, the sequence of states $\(\shat^\t, \qt\)_{\t=1}^{\infty}$ of learning dynamics \eqref{eq:OLS} -- \eqref{eq:br_update} satisfy $\lim_{\t \to \infty} \shat^\t=\sbar \in \S$ and $\lim_{\t \to \infty} \mathrm{dist}\(\qt, \EQ(\sbar)\)=0$ with probability 1, where $u^{\sbar}_i(\qbar) =u^{\sran}_i(\q)$ for all $\i \in \I$ and all $\q \in \Qbar  \subseteq  \EQ(\thetabar_M)$ and $\limitq$ is the limit set of the strategy sequence $\(\qt\)_{\t=1}^{\infty}$. 
\end{corollary}

\section{Concluding Remarks}
In this article, we studied stochastic learning dynamics induced by a set of strategic players who repeatedly play a game with an unknown parameter. We analyzed the convergence of beliefs and strategies induced by the stochastic dynamics, and derived conditions for local and global stability of fixed points. We also provide a simple condition which guarantees the convergence of strategies to complete information equilibrium. 

A future research question of interest is to analyze how players learn the true parameter efficiently by exploring the payoff distributions of non-equilibrium strategies. As we have mentioned in Sec. \ref{sec:discussion}, when there are one or more parameters that are payoff equivalent to the true parameter at the fixed point, learning the true parameter requires players to take non-equilibrium strategies, which will reduce some players' payoffs in some steps. In our problem, whenever a player chooses a non-equilibrium strategy, the information of the unknown parameter acquired from that players' payoff is known by all players since the belief update is public. Under what scenarios players can efficiently explore strategies given that their payoff information is shared among all players is an interesting question, and worth further investigation. 

Another promising extension of our approach is to study multi-agent reinforcement learning problem from a Bayesian viewpoint. Consider the setting, where the unknown parameter changes over time according to a Markov transition  process. Players may have imperfect or no knowledge of the underlying transition kernel. Our approach can be used to analyze how players can learn the belief estimates of payoffs given each unknown parameter realized from the Markov chain, and adaptively adjust their strategies either with the stationary equilibrium strategy or with a best response strategy that accounts for the stationary distribution of the Markov chain. 

\section*{Acknowledgement}
We thank Daron Acemoglu, Tamer Basar, Vivek Borkar, P.R. Kumar, Shankar Sastry, Demosthenis Teneketzis, John Tsitsiklis, Adam Wierman, Leeat Yariv, Muhammet Yildiz, Georges Zaccour for useful discussions.  We are grateful to speakers and participants at IPAM Summer School on Games and Contracts for Cyber-Physical Security (2015); 27th Jerusalem School in Economic Theory: The Theory of Networks (2016); 9th Workshop on Dynamic Games in Management Science at HEC Montréal (2017); 8th IFAC Workshop on Distributed Estimation and Control in Networked Systems (2019); 57th Annual Allerton Conference on Communication, Control, and Computing (2019); 2nd Annual Conference on Learning for Dynamics and Control at Berkeley (2020).  This research was supported in part by Michael Hammer Fellowship, AFOSR project Building Attack Resilience into Complex Networks, and NSF CAREER award (CNS 1453126).  
\bibliographystyle{plainnat}
\bibliography{library.bib}

\appendix
 \section{Supplementary Proofs}\label{apx:proof}
\noindent\emph{\textbf{Proof of Lemma \ref{lemma:residual_zero}}.} For any $\i \in \I$ and any $\t$, the residual term $\xi_i^\t = \gibr(\thetatone, \qt_{\mi}) - \tilde{h}_i(\thetabar, \qt_{\mi})$ is bounded because both $\gibr(\thetatone, \qt)$ and $\tilde{h}_i(\thetabar, \qt_{\mi})$ are feasible strategies of player $\i \in \I$ in the bounded set $\Qi$. Additionally, 
\begin{align*}&\|\xi_i^\t\| =  \|\gibr(\thetatone, \qt_{\mi}) - \tilde{h}_i(\thetabar, \qt_{\mi})\|\\
=& \left\|\gibr(\thetatone, \qt_{\mi})- \argmin_{\qi \in \BRi(\thetabar, \qt_{\mi})} |\qi-\gibr(\thetatone, \qt_{\mi})|\right\| 
=d\(\gibr(\thetatone, \qt_{\mi}), \BRi(\thetabar, \qt_{\mi})\). \end{align*} 

From Assumption \textbf{(A5)} and the fact that $\lim_{\t \to \infty} \thetat=\thetabar$, we know that for any $\delta>0$, we can find a positive integer $K$ such that for any $\t > K$, $\BRi(\thetatone, \qt_{\mi}) \subseteq N_{\delta}\(\BRi(\thetabar, \qt_{\mi})\)$. Since $\gibr(\thetatone, \qt_{\mi}) \in \BRi(\thetatone, \qt_{\mi})$, we must have $\lim_{\t \to \infty} d\(\gibr(\thetatone, \qt_{\mi}), \BRi(\thetabar, \qt_{\mi})\)=0$ with probability 1. Therefore, we can conclude that $\lim_{\t \to \infty}\xi_i^\t=0$ with probability 1. \QEDA

\vspace{0.2cm}

\noindent\textbf{\emph{Proof of Lemma \ref{lemma:consistency_br}.}} We prove that for any $\s \in \S \setminus \Sequiv(\limitq)$, $\thetabar(\s)=0$ with probability 1. In particular, we consider the following two cases:\\
\emph{(Case 1):} The sequence $\(\qt\)_{\t=1}^{\infty}$ converges to a fixed point strategy $\qbar$. \\
In this case, $\limitq=\{\qbar\}$ is a singleton set, and Lemma \ref{lemma:consistency} directly shows that $\thetabar(\s)=0$ for all $\s \in \S\setminus \Sequiv(\qbar)$. \\
\emph{(Case 2):} The sequence $\(\qt\)_{\t=1}^{\infty}$ does not converge.\\
In this case, for any $\s \in \S \setminus \Sequiv(\limitq)$, there must exist a strategy $\qbar^1 \in \limitq$ such that $\s \notin \Sequiv(\qbar^1)$ (Definition \ref{def:set_payoff}). From Definition \ref{def:limit_point}, we know that we can find a subsequence $\(\q^{\tk_1}\)_{j=1}^{\infty}$ that converges to $\qbar^1$.

If $\phi^{\sran}(\cbar|\qbar^1)$ is absolutely continuous in $\phi^{\s}(\cbar|\qbar^1)$, then we further decompose the remaining strategies into converging subsequences. If the remaining strategies form a converging sequence, then we have decomposed $\(\qt\)_{\t=1}^{\infty}$ into two converging subsequences with two limit points $\{\qbar^1, \qbar^2\}$. Otherwise, we can repeat this process of finding converging subsequences until all strategies in $\(\qt\)_{\t=1}^{\infty}$ are assigned to a subsequence. The total number of subsequences is denoted as $M$. For any $m=1, \dots, M$, and any $\t$, we denote $\tau_m^\t$ as the number of strategies in $\(\qj\)_{j=1}^{\t}$ that belongs to the $m$-th subsequence. Therefore, we can write the log-likelihood ratio as follows: 
\begin{align}
&\lim_{\t\to \infty}\frac{1}{\t}\log \(\frac{\Phi^\s(Y^k|Q^k)}{\Phi^{\sran}(Y^k|Q^k)}\)=  \lim_{\t\to \infty}\frac{1}{\t} \sum_{\j=1}^{\t} \log\(\frac{\phibar^{\s}(\cj|\qj)}{\phibar^{\sran}(\cj|\qj)}\)\notag\\
=&\lim_{\t\to \infty}\frac{\tau_1^{\t}}{\t}\(\frac{1}{\tau_1^\t}\sum_{j=1}^{\tau_1^{\t}} \log\(\frac{\phibar^{\s}(\c^{\tk_1}|\q^{\tk_1})}{\phibar^{\sran}(\c^{\tk_1}|\q^{\tk_1})}\)\)+ \sum_{m=2}^{M}\frac{\tau_m^{\t}}{\t} \(\frac{1}{\tau_m^{\t}}\sum_{j=1}^{\tau_m^{\t}} \log\(\frac{\phibar^{\s}(\c^{\tk_m}|\q^{\tk_m})}{\phibar^{\sran}(\c^{\tk_m}|\q^{\tk_m})}\)\)\label{eq:vague}
\end{align}
Analogous to \eqref{eq:hold}, for each $m$, as the subsequence $\(\q^{\tk_m}\)_{j=1}^{\infty}$ converges to $\qbar_m$, we have: 
\begin{align*}
\lim_{\t \to \infty} \frac{1}{\tau_m^{\t}}\sum_{j=1}^{\tau_m^{\t}} \log\(\frac{\phibar^{\s}(\c^{\tk_m}|\q^{\tk_m})}{\phibar^{\sran}(\c^{\tk_m}|\q^{\tk_m})}\) \left\{
\begin{array}{ll}
<0, & \quad \text{if $\s \notin \Sequiv(\qbar_m)$,}\\
=0, & \quad \text{otherwise.}
\end{array}
\right.
\end{align*}
From the way we construct the subsequences, we know that $\s \notin \Sequiv(\qbar_1)$. Therefore, from \eqref{eq:vague} we have $\lim_{\t \to \infty} \log \(\frac{\Phi^\s(Y^k|Q^k)}{\Phi^{\sran}(Y^k|Q^k)}\)=-\infty$. Hence, analogous to the proof of Lemma \ref{lemma:consistency}, we know that $\lim_{\t \to \infty} \thetat(\s)=0$ with probability 1. Note that this procedure can be applied to any $\s \in \S \setminus \Sequiv(\limitq)$. Hence, we can conclude that $\lim_{\t \to \infty} \thetat(\s)=0$ for any $\s \in \S \setminus \Sequiv(\limitq)$ with probability 1.

Finally, consider the situation where $\phi^{\sran}(\cbar|\qbar^1)$ is not absolutely continuous in $\phi^{\s}(\cbar|\qbar^1)$. Analogous to the analysis in case 2 of Lemma \ref{lemma:consistency}, we can apply Borel-Cantalli lemma on this strategy subsequence to show that there exists a finite step $K>0$ such that $\thetat(\s)=0$ for any $\t>K$ with probability 1. Hence, $\lim_{\t\to \infty} \thetat(\s)=0$ with probability 1.\QEDA

\vspace{0.2cm}

\noindent\textbf{\emph{Proof of Theorem \ref{theorem:stability_br}.}}  For any $\loadepfin>$ and any $\thetaepfin>0$, we define $\deltahat \deleq \min\{\delta, \loadepfin\}$. From condition (b), we know that $[\thetabar] \subseteq \Sequiv(\q)$ for any $\q \in \Qhat$. From condition (a), we can find a positive number $\epsilon'$ such that if $\theta \in N_{\epsilon'}(\thetabar)$, then $\EQ(\theta) \subseteq \Qhat$. We define $ \epsilonhat \deleq \min\left\{\thetaepfin, \epsilon, \epsilon'\right\}$. Then, again from condition (b), $\BR(\theta, \q) \subseteq N_{\delta}\(\EQ(\thetabar)\)$ for any $\theta \in \Thetahat$ and any $\q \in N_{\delta}\(\EQ(\thetabar)\)$. 

Analogous to the proof of Lemma \ref{lemma:stopping_time}, we can show that if $\thetazero(\s)< \rhoone$ for all $\s \in \S \setminus [\thetabar]$ and $|\thetazero(\s)-\thetabar(\s)|<\rhoone$ for all $\s \in [\thetabar]$, then $\pro\(\thetat(\s)\leq \rhotwo, ~\forall \s \in \Sbar, ~ \forall \t\)>\gamma$. The thresholds $\rhoone$ and $\rhotwo$ are from \eqref{eq:rhoone} -- \eqref{eq:rhotwo}.  

Next, consider any $\q^1 \in N_{\delta}\(\EQ(\thetabar)\)$ and any $\thetazero$ that satisfies $|\thetazero(\s)- \thetabar(\s)|< \rhothree$ for all $\s \in [\thetabar]$. Since $[\thetabar] \subseteq \Sequiv(\q)$ for any $\q \in N_{\delta}\(\EQ(\thetabar)\)$, we have $[\thetabar] \subseteq \Sequiv(\q^1)$. Analogous to the proof of Lemma \ref{lemma:other_parameters}, we can show that if $|\thetazero(\s)-\thetabar(\s)|<\rhotwo$ for all $\s \in \Sbar$, then $\theta^2(\s) \in \(\thetabar(\s)-\frac{\epsilonhat}{|\S|}, \thetabar(\s)+\frac{\epsilonhat}{|\S|}\)$ for all $\s \in [\thetabar]$. Moreover, if $\theta^2(\s)<\rhotwo$ for all $\s \in \Sbar$, then $\theta^2 \in N_{\epsilonhat}\(\thetabar\)$. From condition (b), we know that $\BR(\theta^2, \q^1) \in N_{\delta}\(\EQ(\thetabar)\)$. Therefore, the updated strategy $\q^2 \in N_{\delta}\(\EQ(\thetabar)\)$. Condition (b) further ensures that $[\thetabar] \subseteq \Sequiv(\q^2)$. 

By using mathematical induction for $\t>2$, we have: 
\begin{align*}
\pro\(\left.\begin{array}{l}
|\thetat(\s)-\thetabar(\s)|<\frac{\epsilonhat}{|\S|}, ~\forall \s \in [\thetabar], ~\forall \t\\
\qt \in N_{\delta}\(\EQ(\thetabar)\), ~\forall \t
\end{array}\right\vert
\thetat(\s)<\rhotwo, ~\forall \s \in \Sbar, ~\forall \t\)=1.
\end{align*} 
Finally, for any $\gamma \in (0, 1)$, and any $\thetaepfin, \loadepfin >0$, we consider $\loadep =\delta$ and $\thetaep = \min\{\rhoone, \rhothree\}$. If $\thetazero \in N_{\epsilon^1}(\thetabar)$ and $\q^1 \in N_{\delta^1}\(\EQ(\thetabar)\)$, then
\begin{align*}
& \pro\(\thetat \in \Thetahat, ~  \qt \in N_{\delta}\(\EQ(\thetabar)\), ~ \forall \t\)
\geq \pro\(\begin{array}{l}
|\thetat(\s)-\thetabar(\s)|<\frac{\epsilonhat}{|\S|}, ~ \forall \s \in [\thetabar], \forall \t\\
\thetat < \rhotwo, \forall \s \in \Sbar,  \qt \in N_{\delta}\(\EQ(\thetabar)\), \forall \t
\end{array}\) \\
 =& \pro\(\thetat(\s)<\rhotwo, \forall \s \in \Sbar, \forall \t\) \\
 &\qquad \cdot \pro\(\left.\begin{array}{l}
|\thetat(\s)-\thetabar(\s)|<\frac{\epsilonhat}{|\S|}, \forall \s \in [\thetabar], \forall \t\\
 \qt \in N_{\delta}\(\EQ(\thetabar)\), \forall \t
\end{array}\right\vert
\thetat(\s)<\rhotwo, \forall \s \in \Sbar, \forall \t\)\\
=&\pro\(\thetat(\s)<\rhotwo, \forall \s \in \Sbar, \forall \t\)>\gamma.
\end{align*}
Furthermore, from Theorem \ref{prop:asynchronous}, if $\thetat \in \Thetahat$ for all $\t$, then $\lim_{\t \to \infty} \thetat \in \Thetahat \subseteq \neighinftheta(\thetabar)$, $\qt$ converges to the equilibrium set associated with a belief in $\Thetahat$. Since for any $\theta \in \Thetahat$, $\EQ(\theta) \subseteq \Qhat \subseteq N_{\loadepfin}(\EQ(\thetabar))$, $\qt$ must converge to the set $N_{\loadepfin}(\EQ(\thetabar))$. Therefore, 
\begin{align*}
\lim_{\t \to \infty} \pro\(\thetat \in \neighinftheta(\thetabar), ~ \qt \in \neighinfload(\EQ(\thetabar))\) \geq  \pro\(\thetat \in \Thetahat, ~  \qt \in N_{\delta}\(\EQ(\thetabar)\), ~ \forall \t\)> \gamma. 
\end{align*}
Thus, we can conclude that $\(\thetabar, \EQ(\thetabar)\)$ is locally stable. \QEDA

\vspace{0.2cm}

 \vspace{0.2cm}
 
\noindent\textbf{\emph{Proof of Proposition \ref{prop:complete}.}} Since $[\thetabar] \subseteq \Sequiv(\q)$ for any $\q \in N_{\delta}\(\EQ(\thetabar)\)$, we have: 
\begin{align}\label{eq:expected}
    \mathbb{E}_{\thetabar}[u_i^{\s}(\q)] = u_i^{\sran}(\q), \quad \forall \q \in N_{\delta}\(\EQ(\thetabar)\), \quad \forall \i \in \I.
\end{align}
For any $\qbar \in \EQ(\thetabar)$, since $\qbar_i$ is a best response to $\qbar_{\mi}$, $\qbar_i$ must be a local maximizer of $\mathbb{E}_{\thetabar}[u_i^{\s}(\qi, \qbar_{\mi})]$. From \eqref{eq:expected}, $\qbar_i$ is a local maximizer of $u_i^{\sran}(\qi, \qbar_{\mi})$. Since the function $u_i^{\sran}(\qi, \qbar_{\mi})$ is concave in $\qi$, $\qbar_i$ is also a global maximizer of $u_i^{\sran}(\qi, \qbar_{\mi})$, and hence is a best response of $\qbar_{\mi}$ with complete information of $\sran$. Since this argument holds for all $\i \in \I$, $\qbar$ is a complete information equilibrium. Hence, we can conclude that $\EQ(\thetabar)=\EQ(\thetasran)$ \QEDA

\vspace{0.2cm}
\noindent\textbf{\emph{Proof of Proposition \ref{prop:map_convergence}}
.} 
First, since the Bayesian belief $\thetat$ given by \eqref{eq:update_belief} converges to a fixed point belief $\thetabar$ and $\thetat_M = \argmax_{\s \in \S} \thetat(\s)$, we know that $\lim_{\t \to \infty} \thetat_M= \thetabar_M$, where $\thetabar_M = \argmax_{\s \in \S} \thetabar(\s)$ with probability 1. Second, analogous to Lemma \ref{lemma:q_eq}, we can show that $\lim_{\t \to \infty} \qt= \qbar=\gwe(\thetabar_M)$ in learning with equilibrium strategies under \textbf{(A1)} -- \textbf{(A2)}. Additionally, we know from Lemma \ref{lemma:consistency} that $\lim_{\t \to \infty}\thetat(\s) = 0$ for $\s \in \S \setminus \Sequiv(\qbar)$. Therefore, $\thetabar_M = \argmax_{\s \in \S} \thetabar(\s)$ must be in the set $\Sequiv(\qbar)$. Finally, analogous to Lemma \ref{lemma:q_br}, we can show that $\lim_{\t \to \infty} d\(\qt, \EQ(\thetabar_M)\)=0$ in learning with best response strategies under \textbf{(A3)} -- \textbf{(A6)}. Additionally, we know from the proof of Lemma \ref{lemma:constrained_set} that $\lim_{\t \to \infty}\thetat(\s) = 0$ for $\s \in \S \setminus \Sequiv(\Qbar)$, where $\Qbar$ is the limit set of sequence $\(\qt\)_{\t=1}^{\infty}$. Therefore, $\thetabar_M = \argmax_{\s \in \S} \thetabar(\s)$ must be in the set $\Sequiv(\Qbar)$. 
\QEDA

\section{Learning in Games with Finite Strategy Set}\label{apx:extension}
Our results in Sections \ref{sec:eq} -- \ref{sec:br} can be extended to learning in games where strategy sets are finite and atomic players play mixed strategies. In this game, each player $\i$'s action set (pure strategies) is a finite set $\Ai$, and the action profile (pure strategy profile) is denoted as $n=\(\ai\)_{\i \in \I} \in \prod_{\i \in \I} \Ai$. Given an environment parameter $\s$ and any action profile $n$, the distribution of players' payoff $\c$ is $\phi^\s\(\c|n\)$. The true parameter $\sran \in \S$ is unknown. 

Players can choose mixed strategies in the game. We denote player $\i$'s strategy as $\qi = \(\qi(\ai)\)_{\ai \in \Ai} \in \Qi = \Delta\(\Ai\)$, where $\qi(\ai)$ is the probability of choosing the action $\ai$. The mixed strategy set $\Qi$ is bounded and convex.  

In each step $\t$, players' action profile $\actiont =\(\ai^{\t}\)_{\i \in \I}$ is realized from the mixed strategy profile $\qt$. Based on $\actiont$ and the realized payoff vector $\ct$, the information system updates the belief of the parameter according to Bayes' rule: 
\begin{align*}
\theta^{\t+1}(\s)&= \frac{\thetat(\s)\phibar^\s(\ct|\actiont)}{\sum_{s' \in \S} \thetat(\s') \phibar^{\s'}(\ct|\actiont)}, \quad \forall \s \in \S.
\end{align*}
Then, players update their mixed strategy profile with equilibrium strategies as in \eqref{eq:eq_update} or with best-response strategies as in \eqref{eq:br_update}. 

A parameter $\s$ is payoff equivalent to $\sran$ given a mixed strategy profile $\q \in \Q$ if the distribution of payoffs under $\s$ is identical to that under $\sran$ for all action profiles that are assigned with positive probability according to $\q$. Therefore, the payoff equivalent parameter set given $\q$ is defined as $\Sequiv(\q) \deleq \left\{\S\left\vert D_{KL}\(\phi^{\s}\(\c|\n\)||\phi^{\sran}\(\c|\n\)\)=0, ~ \forall \n \in [\q]\right.\right\}$, where $[\q]=\{\N|\q(\n)>0\}$ is the support set of the mixed strategy profile $\q$. In addition, the set of payoff equivalent parameters on the strategy set $\Qbar \subseteq \Q$ is $\Sequiv(\Qbar)=\{\S| \s \in \Sequiv\(\q\), ~ \forall \q \in \Qbar\}$. 

The convergence results in Theorems \ref{theorem:convergence_eq} and \ref{prop:asynchronous} can be readily extended to games with finite strategy sets. In both learning dynamics \eqref{eq:update_belief} -- \eqref{eq:eq_update} and \eqref{eq:update_belief} -- \eqref{eq:br_update}, the beliefs $\(\thetat\)_{\t =1}^{\infty}$ converge to a fixed point belief $\thetabar$ that accurately estimates the payoff distribution for all action profiles that are taken with positive probability, and the strategies $\(\qt\)_{\t=1}^{\infty}$ converge to the equilibrium set $\EQ(\thetabar)$ in game $\G$ with belief $\thetabar$. 

Moreover, the results on local and global stability properties in Theorems \ref{theorem:stability_eq} and \ref{theorem:stability_br} and Proposition \ref{prop:global} still hold for games with finite strategy set. In particular, recall from Theorems \ref{theorem:stability_eq} and \ref{theorem:stability_br}, the sufficient condition for locally stability property in both learning dynamics require that parameters $\s \in [\thetabar]$ remain payoff equivalent to $\sran$ for strategies in a small neighborhood of $\EQ(\q)$. In games with finite strategy set, any neighborhood of a strategy profile must contain mixed strategies that assign positive probability on all action profiles. Therefore, if a parameter $\s$ is payoff equivalent to $\sran$ in that neighborhood, then it must have identical payoff distribution to $\sran$ for all action profiles. Then, such parameter $\s$ is equivalent to the true parameter $\sran$. This implies that in games with finite strategy set, any fixed point that satisfies the sufficient condition of local stability must be a complete information fixed point. 

\end{document}